\documentclass[a4paper,10pt]{iopart}
\usepackage[utf8x]{inputenc}
\usepackage{color}
\usepackage{graphicx}
\usepackage{amssymb}
\usepackage[bookmarks, bookmarksopen, bookmarksnumbered]{hyperref}

\newcommand{\beq}{\begin{equation}}
\newcommand{\eeq}{\end{equation}}
\newcommand{\bea}{\begin{eqnarray}}
\newcommand{\eea}{\end{eqnarray}}
\newcommand{\bit}{\begin{itemize}}
\newcommand{\eit}{\end{itemize}}
\newcommand{\bfi}{\begin{figure}}
\newcommand{\efi}{\end{figure}}
\newcommand{\bfic}{\begin{figure*}}
\newcommand{\efic}{\end{figure*}}
\newcommand{\bce}{\begin{center}}
\newcommand{\ece}{\end{center}}
\newcommand{\bt}{\begin{table}}
\newcommand{\et}{\end{table}}
\newcommand{\btb}{\begin{tabular}}
\newcommand{\etb}{\end{tabular}}

\begin{document}

\hspace{4.5in} \mbox{AEI-2013-214}

\title{Solving the Einstein constraints in periodic spaces with a multigrid approach}

\author{Eloisa Bentivegna}
\address{
Max-Planck-Institut f\"ur Gravitationsphysik\\
Albert-Einstein-Institut \\ 
Am M\"uhlenberg 1, D-14476 Golm \\
Germany}

\begin{abstract}
Novel applications of Numerical Relativity demand for more
flexible algorithms and tools.
In this paper, I develop and test a multigrid solver, based
on the infrastructure provided by the Einstein Toolkit, for elliptic 
partial differential equations on spaces with periodic boundary 
conditions. This type of boundary often characterizes
the numerical representation of cosmological models, where
space is assumed to be made up of identical copies 
of a single fiducial domain, so that only a finite volume
(with periodic boundary conditions at its edges) needs to be simulated.
After a few tests and comparisons with existing codes,
I use the solver to generate initial data for an infinite, periodic, 
cubic black-hole lattice. 
\end{abstract}

\pacs{02.60.Lj, 04.20.Ex, 04.25.dg, 98.80.Jk}

\section{Introduction}

In recent years, the techniques of Numerical Relativity have
more and more frequently been used to explore domains outside
of the usual compact-object applications (for a review,
see e.g.~the collection introduced in~\cite{0264-9381-29-24-240301}).
Numerical studies in classical, late-time cosmology, examples of which 
can be found thoughout the history of Numerical Relativity~\cite{
858, Laguna:1991zs, KurkiSuonio:1993fg, PhysRevD.58.064010, PhysRevD.60.064011},
have also seen a surge both in demand~\cite{Zhao:2009yp,Ellis:2011ys,
0264-9381-28-16-164007,Clarkson:2011zq,Roukema:2006yd,Roukema:2009sp,Ostrowski:2011nx} 
and in results~\cite{Yoo:2012jz,Bentivegna:2012ei}. 

This increase in popularity motivates the need for new, more
general tools for solving Einstein's equation in three (or more)
dimensions. In particular, components such as the solvers of
elliptic partial differential equations (PDEs) are typically 
fine-tuned to the geometry and symmetries of the physical
system that one desires to study. The current freely available 
tools~\cite{lorene,Ansorg:2004ds,Grandclement:2010fk}, designed to tackle asymptotically-flat spaces 
containing only a few compact sources, are therefore unsuitable
to solve the elliptic PDEs that constitute the constraint 
part of the Einstein system in a generic slice of a cosmological
spacetime. The difference between compact-object and cosmological spaces 
is particularly acute under one aspect: on asymptotically-flat
spaces, elliptic PDEs are usually solved with Dirichlet boundary conditions, 
whilst on cosmological spaces, one typically models a fiducial cell
with periodic boundary conditions. It will be shown below that
this aspect has repercussions on the formulation of the problem and,
consequently, on the numerical algorithms necessary to solve it.

In this paper, I develop and test a solver based
on the multigrid paradigm. The solver is built as a component of
the open-source Einstein Toolkit~\cite{Loffler:2011ay}, and is
released as free software\footnote{A copy can be obtained from
the \texttt{git} repository at \texttt{git@bitbucket.org:eloisa/cosmology.git}.}.
Due to the power and flexibility of this paradigm, which does not require
special tuning to the individual properties of different problems, 
examples of multigrid solvers appear regularly in the Numerical-Relativity
literature~\cite{Brandt:1997fk,Brown:2004ma,East:2012zn}. None of 
these works, however, mentions periodic boundary conditions;
an explicit treatment of this aspect constitutes the focus
of this paper.

In the following, I will briefly discuss the properties of elliptic
PDEs in conjunction with periodic boundary conditions in section~\ref{sec:pell}.
I will then present an outline of multigrid techniques in section~\ref{sec:mg}.
In section~\ref{sec:tests}, I will illustrate these concepts and test the
implementation with a number of examples, including Poisson's equation, 
Helmholtz's equation and the system of constraints of General Relativity,
in cases where a solution is known exactly or can be obtained via
an independent code.
In section~\ref{sec:lattice}, I will finally use the solver to generate
a solution of the Einstein constraints representing an infinite, cubic
black hole lattice, which can be compared qualitatively to the 
results of~\cite{Yoo:2012jz}. 
The usual $G=c=1$ convention will be used throughout.

\section{Elliptic problems on spaces without boundaries}
\label{sec:pell}
Elliptic problems with periodic boundary conditions (PBCs) can be
underdetermined with respect to their Dirichlet-boundaries counterparts, 
as the former sort of boundary specification is typically weaker than
the latter. From 
this standpoint, these problems are close to elliptic equations
with Neumann boundary conditions. In this section, I will
identify the degrees of freedom that are left undetermined by
PBCs in different cases, and illustrate some ways in which they
can be fixed. Whilst, in principle, any condition on the 
unconstrained modes is sufficient to close the system and lead
to a unique solution, the details of how and when these conditions
are applied in the solution algorithm are rather important from 
the practical standpoint, as the extra information must come 
at the right time so as not to disturb the convergence process.

\subsection{Well-posedness without Dirichlet boundaries}
\label{sec:wellpos}
Let us examine the problem of well-posedness (thereby referring
to the formulation of a problem which has one, and exactly one
solution) in a few concrete cases.

Let us begin with the linear case. A simple example is Poisson's
equation, which reads:
\beq
\label{eq:poisson}
\Delta f + d = 0
\eeq
First off, let us recall that, similarly to the Neumann 
BCs case~\cite{ChoquetBruhat:2006ni, Briggs:2000fk}, PBCs add an integrability condition to this
equation: since the integral of the laplacian operator of any
sufficiently smooth function will vanish on a periodic domain, then the integral of
$d$ must also vanish over the same region, lest the equation
be insoluble. Now, let us assume we have a solution $\bar f$ of (\ref{eq:poisson});
obviously any $f = \bar f + J$ is also a solution of the
same equation, if $J$
is a constant. The elliptic equation does not constrain this
zero-frequency mode, nor do the PBCs. Somewhere in the course of
the solution process, this information must be supplied by hand,
for instance by rescaling $f$ over the whole domain by prescribing
a desired value $f_{\cal O}$ at a particular point ${\cal O}$:
\beq
\label{eq:reset}
f \to f - f({\cal O}) + f_{\cal O}
\eeq
This is equivalent to the choice of $J=f_{\cal O} - \bar f({\cal O})$.

Another example of linear elliptic PDE is Helmholtz's equation,
which reads:
\beq
\Delta f + c f = 0
\eeq
where $c$ is a constant. Here, it is the integral of $f$ that needs
to be zero in order for it to be a viable solution. Again, we notice
that, once we find a solution $\bar f$, we can generate a family
of other solutions by simply taking $f = k \bar f$, where $k$ is a constant. 
Again, we can resolve this
ambiguity by imposing a condition to the value of $f$ at a point, and
rescaling the solution, effectively fixing $k=f_{\cal O}/\bar f({\cal O})$.

Things begin to be non-trivial when both linear terms are included,
in the inhomogeneous Helmoltz's equation:
\beq
\label{eq:iHeq}
\Delta f + c f + d = 0
\eeq
Here, the integral condition reads:
\beq
\int_{{\cal V}} ( c f + d ) \; {\rm d} V= 0
\eeq
where ${\cal V}$ denotes the periodic cell.
Now, let us assume that $\bar f$ is a solution of (\ref{eq:iHeq}), and 
explore the conditions under which $f = k \bar f + J$ is a solution
too. In principle, there are two:
\bea
\Delta f + c f + d &=& 0 \label{eq:sol}\\
\int_{{\cal V}} ( c f + d ) \; {\rm d} V &=& 0 \label{eq:int}
\eea
which could be used to fix both $k$ and $J$. However, using the fact that 
$\bar f$ is a solution, it is easy to show that (\ref{eq:sol}) implies 
(\ref{eq:int}). We thus cannot use these two conditions independently,
and we again need to supplement the system with extra information, perhaps
through the usual condition that $f$ assume a specific value at a certain
point, i.e.~a constraint on the zero-frequency mode. 
Notice that substituting the definition of $f$ into equation (\ref{eq:sol}),
one obtains:
\beq
k(J)=\frac{cJ}{d}+1
\eeq
If $c$, $k$, and $J$ are spatial constants, this implies, in particular, that
$f=k\bar f+J$ is a solution if and only if $d$ is a constant as well. If that 
is the case, it is also straighforward to prove that (i) any solution is 
a superposition of Fourier modes of wave number zero or $\sqrt{c}$, (ii) the 
amplitude of the zero-frequency mode is equal to $-d/c$ and (iii) the amplitude
of the mode of frequency $\sqrt{c}$ is arbitrary, i.e. extra information
has to be supplied to render the problem well-posed. If, on the other
hand, $d$ is not a constant, $f=k\bar f+J$ is not a solution (unless, of course,
$J=0$). These three statements are proven in~\ref{app:csource}.

Two features of this test must be highlighted: first, the integral equation
does not need to be explicitly enforced as it is not independent from the
corresponding PDE, and it will be progressively better
and better satisfied as the algorithm converges to a solution; and second,
the need to supply additional information to equation (\ref{eq:sol}) in
order to single out a solution depends on the nature of the source term $d$:
if $d$ is a constant, there is a rescaling freedom for the non-zero
frequency components of $f$; if it is not, such freedom does not exist.

Carrying out a similar analysis for a non-linear equation is hardly, if
at all, viable.
I thus employ an empirical approach: in section~\ref{sec:testIH}, I will
try and solve equation (\ref{eq:sol}), with constant $d$, without imposing
any additional constraint, and observe the solver's behavior. Not 
surprisingly, the symptom of an unconstrained zero-frequency
mode is a runaway behavior of $f$, with an exponentially increasing
error norm. 
Further down, in the tests involving the Einstein constraints with PBCs,
I will again start out relaxing the equations with no extra conditions:
if the error norm converges to below a tolerance set by the magnitude
of the truncation error (see below for details), I will assume that 
the system has a single solution $\bar f$, or at least that any other solution
is sufficiently far away from $\bar f$ that it does not disturb the
algorithm's convergence to $\bar f$ via relaxation.

A final observation is that, whilst supplying extra information on, say,
the expected zero-frequency mode is not necessary, it may be helpful to factor
it into the initial guess for $f$, so as to speed up the convergence
to a solution.

\section{Implementation of a multigrid solver for spaces without boundaries}
\label{sec:mg}

\subsection{Elements of multigrid theory}
Elliptic PDEs can be solved with a variety of
strategies, that vary in robustness, convergence speed and degree of 
generality. A fairly universal approach is that embodied by \emph{relaxation
methods}, whose basic idea is to transform an elliptic equation like:
\beq
\label{eq:ellPDE}
L f = 0
\eeq
into a parabolic equation like:
\beq
\dot f = L f
\eeq
the dot denoting a derivative with respect to time.
Clearly, if the latter equation admits stationary solutions, those will
also be solutions of the original problem. One can then start with an 
initial guess for $f$ and solve the initial value problem, hoping to
end up in a stationary state after some amount of time.
Whilst this technique is rather dependable in most circumstances, the
time required to drive $\dot f$ to zero within the required tolerance
becomes quickly prohibitive as the dimension of the numerical grid
used to discretize the problem grows.

At this point, it is useful to introduce a few definitions. The 
\emph{residual} of equation (\ref{eq:ellPDE}) is defined as:
\beq
r := - L f
\eeq
If $\bar f$ is a solution of (\ref{eq:ellPDE}), then its residual will be
exactly zero, and $\dot f = 0$. Otherwise, the relaxation process will act on $f$ 
according to:
\beq
\label{eq:dotf}
\dot f = - r
\eeq
which explains why the convergence slows down as one gets closer
to the solution, as $r$ (and thus the time derivative of $f$)
decreases.
Another useful quantity is the \emph{solution error}, defined as the
difference between a solution $\bar f$ of (\ref{eq:ellPDE}) and a given
iterate $f$:
\beq
\label{eq:serr}
e := \bar f - f
\eeq
\emph{Multigrid} solvers are based on relaxation methods, and are designed
to increase the convergence speed. They do so via two 
improvements:
\bit
\item Since the higher-frequency modes of $f$ relax faster than
the low-frequency ones, one can speed up the algorithm's convergence
by discretizing the equation on a hierarchy of grids, each covering
the same physical domain with a different number of points. Each grid
will be responsible to relax a particular section of $f$'s spectrum:
finer grids will smooth the high-frequency portion of $f$ while coarser
ones will take care of the long-range modes.
\item Since the relaxation speed is high when the residual is high,
but decreases as one approaches the solution,
one could conceive stopping the process as soon as $e$ reaches a 
plateau, and start solving for $e$ directly. The equation satisfied
by $e$ is straightforward to derive starting from the 
definitions, and reads
\beq
\label{eq:MGerr}
L(f+e) - L(f) = r
\eeq
One can then relax this equation until the convergence speed drops,
then move to relax the equation for the error of $e$, and keep repeating
until necessary. 
\eit
These two improvements are then combined together: in the so-called
V-cycle, for instance, one solves (\ref{eq:ellPDE}) on a grid,
moves $\psi$ and $e$ to a coarser grid and solves (\ref{eq:MGerr}),
then updates $\psi$ on the original grid and relaxes (\ref{eq:ellPDE}) 
again. More sophisticated schemes exist that improve on this
basic strategy and generalize it to deeper grid hierarchies~\cite{Briggs:2000fk}.

A variant of the multigrid approach, sometimes referred to as 
\emph{multilevel}~\cite{Briggs:2000fk}, involves a hierarchy of grids in which a subset
of levels (typically the coarser ones) cover the entire domain, the
rest being local fine grids in the regions where more resolution
is needed, so as to extend the multigrid paradigm to 
Adaptive-Mesh-Refimenent (AMR) grids. For a discussion of this
method in the context black-hole initial data, see~\cite{Brown:2004ma}.
In the case of interest here, the notable feature of multilevel schemes
is that the global grids will have periodic boundary conditions, while
the local one will have Dirichlet boundary conditions (typically deriving
from the prolongation of boundary points from the next coarser grid).

The detailed recipe used to solve an equation with a multigrid (or multilevel)
algorithm can vary based on the equation's properties (e.g., linearity);
an overview of this topic can be found in~\cite{Briggs:2000fk}. In the following
subsection, I will describe the actual recipe adopted in this work.

\subsection{Implementation details}
In order to solve generic elliptic equations on periodic spaces, I have
implemented a solver based on the multigrid recipe described in the 
previous subsection. The solver also has the capability to treat
AMR grid hierarchies via a basic multilevel implementation, described below,
which will only be used for the test in section~\ref{sec:TP}.

I have implemented the solver as a component for the ET infrastructure, coupled to 
the Carpet AMR package.
The implementation is rather straightforward: I use a nonlinear Gauss-Seidel relaxation 
algorithm to smooth out the solution on each level, and transfer information
between grids using Carpet's prolongation and restriction operators. I use
fourth-order differencing to discretize the equation; prolongation is obtained
by means of fifth-order interpolation and restriction is a point-wise copy.
Given the nonlinear character of the equation, I adopt the Full
Approximation Scheme (FAS) to solve the error equation. Different grids
are accessed following the Full MultiGrid (FMG) prescription.
When local AMR grids are present, $f$ is relaxed there with simple Gauss-Seidel
iterations and Dirichlet boundary conditions interpolated from the coarser grids;
the error equation is not used on these levels.

When integrals are necessary, I calculate them using the midpoint rule.
The corresponding information is
then injected after each relaxation step, after boundary conditions are applied.

\section{Code tests}
\label{sec:tests}

I discuss three test cases in order both to show the effectiveness of
the solver in different scenarios and illustrate some of the subtleties 
involved in the solution of periodic problems with a multigrid scheme.
The first two examples are Poisson's and Helmholtz's equations with a source term,
i.e.~the first and third case discussed in section~\ref{sec:wellpos}.
I then tackle the Einstein constraint system directly, first using a smooth
solution corresponding to a spherically-symmetric stress-energy source,
and then reproducing the initial data for a pair of equal-mass, non-spinning
black holes in quasicircular orbit. For each problem (with the exception of 
the binary black-hole case), I first discuss the 
corresponding Dirichlet version, to verify the solver's behavior
on familiar grounds, and then solve the PBCs counterpart, highlighting 
the differences between the two cases.

\subsection{Poisson's equation}

Poisson's equation simply reads:
\beq
\Delta f + d = 0
\eeq
where $d$ has no functional dependence on $f$ but is an otherwise arbitrary
function of the spatial coordinates.
We can, for instance, choose it so that a certain function $\bar f$ is a solution:
\beq
d := -\Delta \bar f
\eeq
Let us set, for example:
\bea
\bar f &=& e^{-r^2/2\sigma^2} \\
d &=& -\Delta \bar f = \left(\frac{3}{\sigma^2}-\frac{r^2}{\sigma^4}\right) e^{-r^2/2\sigma^2}
\eea
The corresponding error equation reads:
\beq
\Delta e - r = 0
\eeq
First, let us solve this problem with Dirichlet boundary conditions (derived from the
exact solution) on a cubic domain with outer boundaries at $x=\pm 1$, $y=\pm 1$, 
and $z=\pm 1$ (we will use this domain for all of the examples below). To initialize 
the scheme, I set $\bar f=0$ and $e=0$, and then perform a
FMG cycle on a hierarchy of five refinement levels, all covering the entire domain,
with grid spacings $\Delta = 0.2, 0.1, 0.05, 0.025, 0.0125$. The Gauss-Seidel 
step is applied 50 times on the coarsest level, 10 on the finest level, 50 times on the other levels on 
the downward parts of the cycle, and 10 times otherwise. The error
after each step is illustrated in Figure~\ref{fig:PD}. 
Notice that, as the discretized equation represents the continuum equation
only up to the truncation error, this quantity constitutes an estimate of 
the lowest solution error attainable with a given grid configuration. 
In the following, for all the test cases for which a solution is known,
I simply define the truncation error as the difference between the 
continuum laplacian operator applied to the exact solution and its
finite-difference version. I plot the truncation error on the five grids
along with the solution error for each test, to provide a reference
for when the convergence of the solver is expected to stop. As is
evident from the plots below, this limit is always achieved.

\bfi
\bce
\begingroup
  \makeatletter
  \providecommand\color[2][]{%
    \GenericError{(gnuplot) \space\space\space\@spaces}{%
      Package color not loaded in conjunction with
      terminal option `colourtext'%
    }{See the gnuplot documentation for explanation.%
    }{Either use 'blacktext' in gnuplot or load the package
      color.sty in LaTeX.}%
    \renewcommand\color[2][]{}%
  }%
  \providecommand\includegraphics[2][]{%
    \GenericError{(gnuplot) \space\space\space\@spaces}{%
      Package graphicx or graphics not loaded%
    }{See the gnuplot documentation for explanation.%
    }{The gnuplot epslatex terminal needs graphicx.sty or graphics.sty.}%
    \renewcommand\includegraphics[2][]{}%
  }%
  \providecommand\rotatebox[2]{#2}%
  \@ifundefined{ifGPcolor}{%
    \newif\ifGPcolor
    \GPcolortrue
  }{}%
  \@ifundefined{ifGPblacktext}{%
    \newif\ifGPblacktext
    \GPblacktexttrue
  }{}%
  \let\gplgaddtomacro\g@addto@macro
  \gdef\gplbacktext{}%
  \gdef\gplfronttext{}%
  \makeatother
  \ifGPblacktext
    \def\colorrgb#1{}%
    \def\colorgray#1{}%
  \else
    \ifGPcolor
      \def\colorrgb#1{\color[rgb]{#1}}%
      \def\colorgray#1{\color[gray]{#1}}%
      \expandafter\def\csname LTw\endcsname{\color{white}}%
      \expandafter\def\csname LTb\endcsname{\color{black}}%
      \expandafter\def\csname LTa\endcsname{\color{black}}%
      \expandafter\def\csname LT0\endcsname{\color[rgb]{1,0,0}}%
      \expandafter\def\csname LT1\endcsname{\color[rgb]{0,1,0}}%
      \expandafter\def\csname LT2\endcsname{\color[rgb]{0,0,1}}%
      \expandafter\def\csname LT3\endcsname{\color[rgb]{1,0,1}}%
      \expandafter\def\csname LT4\endcsname{\color[rgb]{0,1,1}}%
      \expandafter\def\csname LT5\endcsname{\color[rgb]{1,1,0}}%
      \expandafter\def\csname LT6\endcsname{\color[rgb]{0,0,0}}%
      \expandafter\def\csname LT7\endcsname{\color[rgb]{1,0.3,0}}%
      \expandafter\def\csname LT8\endcsname{\color[rgb]{0.5,0.5,0.5}}%
    \else
      \def\colorrgb#1{\color{black}}%
      \def\colorgray#1{\color[gray]{#1}}%
      \expandafter\def\csname LTw\endcsname{\color{white}}%
      \expandafter\def\csname LTb\endcsname{\color{black}}%
      \expandafter\def\csname LTa\endcsname{\color{black}}%
      \expandafter\def\csname LT0\endcsname{\color{black}}%
      \expandafter\def\csname LT1\endcsname{\color{black}}%
      \expandafter\def\csname LT2\endcsname{\color{black}}%
      \expandafter\def\csname LT3\endcsname{\color{black}}%
      \expandafter\def\csname LT4\endcsname{\color{black}}%
      \expandafter\def\csname LT5\endcsname{\color{black}}%
      \expandafter\def\csname LT6\endcsname{\color{black}}%
      \expandafter\def\csname LT7\endcsname{\color{black}}%
      \expandafter\def\csname LT8\endcsname{\color{black}}%
    \fi
  \fi
  \setlength{\unitlength}{0.0500bp}%
  \begin{picture}(7200.00,5040.00)%
    \gplgaddtomacro\gplbacktext{%
      \csname LTb\endcsname%
      \put(1342,2254){\makebox(0,0)[r]{\strut{} 0.1}}%
      \put(1342,2464){\makebox(0,0)[r]{\strut{} 0.2}}%
      \put(1342,2674){\makebox(0,0)[r]{\strut{} 0.3}}%
      \put(1342,2884){\makebox(0,0)[r]{\strut{} 0.4}}%
      \put(1342,3094){\makebox(0,0)[r]{\strut{} 0.5}}%
      \put(1342,3304){\makebox(0,0)[r]{\strut{} 0.6}}%
      \put(1342,3514){\makebox(0,0)[r]{\strut{} 0.7}}%
      \put(1342,3724){\makebox(0,0)[r]{\strut{} 0.8}}%
      \put(1342,3934){\makebox(0,0)[r]{\strut{} 0.9}}%
      \put(1342,4144){\makebox(0,0)[r]{\strut{} 1}}%
      \put(1342,4354){\makebox(0,0)[r]{\strut{} 1.1}}%
      \put(1474,4574){\makebox(0,0){\strut{}-1}}%
      \put(2140,4574){\makebox(0,0){\strut{}-0.5}}%
      \put(2807,4574){\makebox(0,0){\strut{} 0}}%
      \put(3473,4574){\makebox(0,0){\strut{} 0.5}}%
      \put(4139,4574){\makebox(0,0){\strut{} 1}}%
      \put(572,3304){\rotatebox{-270}{\makebox(0,0){\strut{}$f$}}}%
      \put(2806,4903){\makebox(0,0){\strut{}$x$}}%
    }%
    \gplgaddtomacro\gplfronttext{%
    }%
    \gplgaddtomacro\gplbacktext{%
      \csname LTb\endcsname%
      \put(7108,2254){\makebox(0,0)[l]{\strut{}-3.5}}%
      \put(7108,2554){\makebox(0,0)[l]{\strut{}-3}}%
      \put(7108,2854){\makebox(0,0)[l]{\strut{}-2.5}}%
      \put(7108,3154){\makebox(0,0)[l]{\strut{}-2}}%
      \put(7108,3454){\makebox(0,0)[l]{\strut{}-1.5}}%
      \put(7108,3754){\makebox(0,0)[l]{\strut{}-1}}%
      \put(7108,4054){\makebox(0,0)[l]{\strut{}-0.5}}%
      \put(7108,4354){\makebox(0,0)[l]{\strut{} 0}}%
      \put(4140,4574){\makebox(0,0){\strut{}-1}}%
      \put(4849,4574){\makebox(0,0){\strut{}-0.5}}%
      \put(5558,4574){\makebox(0,0){\strut{} 0}}%
      \put(6267,4574){\makebox(0,0){\strut{} 0.5}}%
      \put(6976,4574){\makebox(0,0){\strut{} 1}}%
      \put(7745,3304){\rotatebox{-270}{\makebox(0,0){\strut{}$e \cdot 10^5$}}}%
      \put(5558,4903){\makebox(0,0){\strut{}$x$}}%
    }%
    \gplgaddtomacro\gplfronttext{%
    }%
    \gplgaddtomacro\gplbacktext{%
      \csname LTb\endcsname%
      \put(1342,897){\makebox(0,0)[r]{\strut{}$10^{-6}$}}%
      \put(1342,1123){\makebox(0,0)[r]{\strut{}$10^{-5}$}}%
      \put(1342,1350){\makebox(0,0)[r]{\strut{}$10^{-4}$}}%
      \put(1342,1576){\makebox(0,0)[r]{\strut{}$10^{-3}$}}%
      \put(1342,1803){\makebox(0,0)[r]{\strut{}$10^{-2}$}}%
      \put(1342,2029){\makebox(0,0)[r]{\strut{}$10^{-1}$}}%
      \put(1474,484){\makebox(0,0){\strut{} 0}}%
      \put(2260,484){\makebox(0,0){\strut{} 100}}%
      \put(3046,484){\makebox(0,0){\strut{} 200}}%
      \put(3832,484){\makebox(0,0){\strut{} 300}}%
      \put(4617,484){\makebox(0,0){\strut{} 400}}%
      \put(5403,484){\makebox(0,0){\strut{} 500}}%
      \put(6189,484){\makebox(0,0){\strut{} 600}}%
      \put(6975,484){\makebox(0,0){\strut{} 700}}%
      \put(572,1480){\rotatebox{-270}{\makebox(0,0){\strut{}$|e|$}}}%
      \put(4224,154){\makebox(0,0){\strut{}Iteration}}%
    }%
    \gplgaddtomacro\gplfronttext{%
    }%
    \gplbacktext
    \put(0,0){\includegraphics{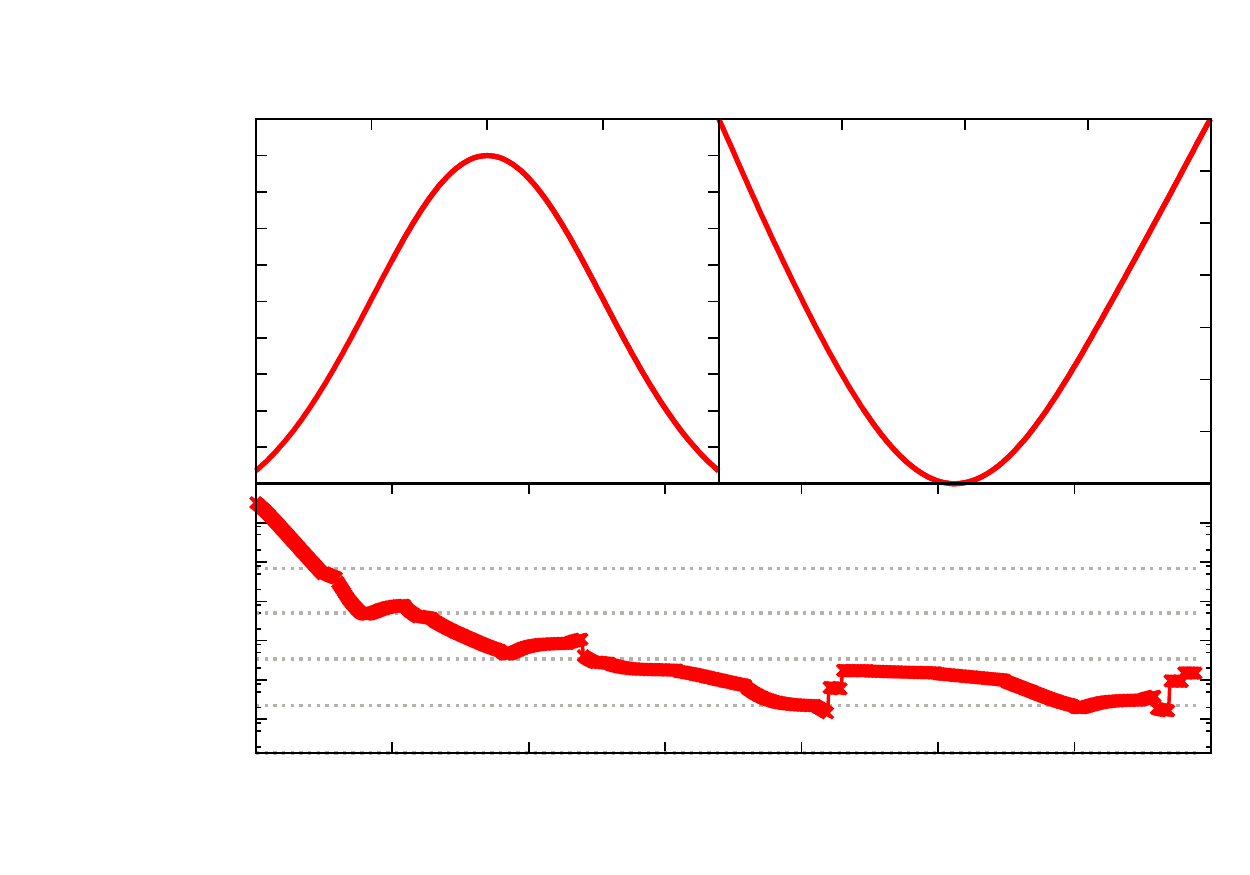}}%
    \gplfronttext
  \end{picture}%
\endgroup

\caption{Top: final solution and associated error for Poisson's equation with
Dirichlet boundary conditions. Bottom: $L_1$ norm of the solution error. 
The dashed grey lines denote the norm of the truncation error on the five grids (the topmost
corresponding to the coarsest grid).
\label{fig:PD}}
\ece
\efi

Let us turn to the PBCs case. Our considerations in the previous section
imply that, in order for this equation to admit a solution, $d$ must integrate to 
zero over the periodic cell. Since $d$ 
is a given, fixed function, it suffices to guarantee that this condition is satisfied 
at the beginning.

As an exact solution to the problem, for instance, one could take:
\beq
\bar f = \cos \pi x \; \cos \pi y \; \cos \pi z
\eeq
provided that $d = \Delta \bar f = -3\pi^2\bar f$, which does indeed have a zero 
integral over the periodic cell.

Let us attempt to obtain a solution with the multigrid solver. 
I again initialize both $f$ and $e$ to zero and begin a FMG cycle. After each
relaxation sweep, I perform a rescaling so that:
\bea
f &\to& f - f({\cal O}) + 1\\
e &\to& e - e({\cal O})
\eea
where $\cal O$ in this case denotes the origin.
As discussed in the previous section,
this rescaling is necessary since the linear equation in combination with PBCs
does not constrain zero-frequency modes in either $f$ or $e$,
so it is necessary to set these constants by hand. 

In this case, there is in fact an additional reason for the rescaling: any constant
mode that creeps into either $f$ or $e$ due to numerical error will grow linearly
in absolute value, since both the equation for $f$ and that for the error take
the form
\beq
\dot f = C
\eeq
for spatially-constant $f$. The scaling is thus necessary not only to make the 
problem well-posed, but also to counteract the danger of runaway behavior.

Inside the FMG cycle, I again relax 50 times in the downward portions and 10 times
on the upward ones, with 50 iterations being executed on the coarsest level and
10 on the finest.
In Figure~\ref{fig:PP}, I show the norm of the true error $e$
as a function of the iteration, along with a plot of the final profiles
for $f$ and $e$.

\bfi
\bce
\begingroup
  \makeatletter
  \providecommand\color[2][]{%
    \GenericError{(gnuplot) \space\space\space\@spaces}{%
      Package color not loaded in conjunction with
      terminal option `colourtext'%
    }{See the gnuplot documentation for explanation.%
    }{Either use 'blacktext' in gnuplot or load the package
      color.sty in LaTeX.}%
    \renewcommand\color[2][]{}%
  }%
  \providecommand\includegraphics[2][]{%
    \GenericError{(gnuplot) \space\space\space\@spaces}{%
      Package graphicx or graphics not loaded%
    }{See the gnuplot documentation for explanation.%
    }{The gnuplot epslatex terminal needs graphicx.sty or graphics.sty.}%
    \renewcommand\includegraphics[2][]{}%
  }%
  \providecommand\rotatebox[2]{#2}%
  \@ifundefined{ifGPcolor}{%
    \newif\ifGPcolor
    \GPcolortrue
  }{}%
  \@ifundefined{ifGPblacktext}{%
    \newif\ifGPblacktext
    \GPblacktexttrue
  }{}%
  \let\gplgaddtomacro\g@addto@macro
  \gdef\gplbacktext{}%
  \gdef\gplfronttext{}%
  \makeatother
  \ifGPblacktext
    \def\colorrgb#1{}%
    \def\colorgray#1{}%
  \else
    \ifGPcolor
      \def\colorrgb#1{\color[rgb]{#1}}%
      \def\colorgray#1{\color[gray]{#1}}%
      \expandafter\def\csname LTw\endcsname{\color{white}}%
      \expandafter\def\csname LTb\endcsname{\color{black}}%
      \expandafter\def\csname LTa\endcsname{\color{black}}%
      \expandafter\def\csname LT0\endcsname{\color[rgb]{1,0,0}}%
      \expandafter\def\csname LT1\endcsname{\color[rgb]{0,1,0}}%
      \expandafter\def\csname LT2\endcsname{\color[rgb]{0,0,1}}%
      \expandafter\def\csname LT3\endcsname{\color[rgb]{1,0,1}}%
      \expandafter\def\csname LT4\endcsname{\color[rgb]{0,1,1}}%
      \expandafter\def\csname LT5\endcsname{\color[rgb]{1,1,0}}%
      \expandafter\def\csname LT6\endcsname{\color[rgb]{0,0,0}}%
      \expandafter\def\csname LT7\endcsname{\color[rgb]{1,0.3,0}}%
      \expandafter\def\csname LT8\endcsname{\color[rgb]{0.5,0.5,0.5}}%
    \else
      \def\colorrgb#1{\color{black}}%
      \def\colorgray#1{\color[gray]{#1}}%
      \expandafter\def\csname LTw\endcsname{\color{white}}%
      \expandafter\def\csname LTb\endcsname{\color{black}}%
      \expandafter\def\csname LTa\endcsname{\color{black}}%
      \expandafter\def\csname LT0\endcsname{\color{black}}%
      \expandafter\def\csname LT1\endcsname{\color{black}}%
      \expandafter\def\csname LT2\endcsname{\color{black}}%
      \expandafter\def\csname LT3\endcsname{\color{black}}%
      \expandafter\def\csname LT4\endcsname{\color{black}}%
      \expandafter\def\csname LT5\endcsname{\color{black}}%
      \expandafter\def\csname LT6\endcsname{\color{black}}%
      \expandafter\def\csname LT7\endcsname{\color{black}}%
      \expandafter\def\csname LT8\endcsname{\color{black}}%
    \fi
  \fi
  \setlength{\unitlength}{0.0500bp}%
  \begin{picture}(7200.00,5040.00)%
    \gplgaddtomacro\gplbacktext{%
      \csname LTb\endcsname%
      \put(1122,2254){\makebox(0,0)[r]{\strut{}-1}}%
      \put(1122,2779){\makebox(0,0)[r]{\strut{}-0.5}}%
      \put(1122,3304){\makebox(0,0)[r]{\strut{} 0}}%
      \put(1122,3829){\makebox(0,0)[r]{\strut{} 0.5}}%
      \put(1122,4354){\makebox(0,0)[r]{\strut{} 1}}%
      \put(1254,4574){\makebox(0,0){\strut{}-1}}%
      \put(1975,4574){\makebox(0,0){\strut{}-0.5}}%
      \put(2697,4574){\makebox(0,0){\strut{} 0}}%
      \put(3418,4574){\makebox(0,0){\strut{} 0.5}}%
      \put(4139,4574){\makebox(0,0){\strut{} 1}}%
      \put(616,3304){\rotatebox{-270}{\makebox(0,0){\strut{}$f$}}}%
      \put(2696,4903){\makebox(0,0){\strut{}$x$}}%
    }%
    \gplgaddtomacro\gplfronttext{%
    }%
    \gplgaddtomacro\gplbacktext{%
      \csname LTb\endcsname%
      \put(7189,2254){\makebox(0,0)[l]{\strut{}-1}}%
      \put(7189,2517){\makebox(0,0)[l]{\strut{} 0}}%
      \put(7189,2779){\makebox(0,0)[l]{\strut{} 1}}%
      \put(7189,3042){\makebox(0,0)[l]{\strut{} 2}}%
      \put(7189,3304){\makebox(0,0)[l]{\strut{} 3}}%
      \put(7189,3567){\makebox(0,0)[l]{\strut{} 4}}%
      \put(7189,3829){\makebox(0,0)[l]{\strut{} 5}}%
      \put(7189,4092){\makebox(0,0)[l]{\strut{} 6}}%
      \put(7189,4354){\makebox(0,0)[l]{\strut{} 7}}%
      \put(4140,4574){\makebox(0,0){\strut{}-1}}%
      \put(4869,4574){\makebox(0,0){\strut{}-0.5}}%
      \put(5599,4574){\makebox(0,0){\strut{} 0}}%
      \put(6328,4574){\makebox(0,0){\strut{} 0.5}}%
      \put(7057,4574){\makebox(0,0){\strut{} 1}}%
      \put(7694,3304){\rotatebox{-270}{\makebox(0,0){\strut{}$e \cdot 10^6$}}}%
      \put(5598,4903){\makebox(0,0){\strut{}$x$}}%
    }%
    \gplgaddtomacro\gplfronttext{%
    }%
    \gplgaddtomacro\gplbacktext{%
      \csname LTb\endcsname%
      \put(1126,861){\makebox(0,0)[r]{\strut{}$10^{-6}$}}%
      \put(1126,1084){\makebox(0,0)[r]{\strut{}$10^{-5}$}}%
      \put(1126,1307){\makebox(0,0)[r]{\strut{}$10^{-4}$}}%
      \put(1126,1530){\makebox(0,0)[r]{\strut{}$10^{-3}$}}%
      \put(1126,1753){\makebox(0,0)[r]{\strut{}$10^{-2}$}}%
      \put(1126,1976){\makebox(0,0)[r]{\strut{}$10^{-1}$}}%
      \put(1258,484){\makebox(0,0){\strut{} 0}}%
      \put(2086,484){\makebox(0,0){\strut{} 100}}%
      \put(2914,484){\makebox(0,0){\strut{} 200}}%
      \put(3742,484){\makebox(0,0){\strut{} 300}}%
      \put(4570,484){\makebox(0,0){\strut{} 400}}%
      \put(5398,484){\makebox(0,0){\strut{} 500}}%
      \put(6226,484){\makebox(0,0){\strut{} 600}}%
      \put(7054,484){\makebox(0,0){\strut{} 700}}%
      \put(356,1480){\rotatebox{-270}{\makebox(0,0){\strut{}$|e|$}}}%
      \put(4156,154){\makebox(0,0){\strut{}Iteration}}%
    }%
    \gplgaddtomacro\gplfronttext{%
    }%
    \gplbacktext
    \put(0,0){\includegraphics{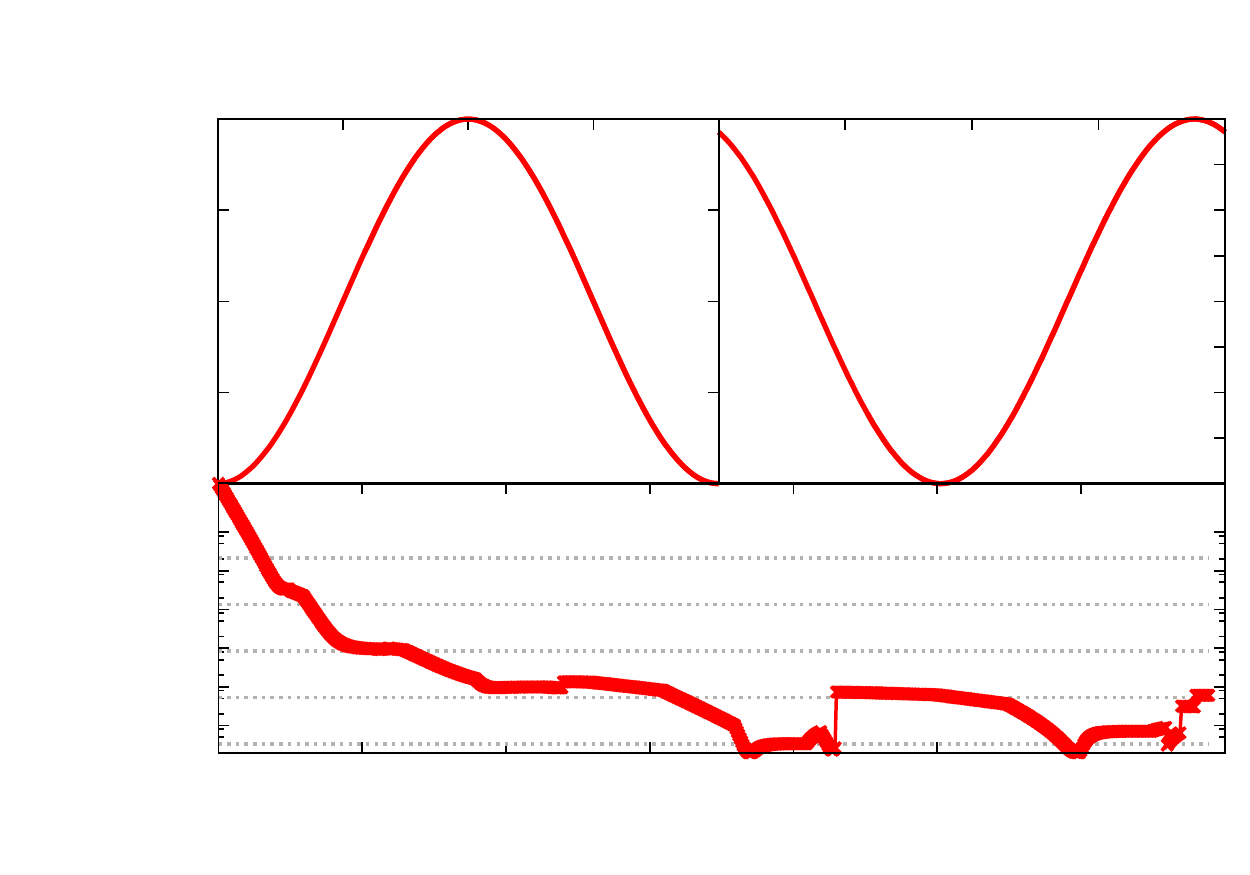}}%
    \gplfronttext
  \end{picture}%
\endgroup

\caption{Top: final solution and associated error for Poisson's equation with
PBCs. Bottom: $L_1$ norm of the solution error.
The dashed grey lines denote the norm of the truncation error on the five grids (the topmost
corresponding to the coarsest grid).
\label{fig:PP}}
\ece
\efi

\subsection{Inhomogeneous Helmholtz's equation}
\label{sec:testIH}

The inhomogeneous Helmholtz's equation is another linear elliptic PDE:
\beq
\Delta f + c f + d = 0\\
\eeq
In the Dirichlet BCs case, we can proceed as in the Poisson's equation's case and set:
\bea
\bar f &=& e^{-r^2/2\sigma^2} \\
d &=& -\Delta \bar f - c \bar f= \left(\frac{3}{\sigma^2}-\frac{r^2}{\sigma^4}-c \right) e^{-r^2/2\sigma^2}
\eea
Let us choose, for instance, $c=-1$ (notice that a positive $c$ would be problematic during
the Gauss-Seidel relaxation as the equation would have exponentially growing modes).
The value of $c$ (implying that of $d$) and the boundary conditions derived from the exact solution are
sufficient to determine the solution. As illustrated in Figure~\ref{fig:HD}, the solver
finds no problem converging to this solution.

\bfi
\bce
\begingroup
  \makeatletter
  \providecommand\color[2][]{%
    \GenericError{(gnuplot) \space\space\space\@spaces}{%
      Package color not loaded in conjunction with
      terminal option `colourtext'%
    }{See the gnuplot documentation for explanation.%
    }{Either use 'blacktext' in gnuplot or load the package
      color.sty in LaTeX.}%
    \renewcommand\color[2][]{}%
  }%
  \providecommand\includegraphics[2][]{%
    \GenericError{(gnuplot) \space\space\space\@spaces}{%
      Package graphicx or graphics not loaded%
    }{See the gnuplot documentation for explanation.%
    }{The gnuplot epslatex terminal needs graphicx.sty or graphics.sty.}%
    \renewcommand\includegraphics[2][]{}%
  }%
  \providecommand\rotatebox[2]{#2}%
  \@ifundefined{ifGPcolor}{%
    \newif\ifGPcolor
    \GPcolortrue
  }{}%
  \@ifundefined{ifGPblacktext}{%
    \newif\ifGPblacktext
    \GPblacktexttrue
  }{}%
  \let\gplgaddtomacro\g@addto@macro
  \gdef\gplbacktext{}%
  \gdef\gplfronttext{}%
  \makeatother
  \ifGPblacktext
    \def\colorrgb#1{}%
    \def\colorgray#1{}%
  \else
    \ifGPcolor
      \def\colorrgb#1{\color[rgb]{#1}}%
      \def\colorgray#1{\color[gray]{#1}}%
      \expandafter\def\csname LTw\endcsname{\color{white}}%
      \expandafter\def\csname LTb\endcsname{\color{black}}%
      \expandafter\def\csname LTa\endcsname{\color{black}}%
      \expandafter\def\csname LT0\endcsname{\color[rgb]{1,0,0}}%
      \expandafter\def\csname LT1\endcsname{\color[rgb]{0,1,0}}%
      \expandafter\def\csname LT2\endcsname{\color[rgb]{0,0,1}}%
      \expandafter\def\csname LT3\endcsname{\color[rgb]{1,0,1}}%
      \expandafter\def\csname LT4\endcsname{\color[rgb]{0,1,1}}%
      \expandafter\def\csname LT5\endcsname{\color[rgb]{1,1,0}}%
      \expandafter\def\csname LT6\endcsname{\color[rgb]{0,0,0}}%
      \expandafter\def\csname LT7\endcsname{\color[rgb]{1,0.3,0}}%
      \expandafter\def\csname LT8\endcsname{\color[rgb]{0.5,0.5,0.5}}%
    \else
      \def\colorrgb#1{\color{black}}%
      \def\colorgray#1{\color[gray]{#1}}%
      \expandafter\def\csname LTw\endcsname{\color{white}}%
      \expandafter\def\csname LTb\endcsname{\color{black}}%
      \expandafter\def\csname LTa\endcsname{\color{black}}%
      \expandafter\def\csname LT0\endcsname{\color{black}}%
      \expandafter\def\csname LT1\endcsname{\color{black}}%
      \expandafter\def\csname LT2\endcsname{\color{black}}%
      \expandafter\def\csname LT3\endcsname{\color{black}}%
      \expandafter\def\csname LT4\endcsname{\color{black}}%
      \expandafter\def\csname LT5\endcsname{\color{black}}%
      \expandafter\def\csname LT6\endcsname{\color{black}}%
      \expandafter\def\csname LT7\endcsname{\color{black}}%
      \expandafter\def\csname LT8\endcsname{\color{black}}%
    \fi
  \fi
  \setlength{\unitlength}{0.0500bp}%
  \begin{picture}(7200.00,5040.00)%
    \gplgaddtomacro\gplbacktext{%
      \csname LTb\endcsname%
      \put(1117,2532){\makebox(0,0)[r]{\strut{} 0.25}}%
      \put(1117,3140){\makebox(0,0)[r]{\strut{} 0.5}}%
      \put(1117,3747){\makebox(0,0)[r]{\strut{} 0.75}}%
      \put(1117,4354){\makebox(0,0)[r]{\strut{} 1}}%
      \put(1249,4574){\makebox(0,0){\strut{}-1}}%
      \put(1973,4574){\makebox(0,0){\strut{}-0.5}}%
      \put(2698,4574){\makebox(0,0){\strut{} 0}}%
      \put(3422,4574){\makebox(0,0){\strut{} 0.5}}%
      \put(4146,4574){\makebox(0,0){\strut{} 1}}%
      \put(479,3304){\rotatebox{-270}{\makebox(0,0){\strut{}$f$}}}%
      \put(2697,4903){\makebox(0,0){\strut{}$x$}}%
    }%
    \gplgaddtomacro\gplfronttext{%
    }%
    \gplgaddtomacro\gplbacktext{%
      \csname LTb\endcsname%
      \put(7331,2254){\makebox(0,0)[l]{\strut{}-10}}%
      \put(7331,2464){\makebox(0,0)[l]{\strut{}-9}}%
      \put(7331,2674){\makebox(0,0)[l]{\strut{}-8}}%
      \put(7331,2884){\makebox(0,0)[l]{\strut{}-7}}%
      \put(7331,3094){\makebox(0,0)[l]{\strut{}-6}}%
      \put(7331,3304){\makebox(0,0)[l]{\strut{}-5}}%
      \put(7331,3514){\makebox(0,0)[l]{\strut{}-4}}%
      \put(7331,3724){\makebox(0,0)[l]{\strut{}-3}}%
      \put(7331,3934){\makebox(0,0)[l]{\strut{}-2}}%
      \put(7331,4144){\makebox(0,0)[l]{\strut{}-1}}%
      \put(7331,4354){\makebox(0,0)[l]{\strut{} 0}}%
      \put(4140,4574){\makebox(0,0){\strut{}-1}}%
      \put(4905,4574){\makebox(0,0){\strut{}-0.5}}%
      \put(5670,4574){\makebox(0,0){\strut{} 0}}%
      \put(6434,4574){\makebox(0,0){\strut{} 0.5}}%
      \put(7199,4574){\makebox(0,0){\strut{} 1}}%
      \put(7968,3304){\rotatebox{-270}{\makebox(0,0){\strut{}$e \cdot 10^6$}}}%
      \put(5669,4903){\makebox(0,0){\strut{}$x$}}%
    }%
    \gplgaddtomacro\gplfronttext{%
    }%
    \gplgaddtomacro\gplbacktext{%
      \csname LTb\endcsname%
      \put(1118,906){\makebox(0,0)[r]{\strut{}$10^{-6}$}}%
      \put(1118,1143){\makebox(0,0)[r]{\strut{}$10^{-5}$}}%
      \put(1118,1380){\makebox(0,0)[r]{\strut{}$10^{-4}$}}%
      \put(1118,1616){\makebox(0,0)[r]{\strut{}$10^{-3}$}}%
      \put(1118,1853){\makebox(0,0)[r]{\strut{}$10^{-2}$}}%
      \put(1118,2090){\makebox(0,0)[r]{\strut{}$10^{-1}$}}%
      \put(1250,484){\makebox(0,0){\strut{} 0}}%
      \put(2099,484){\makebox(0,0){\strut{} 100}}%
      \put(2947,484){\makebox(0,0){\strut{} 200}}%
      \put(3796,484){\makebox(0,0){\strut{} 300}}%
      \put(4645,484){\makebox(0,0){\strut{} 400}}%
      \put(5494,484){\makebox(0,0){\strut{} 500}}%
      \put(6342,484){\makebox(0,0){\strut{} 600}}%
      \put(7191,484){\makebox(0,0){\strut{} 700}}%
      \put(348,1480){\rotatebox{-270}{\makebox(0,0){\strut{}$|e|$}}}%
      \put(4220,154){\makebox(0,0){\strut{}Iteration}}%
    }%
    \gplgaddtomacro\gplfronttext{%
    }%
    \gplbacktext
    \put(0,0){\includegraphics{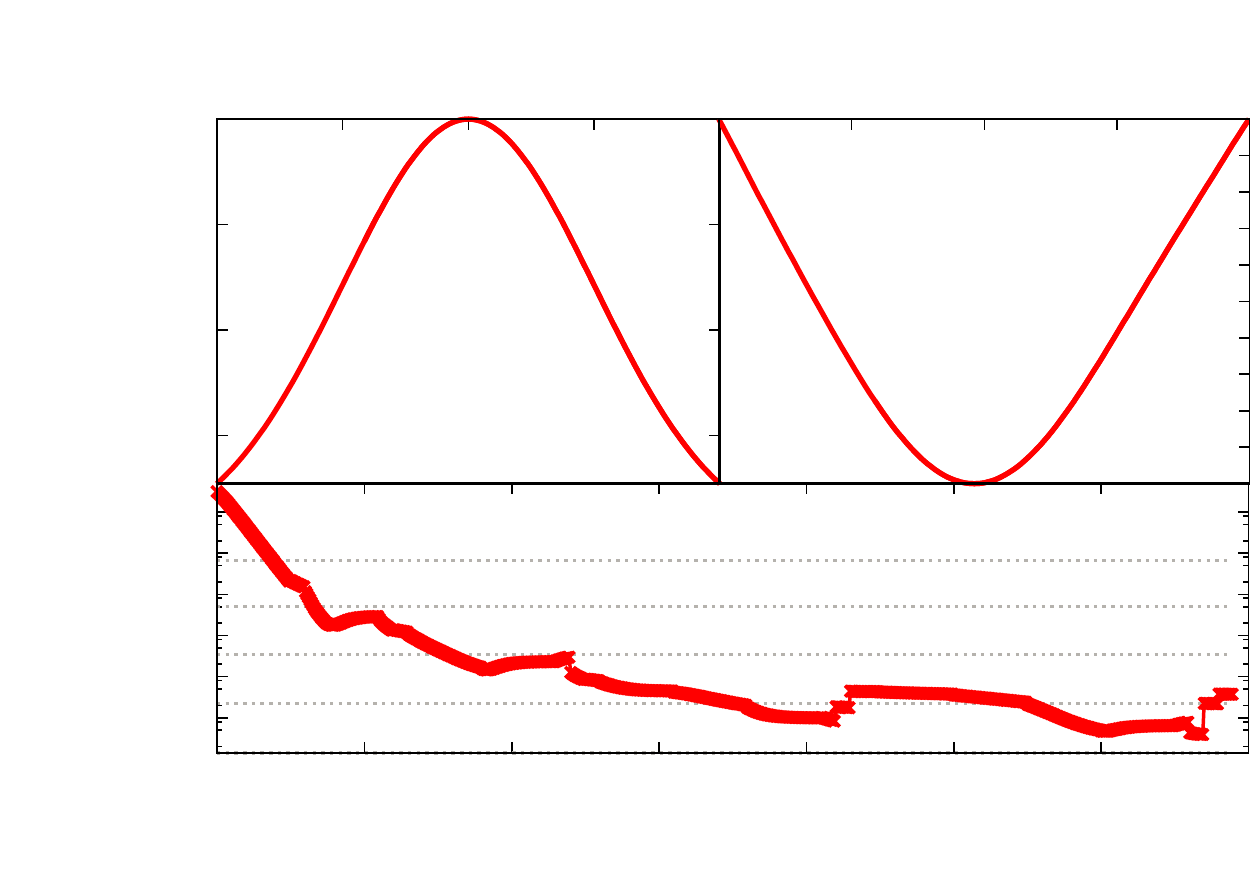}}%
    \gplfronttext
  \end{picture}%
\endgroup

\caption{Top: final solution and associated error for Helmholtz's equation with  
Dirichlet boundary conditions. Bottom: $L_1$ norm of the solution error.
The dashed grey lines denote the norm of the truncation error on the five grids (the topmost
corresponding to the coarsest grid).
\label{fig:HD}}
\ece
\efi

Let us turn to the periodic case: we can again find a solution to this problem by choosing, 
for instance:
\beq
\bar f = 5+\cos \pi x \; \cos \pi y \; \cos \pi z
\eeq
again with $c=-1$ and:
\beq
d = -\Delta \bar f + \bar f= (1 - 3 \pi^2) \cos \pi x \; \cos \pi y \; \cos \pi z + 5 
\eeq
Since $d$ is not constant, according to the discussion in section~\ref{sec:wellpos}
we do not need to worry about specifying a closure relation. 
Figure~\ref{fig:HP} shows that the solver can slowly converge to the exact
solution without aid. However, as discussed above, enforcing the integral 
constraint preconditions the iterate in a way that speeds up convergence
tremendously, as shown in Figure~\ref{fig:HP}. Therefore, here and below,
I will always apply this step even when not strictly necessary from the
wellposedness standpoint. Notice also that the integral condition, which 
in this case reads:
\beq
\int (cf+d) {\rm d}V = 0
\eeq
can be enforced in at least two ways: tuning $c$, or setting the scale of
$f$. In Figure~\ref{fig:HP}, I show the result for both methods, showing that 
rescaling $f$ leads to a slightly superior performance. This observation
also holds true qualitatively for the non-linear examples below.
I conjecture that this is because, if we set the value of $c$ 
after each relaxation step, we are concurrently relaxing
the solution \emph{and} the equation, as $c$ keeps changing too. This makes it harder
for the algorithm to settle to a solution, since the equation itself oscillates between an iteration
and the next. Furthermore, it is arguably more natural to specify the equation's coefficients (including $c$)
and leave the algorithm to find the appropriate scale for the solution, than specifying 
the scale a priori and find out at the end which equation we were trying to solve.

The result for $c=-1$ is to compare to the case when $d$ is constant, 
i.e.~$c=3 \pi^2$, shown in Figure~\ref{fig:HPC}: in this case, the iterate
simply runs away in an exponential fashion, as the residual is insensitive
to the amplitude of the oscillatory part of $f$, and equation (\ref{eq:dotf})
never reaches a stationary state.

\bfi
\bce
\begingroup
  \makeatletter
  \providecommand\color[2][]{%
    \GenericError{(gnuplot) \space\space\space\@spaces}{%
      Package color not loaded in conjunction with
      terminal option `colourtext'%
    }{See the gnuplot documentation for explanation.%
    }{Either use 'blacktext' in gnuplot or load the package
      color.sty in LaTeX.}%
    \renewcommand\color[2][]{}%
  }%
  \providecommand\includegraphics[2][]{%
    \GenericError{(gnuplot) \space\space\space\@spaces}{%
      Package graphicx or graphics not loaded%
    }{See the gnuplot documentation for explanation.%
    }{The gnuplot epslatex terminal needs graphicx.sty or graphics.sty.}%
    \renewcommand\includegraphics[2][]{}%
  }%
  \providecommand\rotatebox[2]{#2}%
  \@ifundefined{ifGPcolor}{%
    \newif\ifGPcolor
    \GPcolortrue
  }{}%
  \@ifundefined{ifGPblacktext}{%
    \newif\ifGPblacktext
    \GPblacktexttrue
  }{}%
  \let\gplgaddtomacro\g@addto@macro
  \gdef\gplbacktext{}%
  \gdef\gplfronttext{}%
  \makeatother
  \ifGPblacktext
    \def\colorrgb#1{}%
    \def\colorgray#1{}%
  \else
    \ifGPcolor
      \def\colorrgb#1{\color[rgb]{#1}}%
      \def\colorgray#1{\color[gray]{#1}}%
      \expandafter\def\csname LTw\endcsname{\color{white}}%
      \expandafter\def\csname LTb\endcsname{\color{black}}%
      \expandafter\def\csname LTa\endcsname{\color{black}}%
      \expandafter\def\csname LT0\endcsname{\color[rgb]{1,0,0}}%
      \expandafter\def\csname LT1\endcsname{\color[rgb]{0,1,0}}%
      \expandafter\def\csname LT2\endcsname{\color[rgb]{0,0,1}}%
      \expandafter\def\csname LT3\endcsname{\color[rgb]{1,0,1}}%
      \expandafter\def\csname LT4\endcsname{\color[rgb]{0,1,1}}%
      \expandafter\def\csname LT5\endcsname{\color[rgb]{1,1,0}}%
      \expandafter\def\csname LT6\endcsname{\color[rgb]{0,0,0}}%
      \expandafter\def\csname LT7\endcsname{\color[rgb]{1,0.3,0}}%
      \expandafter\def\csname LT8\endcsname{\color[rgb]{0.5,0.5,0.5}}%
    \else
      \def\colorrgb#1{\color{black}}%
      \def\colorgray#1{\color[gray]{#1}}%
      \expandafter\def\csname LTw\endcsname{\color{white}}%
      \expandafter\def\csname LTb\endcsname{\color{black}}%
      \expandafter\def\csname LTa\endcsname{\color{black}}%
      \expandafter\def\csname LT0\endcsname{\color{black}}%
      \expandafter\def\csname LT1\endcsname{\color{black}}%
      \expandafter\def\csname LT2\endcsname{\color{black}}%
      \expandafter\def\csname LT3\endcsname{\color{black}}%
      \expandafter\def\csname LT4\endcsname{\color{black}}%
      \expandafter\def\csname LT5\endcsname{\color{black}}%
      \expandafter\def\csname LT6\endcsname{\color{black}}%
      \expandafter\def\csname LT7\endcsname{\color{black}}%
      \expandafter\def\csname LT8\endcsname{\color{black}}%
    \fi
  \fi
  \setlength{\unitlength}{0.0500bp}%
  \begin{picture}(7200.00,6048.00)%
    \gplgaddtomacro\gplbacktext{%
      \csname LTb\endcsname%
      \put(1122,3262){\makebox(0,0)[r]{\strut{} 4}}%
      \put(1122,3787){\makebox(0,0)[r]{\strut{} 4.5}}%
      \put(1122,4312){\makebox(0,0)[r]{\strut{} 5}}%
      \put(1122,4837){\makebox(0,0)[r]{\strut{} 5.5}}%
      \put(1122,5362){\makebox(0,0)[r]{\strut{} 6}}%
      \put(1254,5582){\makebox(0,0){\strut{}-1}}%
      \put(1975,5582){\makebox(0,0){\strut{}-0.5}}%
      \put(2697,5582){\makebox(0,0){\strut{} 0}}%
      \put(3418,5582){\makebox(0,0){\strut{} 0.5}}%
      \put(4139,5582){\makebox(0,0){\strut{} 1}}%
      \put(616,4312){\rotatebox{-270}{\makebox(0,0){\strut{}$f$}}}%
      \put(2696,5911){\makebox(0,0){\strut{}$x$}}%
    }%
    \gplgaddtomacro\gplfronttext{%
    }%
    \gplgaddtomacro\gplbacktext{%
      \csname LTb\endcsname%
      \put(7333,3262){\makebox(0,0)[l]{\strut{}-5}}%
      \put(7333,3525){\makebox(0,0)[l]{\strut{}-4}}%
      \put(7333,3787){\makebox(0,0)[l]{\strut{}-3}}%
      \put(7333,4050){\makebox(0,0)[l]{\strut{}-2}}%
      \put(7333,4312){\makebox(0,0)[l]{\strut{}-1}}%
      \put(7333,4575){\makebox(0,0)[l]{\strut{} 0}}%
      \put(7333,4837){\makebox(0,0)[l]{\strut{} 1}}%
      \put(7333,5100){\makebox(0,0)[l]{\strut{} 2}}%
      \put(7333,5362){\makebox(0,0)[l]{\strut{} 3}}%
      \put(4140,5582){\makebox(0,0){\strut{}-1}}%
      \put(4905,5582){\makebox(0,0){\strut{}-0.5}}%
      \put(5671,5582){\makebox(0,0){\strut{} 0}}%
      \put(6436,5582){\makebox(0,0){\strut{} 0.5}}%
      \put(7201,5582){\makebox(0,0){\strut{} 1}}%
      \put(7838,4312){\rotatebox{-270}{\makebox(0,0){\strut{}$e \cdot 10^7$}}}%
      \put(5670,5911){\makebox(0,0){\strut{}$x$}}%
    }%
    \gplgaddtomacro\gplfronttext{%
    }%
    \gplgaddtomacro\gplbacktext{%
      \csname LTb\endcsname%
      \put(1126,847){\makebox(0,0)[r]{\strut{}$10^{-7}$}}%
      \put(1126,1116){\makebox(0,0)[r]{\strut{}$10^{-6}$}}%
      \put(1126,1384){\makebox(0,0)[r]{\strut{}$10^{-5}$}}%
      \put(1126,1653){\makebox(0,0)[r]{\strut{}$10^{-4}$}}%
      \put(1126,1921){\makebox(0,0)[r]{\strut{}$10^{-3}$}}%
      \put(1126,2189){\makebox(0,0)[r]{\strut{}$10^{-2}$}}%
      \put(1126,2458){\makebox(0,0)[r]{\strut{}$10^{-1}$}}%
      \put(1126,2726){\makebox(0,0)[r]{\strut{}$10^{0}$}}%
      \put(1258,484){\makebox(0,0){\strut{} 0}}%
      \put(2105,484){\makebox(0,0){\strut{} 100}}%
      \put(2951,484){\makebox(0,0){\strut{} 200}}%
      \put(3798,484){\makebox(0,0){\strut{} 300}}%
      \put(4644,484){\makebox(0,0){\strut{} 400}}%
      \put(5491,484){\makebox(0,0){\strut{} 500}}%
      \put(6337,484){\makebox(0,0){\strut{} 600}}%
      \put(7184,484){\makebox(0,0){\strut{} 700}}%
      \put(356,1983){\rotatebox{-270}{\makebox(0,0){\strut{}$|e|$}}}%
      \put(4221,154){\makebox(0,0){\strut{}Iteration}}%
    }%
    \gplgaddtomacro\gplfronttext{%
      \csname LTb\endcsname%
      \put(6197,3090){\makebox(0,0)[r]{\strut{}without constraint}}%
      \csname LTb\endcsname%
      \put(6197,2870){\makebox(0,0)[r]{\strut{}with constraint ($f$)}}%
      \csname LTb\endcsname%
      \put(6197,2650){\makebox(0,0)[r]{\strut{}with constraint ($c$)}}%
    }%
    \gplbacktext
    \put(0,0){\includegraphics{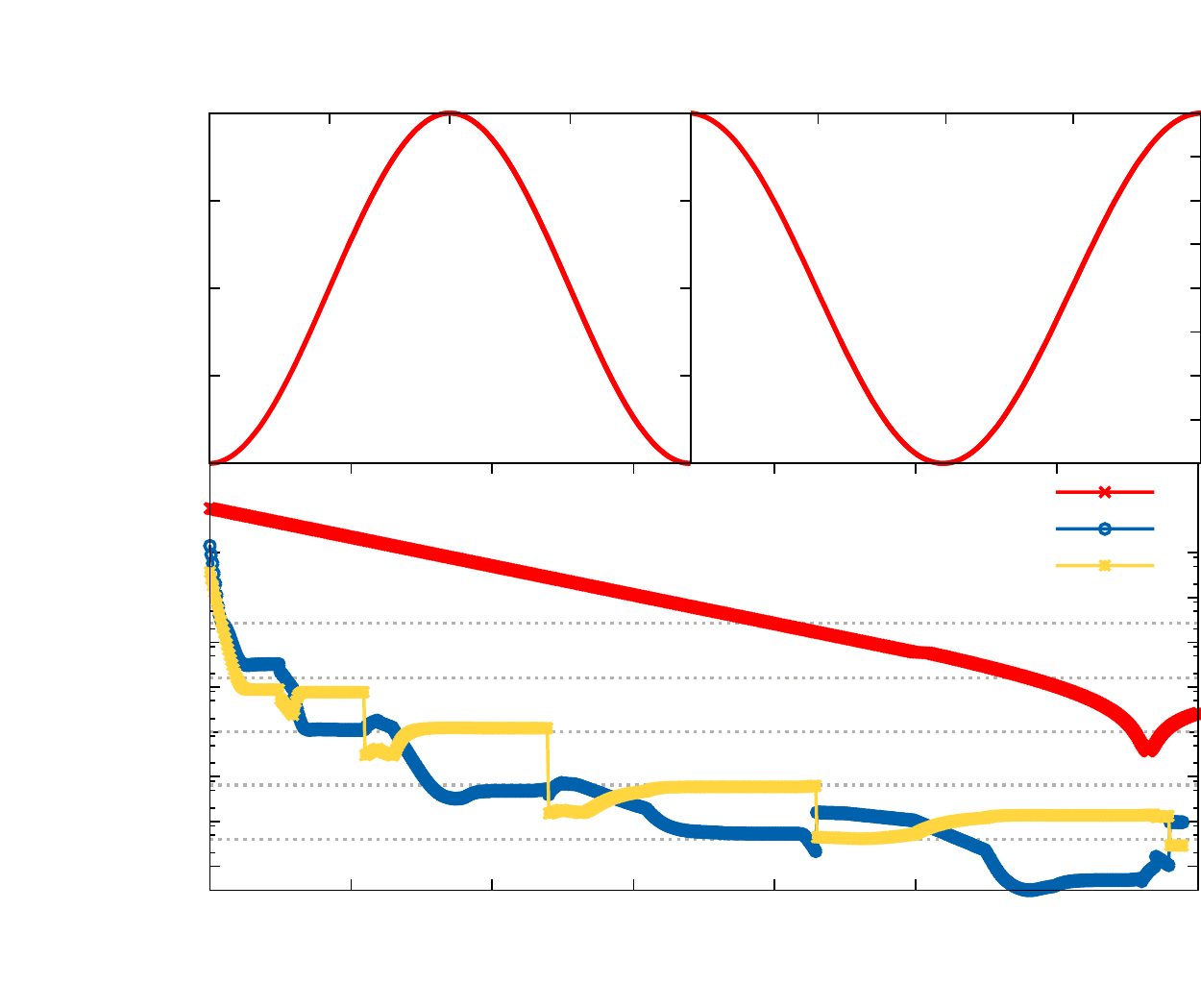}}%
    \gplfronttext
  \end{picture}%
\endgroup

\caption{Top: final solution and associated error for Helmholtz's equation with  
$c=-1$ and periodic boundary conditions. Bottom: $L_1$ norm of the solution error,
in a run that does not enforce the integral constraint (red), and in two runs that do
(blue via $f$ and yellow via $c$).
The dashed grey lines denote the norm of the truncation error on the five grids (the topmost
corresponding to the coarsest grid).
\label{fig:HP}}
\ece
\efi
\bfi
\bce
\begingroup
  \makeatletter
  \providecommand\color[2][]{%
    \GenericError{(gnuplot) \space\space\space\@spaces}{%
      Package color not loaded in conjunction with
      terminal option `colourtext'%
    }{See the gnuplot documentation for explanation.%
    }{Either use 'blacktext' in gnuplot or load the package
      color.sty in LaTeX.}%
    \renewcommand\color[2][]{}%
  }%
  \providecommand\includegraphics[2][]{%
    \GenericError{(gnuplot) \space\space\space\@spaces}{%
      Package graphicx or graphics not loaded%
    }{See the gnuplot documentation for explanation.%
    }{The gnuplot epslatex terminal needs graphicx.sty or graphics.sty.}%
    \renewcommand\includegraphics[2][]{}%
  }%
  \providecommand\rotatebox[2]{#2}%
  \@ifundefined{ifGPcolor}{%
    \newif\ifGPcolor
    \GPcolortrue
  }{}%
  \@ifundefined{ifGPblacktext}{%
    \newif\ifGPblacktext
    \GPblacktexttrue
  }{}%
  \let\gplgaddtomacro\g@addto@macro
  \gdef\gplbacktext{}%
  \gdef\gplfronttext{}%
  \makeatother
  \ifGPblacktext
    \def\colorrgb#1{}%
    \def\colorgray#1{}%
  \else
    \ifGPcolor
      \def\colorrgb#1{\color[rgb]{#1}}%
      \def\colorgray#1{\color[gray]{#1}}%
      \expandafter\def\csname LTw\endcsname{\color{white}}%
      \expandafter\def\csname LTb\endcsname{\color{black}}%
      \expandafter\def\csname LTa\endcsname{\color{black}}%
      \expandafter\def\csname LT0\endcsname{\color[rgb]{1,0,0}}%
      \expandafter\def\csname LT1\endcsname{\color[rgb]{0,1,0}}%
      \expandafter\def\csname LT2\endcsname{\color[rgb]{0,0,1}}%
      \expandafter\def\csname LT3\endcsname{\color[rgb]{1,0,1}}%
      \expandafter\def\csname LT4\endcsname{\color[rgb]{0,1,1}}%
      \expandafter\def\csname LT5\endcsname{\color[rgb]{1,1,0}}%
      \expandafter\def\csname LT6\endcsname{\color[rgb]{0,0,0}}%
      \expandafter\def\csname LT7\endcsname{\color[rgb]{1,0.3,0}}%
      \expandafter\def\csname LT8\endcsname{\color[rgb]{0.5,0.5,0.5}}%
    \else
      \def\colorrgb#1{\color{black}}%
      \def\colorgray#1{\color[gray]{#1}}%
      \expandafter\def\csname LTw\endcsname{\color{white}}%
      \expandafter\def\csname LTb\endcsname{\color{black}}%
      \expandafter\def\csname LTa\endcsname{\color{black}}%
      \expandafter\def\csname LT0\endcsname{\color{black}}%
      \expandafter\def\csname LT1\endcsname{\color{black}}%
      \expandafter\def\csname LT2\endcsname{\color{black}}%
      \expandafter\def\csname LT3\endcsname{\color{black}}%
      \expandafter\def\csname LT4\endcsname{\color{black}}%
      \expandafter\def\csname LT5\endcsname{\color{black}}%
      \expandafter\def\csname LT6\endcsname{\color{black}}%
      \expandafter\def\csname LT7\endcsname{\color{black}}%
      \expandafter\def\csname LT8\endcsname{\color{black}}%
    \fi
  \fi
  \setlength{\unitlength}{0.0500bp}%
  \begin{picture}(7401.60,3528.00)%
    \gplgaddtomacro\gplbacktext{%
      \csname LTb\endcsname%
      \put(1434,704){\makebox(0,0)[r]{\strut{}$10^{0}\;\;$}}%
      \put(1434,1070){\makebox(0,0)[r]{\strut{}$10^{5}\;\;$}}%
      \put(1434,1435){\makebox(0,0)[r]{\strut{}$10^{10}\;\;$}}%
      \put(1434,1801){\makebox(0,0)[r]{\strut{}$10^{15}\;\;$}}%
      \put(1434,2166){\makebox(0,0)[r]{\strut{}$10^{20}\;\;$}}%
      \put(1434,2532){\makebox(0,0)[r]{\strut{}$10^{25}\;\;$}}%
      \put(1434,2897){\makebox(0,0)[r]{\strut{}$10^{30}\;\;$}}%
      \put(1434,3263){\makebox(0,0)[r]{\strut{}$10^{35}\;\;$}}%
      \put(1566,484){\makebox(0,0){\strut{} 0}}%
      \put(2369,484){\makebox(0,0){\strut{} 100}}%
      \put(3171,484){\makebox(0,0){\strut{} 200}}%
      \put(3974,484){\makebox(0,0){\strut{} 300}}%
      \put(4776,484){\makebox(0,0){\strut{} 400}}%
      \put(5579,484){\makebox(0,0){\strut{} 500}}%
      \put(6381,484){\makebox(0,0){\strut{} 600}}%
      \put(7184,484){\makebox(0,0){\strut{} 700}}%
      \put(664,1983){\rotatebox{-270}{\makebox(0,0){\strut{}$|e|$}}}%
      \put(4375,154){\makebox(0,0){\strut{}Iteration}}%
    }%
    \gplgaddtomacro\gplfronttext{%
    }%
    \gplbacktext
    \put(0,0){\includegraphics{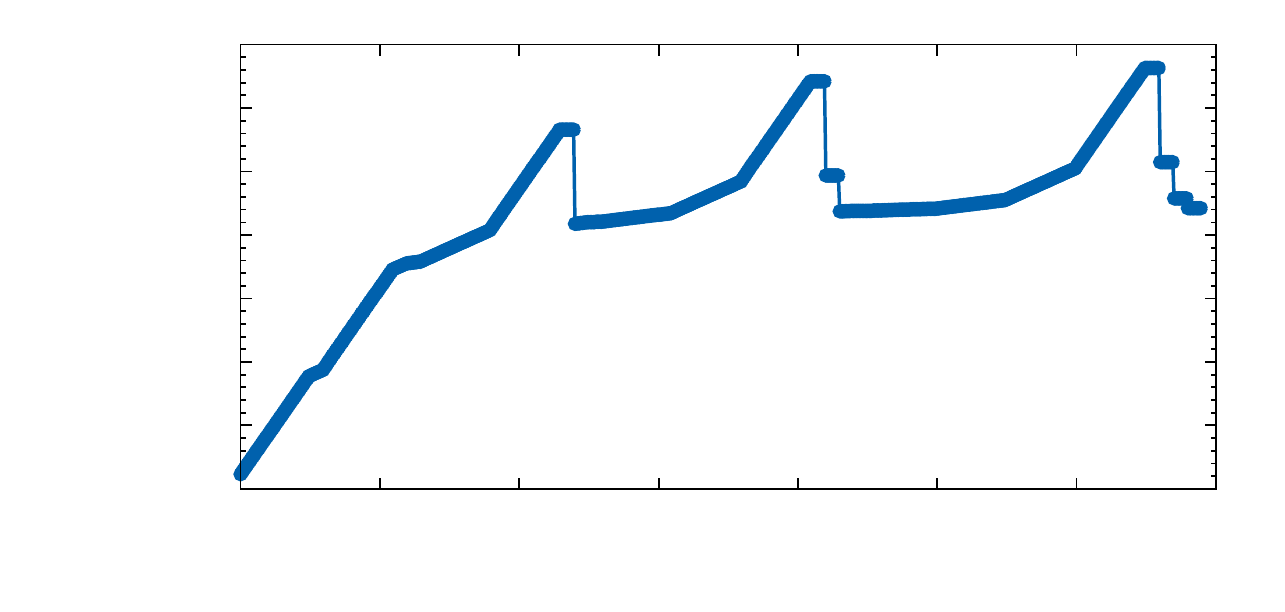}}%
    \gplfronttext
  \end{picture}%
\endgroup

\caption{
$L_1$ norm of the solution error for Helmholtz's equation with $c=3\pi^2$
and periodic boundary conditions. This system is underdetermined, and the relaxation
process does not converge.
\label{fig:HPC}}
\ece
\efi

\subsection{Einstein constraints}
In this section, we focus on the solution of the system of the four Einstein constraints,
in the Lichnerowicz-York form:
\begin{eqnarray}
\label{eq:LY}
 \tilde \Delta \psi - \frac{\tilde R}{8}\,\psi - \frac{K^2}{12}\,\psi^5 + \frac{1}{8} {\tilde A}_{ij} {\tilde A}^{ij} \psi^{-7} = - 2 \pi \rho \psi^5 \\
 \tilde D_i \tilde A^{ij} - \frac{2}{3} \psi^6 \tilde \gamma^{ij} \tilde D_i K = 8 \pi j^i \psi^{10}
\end{eqnarray}
where the tilde denotes conformal quantities, the conformal metric being related to the 
physical one by $\tilde \gamma_{ij} = \psi^{-4} \gamma_{ij}$.
If one assumes $\tilde \gamma_{ij}$ to be flat, and further decomposes $\tilde A_{ij}=\tilde D_i X_j + \tilde D_j X_i + 2/3 \; \tilde \gamma_{ij} \tilde D_k X^k$,
the following system is obtained (dropping the tilde from now on): 
\begin{eqnarray}
\label{eq:CTT}
 \Delta \psi - \frac{K^2}{12}\,\psi^5 + \frac{1}{8} {A}_{ij} {A}^{ij} \psi^{-7} = - 2 \pi \rho \psi^5 \\
 \Delta X_i + D_i D_j X^j - \frac{2}{3} \psi^6 \delta^{ij} D_i K = 8 \pi j^i \psi^{10}
\end{eqnarray}
In the general case, this is a coupled system of one non-linear and three linear elliptic PDEs.
As in the previous sections, one can trade some of the free data (say, $\rho$ and $j^i$) in exchange
for the possibility to work with an exact solution; this is accomplished by choosing a form $\bar \psi$ and $\bar X^i$
for the conformal factor and the vector $X^i$,
applying the right-hand sides of (\ref{eq:CTT}) to these functions, and setting the left-hand sides of the same system to the result:
\bea
 \Delta \psi - \frac{K^2}{12}\,\psi^5 + \frac{1}{8} {A}_{ij} {A}^{ij} \psi^{-7} &=& s \\
 \Delta X_i + D_i D_j X^j - \frac{2}{3} \psi^6 \delta^{ij} D_i K &=& s^i
\eea
where:
\bea
\label{eq:sources}
 s &=& \Delta \bar \psi - \frac{K^2}{12}\,\bar \psi^5 + \frac{1}{8} {A}_{ij} {A}^{ij} \bar \psi^{-7} \\ 
 s^i &=& \Delta \bar X_i + D_i D_j \bar X^j - \frac{2}{3} \bar \psi^6 \delta^{ij} D_i K
\eea
Since, naturally, there is no physical control or insight into this solution, it will be only used as a tool for code testing, 
in sections~\ref{sec:spher} and~\ref{sec:spherp}. I will then use this form for the Hamiltonian constraint,
along with an analytical prescription for $A_{ij}$, to solve for the initial data of
two non-spinning puncture in quasicircular orbit in section~\ref{sec:TP}.

\subsubsection{Generic spherical source}
\label{sec:spher}
A simple form for $\psi$ and $X^i$ can be:
\bea
\bar \psi &=& e^{-r^2/2\sigma^2} \\
\bar X^i &=& e^{-r_{(i)}^2/2\sigma^2}
\eea
where $r_{(i)}=|x^i-0.5|$.
The idea is to represent the gravitational field of a smooth, isolated mass distribution,
the nature of which plays no role in this test. I also choose:
\beq
K = e^{-r^2/2\sigma^2}
\eeq
and set the sources according to (\ref{eq:sources}). 
Solving this equation on the same grid setup presented before, one obtains the 
results illustrated in Figure~\ref{fig:SD}

\bfi
\bce
\begingroup
  \makeatletter
  \providecommand\color[2][]{%
    \GenericError{(gnuplot) \space\space\space\@spaces}{%
      Package color not loaded in conjunction with
      terminal option `colourtext'%
    }{See the gnuplot documentation for explanation.%
    }{Either use 'blacktext' in gnuplot or load the package
      color.sty in LaTeX.}%
    \renewcommand\color[2][]{}%
  }%
  \providecommand\includegraphics[2][]{%
    \GenericError{(gnuplot) \space\space\space\@spaces}{%
      Package graphicx or graphics not loaded%
    }{See the gnuplot documentation for explanation.%
    }{The gnuplot epslatex terminal needs graphicx.sty or graphics.sty.}%
    \renewcommand\includegraphics[2][]{}%
  }%
  \providecommand\rotatebox[2]{#2}%
  \@ifundefined{ifGPcolor}{%
    \newif\ifGPcolor
    \GPcolortrue
  }{}%
  \@ifundefined{ifGPblacktext}{%
    \newif\ifGPblacktext
    \GPblacktexttrue
  }{}%
  \let\gplgaddtomacro\g@addto@macro
  \gdef\gplbacktext{}%
  \gdef\gplfronttext{}%
  \makeatother
  \ifGPblacktext
    \def\colorrgb#1{}%
    \def\colorgray#1{}%
  \else
    \ifGPcolor
      \def\colorrgb#1{\color[rgb]{#1}}%
      \def\colorgray#1{\color[gray]{#1}}%
      \expandafter\def\csname LTw\endcsname{\color{white}}%
      \expandafter\def\csname LTb\endcsname{\color{black}}%
      \expandafter\def\csname LTa\endcsname{\color{black}}%
      \expandafter\def\csname LT0\endcsname{\color[rgb]{1,0,0}}%
      \expandafter\def\csname LT1\endcsname{\color[rgb]{0,1,0}}%
      \expandafter\def\csname LT2\endcsname{\color[rgb]{0,0,1}}%
      \expandafter\def\csname LT3\endcsname{\color[rgb]{1,0,1}}%
      \expandafter\def\csname LT4\endcsname{\color[rgb]{0,1,1}}%
      \expandafter\def\csname LT5\endcsname{\color[rgb]{1,1,0}}%
      \expandafter\def\csname LT6\endcsname{\color[rgb]{0,0,0}}%
      \expandafter\def\csname LT7\endcsname{\color[rgb]{1,0.3,0}}%
      \expandafter\def\csname LT8\endcsname{\color[rgb]{0.5,0.5,0.5}}%
    \else
      \def\colorrgb#1{\color{black}}%
      \def\colorgray#1{\color[gray]{#1}}%
      \expandafter\def\csname LTw\endcsname{\color{white}}%
      \expandafter\def\csname LTb\endcsname{\color{black}}%
      \expandafter\def\csname LTa\endcsname{\color{black}}%
      \expandafter\def\csname LT0\endcsname{\color{black}}%
      \expandafter\def\csname LT1\endcsname{\color{black}}%
      \expandafter\def\csname LT2\endcsname{\color{black}}%
      \expandafter\def\csname LT3\endcsname{\color{black}}%
      \expandafter\def\csname LT4\endcsname{\color{black}}%
      \expandafter\def\csname LT5\endcsname{\color{black}}%
      \expandafter\def\csname LT6\endcsname{\color{black}}%
      \expandafter\def\csname LT7\endcsname{\color{black}}%
      \expandafter\def\csname LT8\endcsname{\color{black}}%
    \fi
  \fi
  \setlength{\unitlength}{0.0500bp}%
  \begin{picture}(7200.00,7560.00)%
    \gplgaddtomacro\gplbacktext{%
      \csname LTb\endcsname%
      \put(1126,5253){\makebox(0,0)[r]{\strut{} 1}}%
      \put(1126,5603){\makebox(0,0)[r]{\strut{} 1.2}}%
      \put(1126,5953){\makebox(0,0)[r]{\strut{} 1.4}}%
      \put(1126,6303){\makebox(0,0)[r]{\strut{} 1.6}}%
      \put(1126,6652){\makebox(0,0)[r]{\strut{} 1.8}}%
      \put(1126,7002){\makebox(0,0)[r]{\strut{} 2}}%
      \put(1126,7352){\makebox(0,0)[r]{\strut{} 2.2}}%
      \put(1258,7572){\makebox(0,0){\strut{}-1}}%
      \put(1950,7572){\makebox(0,0){\strut{}-0.5}}%
      \put(2641,7572){\makebox(0,0){\strut{} 0}}%
      \put(3333,7572){\makebox(0,0){\strut{} 0.5}}%
      \put(4024,7572){\makebox(0,0){\strut{} 1}}%
      \put(356,6302){\rotatebox{-270}{\makebox(0,0){\strut{}$\psi$}}}%
      \put(2641,7901){\makebox(0,0){\strut{}$x$}}%
    }%
    \gplgaddtomacro\gplfronttext{%
    }%
    \gplgaddtomacro\gplbacktext{%
      \csname LTb\endcsname%
      \put(1122,2758){\makebox(0,0)[r]{\strut{}-0.1}}%
      \put(1122,3008){\makebox(0,0)[r]{\strut{}-0.08}}%
      \put(1122,3257){\makebox(0,0)[r]{\strut{}-0.06}}%
      \put(1122,3507){\makebox(0,0)[r]{\strut{}-0.04}}%
      \put(1122,3756){\makebox(0,0)[r]{\strut{}-0.02}}%
      \put(1122,4006){\makebox(0,0)[r]{\strut{} 0}}%
      \put(1122,4256){\makebox(0,0)[r]{\strut{} 0.02}}%
      \put(1122,4505){\makebox(0,0)[r]{\strut{} 0.04}}%
      \put(1122,4755){\makebox(0,0)[r]{\strut{} 0.06}}%
      \put(1122,5004){\makebox(0,0)[r]{\strut{} 0.08}}%
      \put(1122,5254){\makebox(0,0)[r]{\strut{} 0.1}}%
      \put(220,4006){\rotatebox{-270}{\makebox(0,0){\strut{}$X^x$}}}%
    }%
    \gplgaddtomacro\gplfronttext{%
    }%
    \gplgaddtomacro\gplbacktext{%
      \csname LTb\endcsname%
      \put(7158,5565){\makebox(0,0)[l]{\strut{}-6}}%
      \put(7158,5868){\makebox(0,0)[l]{\strut{}-5}}%
      \put(7158,6171){\makebox(0,0)[l]{\strut{}-4}}%
      \put(7158,6474){\makebox(0,0)[l]{\strut{}-3}}%
      \put(7158,6777){\makebox(0,0)[l]{\strut{}-2}}%
      \put(7158,7080){\makebox(0,0)[l]{\strut{}-1}}%
      \put(4023,7572){\makebox(0,0){\strut{}-1}}%
      \put(4774,7572){\makebox(0,0){\strut{}-0.5}}%
      \put(5525,7572){\makebox(0,0){\strut{} 0}}%
      \put(6275,7572){\makebox(0,0){\strut{} 0.5}}%
      \put(7026,7572){\makebox(0,0){\strut{} 1}}%
      \put(7663,6302){\rotatebox{-270}{\makebox(0,0){\strut{}$e_{\psi} \cdot 10^5$}}}%
      \put(5524,7901){\makebox(0,0){\strut{}$x$}}%
    }%
    \gplgaddtomacro\gplfronttext{%
    }%
    \gplgaddtomacro\gplbacktext{%
      \csname LTb\endcsname%
      \put(7158,2993){\makebox(0,0)[l]{\strut{}-10}}%
      \put(7158,3224){\makebox(0,0)[l]{\strut{}-8}}%
      \put(7158,3454){\makebox(0,0)[l]{\strut{}-6}}%
      \put(7158,3684){\makebox(0,0)[l]{\strut{}-4}}%
      \put(7158,3914){\makebox(0,0)[l]{\strut{}-2}}%
      \put(7158,4145){\makebox(0,0)[l]{\strut{} 0}}%
      \put(7158,4375){\makebox(0,0)[l]{\strut{} 2}}%
      \put(7158,4605){\makebox(0,0)[l]{\strut{} 4}}%
      \put(7158,4835){\makebox(0,0)[l]{\strut{} 6}}%
      \put(7158,5066){\makebox(0,0)[l]{\strut{} 8}}%
      \put(7663,4006){\rotatebox{-270}{\makebox(0,0){\strut{}$e_{X^x} \cdot 10^6$}}}%
    }%
    \gplgaddtomacro\gplfronttext{%
    }%
    \gplgaddtomacro\gplbacktext{%
      \csname LTb\endcsname%
      \put(1116,721){\makebox(0,0)[r]{\strut{}$10^{-6}$}}%
      \put(1116,1061){\makebox(0,0)[r]{\strut{}$10^{-5}$}}%
      \put(1116,1401){\makebox(0,0)[r]{\strut{}$10^{-4}$}}%
      \put(1116,1740){\makebox(0,0)[r]{\strut{}$10^{-3}$}}%
      \put(1116,2080){\makebox(0,0)[r]{\strut{}$10^{-2}$}}%
      \put(1116,2419){\makebox(0,0)[r]{\strut{}$10^{-1}$}}%
      \put(1248,484){\makebox(0,0){\strut{} 0}}%
      \put(2074,484){\makebox(0,0){\strut{} 100}}%
      \put(2901,484){\makebox(0,0){\strut{} 200}}%
      \put(3727,484){\makebox(0,0){\strut{} 300}}%
      \put(4554,484){\makebox(0,0){\strut{} 400}}%
      \put(5380,484){\makebox(0,0){\strut{} 500}}%
      \put(6207,484){\makebox(0,0){\strut{} 600}}%
      \put(7033,484){\makebox(0,0){\strut{} 700}}%
      \put(478,1731){\rotatebox{-270}{\makebox(0,0){\strut{}$|e|$}}}%
      \put(4140,154){\makebox(0,0){\strut{}Iteration}}%
    }%
    \gplgaddtomacro\gplfronttext{%
      \csname LTb\endcsname%
      \put(6046,2586){\makebox(0,0)[r]{\strut{}$|e_{\psi}|$}}%
      \csname LTb\endcsname%
      \put(6046,2366){\makebox(0,0)[r]{\strut{}$|e_{X^x}|$}}%
    }%
    \gplbacktext
    \put(0,0){\includegraphics{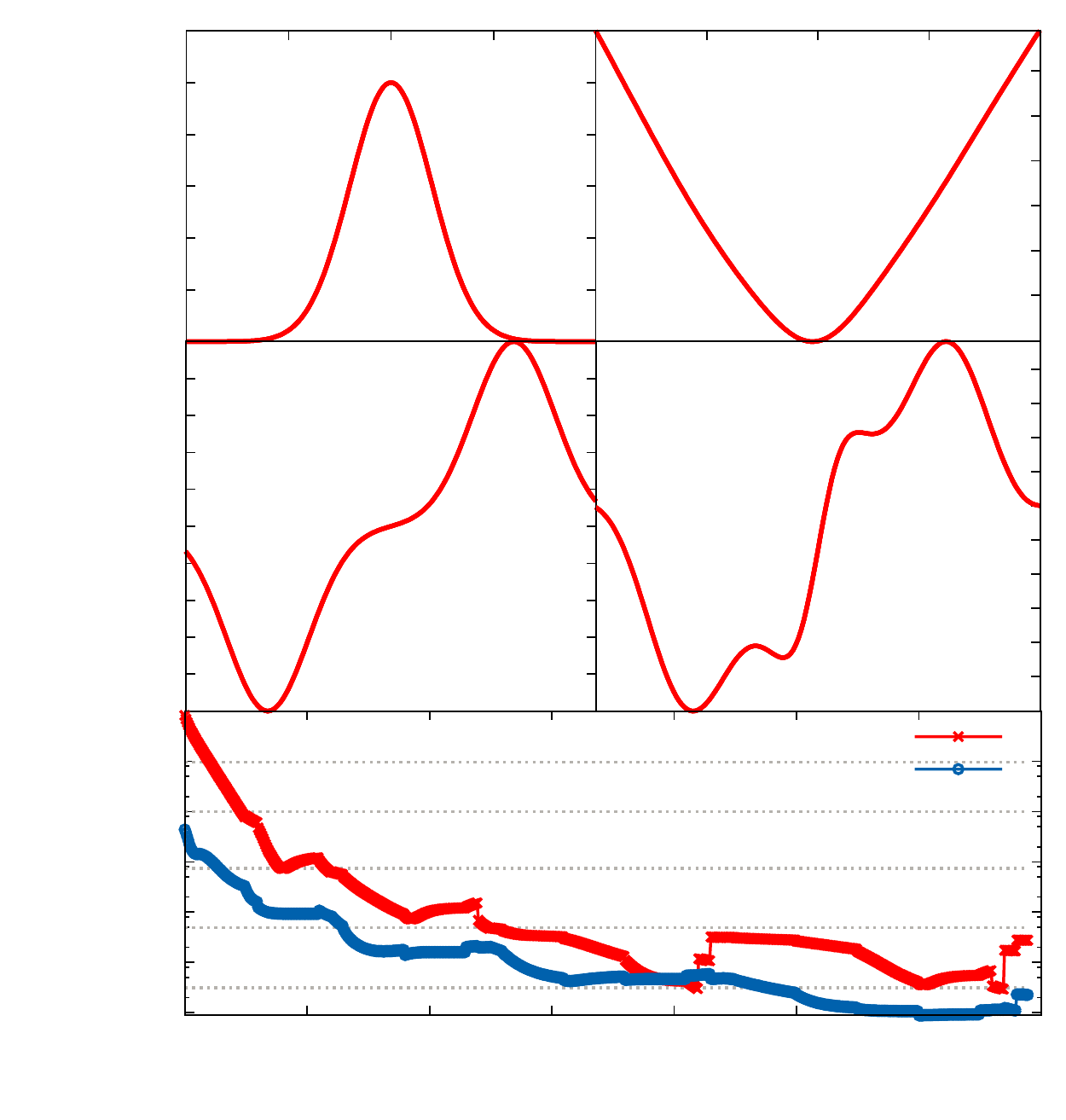}}%
    \gplfronttext
  \end{picture}%
\endgroup

\caption{Top: final solution and associated error for the Hamiltonian constraint equation 
with Dirichlet boundary conditions and a spherical source. Middle: final solution and associated error for the $x$-component
of the momentum constraint equation. Bottom: $L_1$ norm of the solution error for both quantities.
The dashed grey lines denote the norm of the truncation error on the five grids (the topmost
corresponding to the coarsest grid).
\label{fig:SD}}
\ece
\efi

\subsubsection{Sinusoidal source}
\label{sec:spherp}
It is again instructive to study the same system in conjunction with PBCs. 
I now assume a solution of the form:
\bea
\bar \psi &=& 2+\cos \pi x \; \cos \pi y \; \cos \pi z \\
\bar X^i &=& \sin \pi x \; \sin \pi y \; \sin \pi z \qquad \qquad i=1,2,3
\eea
with a constant $K=-0.1$. 
Here, the integrability condition due to the periodic boundaries reads:
\beq
\int \left(- \frac{K^2}{12}\,\psi^5 + \frac{1}{8} {A}_{ij} {A}^{ij} \psi^{-7}-s\right) {\rm d} V = 0
\eeq
As for Helmholtz's equation, the solver does not need this step in order
to converge to a solution, but enforcing this condition improves the 
convergence rate.
This can be done by finding a root of:
\beq
f(A) = K^2 \int (\psi+A)^5 {\rm d}V - \frac{1}{8}\int A_{ij} A^{ij} (\psi+A)^{-7} {\rm d}V - \int s {\rm d}V
\eeq 
which can easily be solved via the Newton-Raphson method after 
each Gauss-Seidel iteration.
A test performed on the configurations described in this section shows that this improvement
leads to a reduction of the final residual of more than order of magnitude
with respect to the case where $K$ is updated.
It is worth noting that, in either case (updating $K$ or updating $\psi$), 
this step produces an additional numerical error that will only
scale with the second power of the grid spacing, since it involves
an integral calculated with the midpoint rule.

Notice that there 
is no integrability condition associated to the momentum constraint since the first two terms
are divergences and the third has a zero integral over the cell due to the system's symmetry 
under parity transformations $x \to -x$, $y \to -y$ and $z \to -z$.

\bfi
\bce
\begingroup
  \makeatletter
  \providecommand\color[2][]{%
    \GenericError{(gnuplot) \space\space\space\@spaces}{%
      Package color not loaded in conjunction with
      terminal option `colourtext'%
    }{See the gnuplot documentation for explanation.%
    }{Either use 'blacktext' in gnuplot or load the package
      color.sty in LaTeX.}%
    \renewcommand\color[2][]{}%
  }%
  \providecommand\includegraphics[2][]{%
    \GenericError{(gnuplot) \space\space\space\@spaces}{%
      Package graphicx or graphics not loaded%
    }{See the gnuplot documentation for explanation.%
    }{The gnuplot epslatex terminal needs graphicx.sty or graphics.sty.}%
    \renewcommand\includegraphics[2][]{}%
  }%
  \providecommand\rotatebox[2]{#2}%
  \@ifundefined{ifGPcolor}{%
    \newif\ifGPcolor
    \GPcolortrue
  }{}%
  \@ifundefined{ifGPblacktext}{%
    \newif\ifGPblacktext
    \GPblacktexttrue
  }{}%
  \let\gplgaddtomacro\g@addto@macro
  \gdef\gplbacktext{}%
  \gdef\gplfronttext{}%
  \makeatother
  \ifGPblacktext
    \def\colorrgb#1{}%
    \def\colorgray#1{}%
  \else
    \ifGPcolor
      \def\colorrgb#1{\color[rgb]{#1}}%
      \def\colorgray#1{\color[gray]{#1}}%
      \expandafter\def\csname LTw\endcsname{\color{white}}%
      \expandafter\def\csname LTb\endcsname{\color{black}}%
      \expandafter\def\csname LTa\endcsname{\color{black}}%
      \expandafter\def\csname LT0\endcsname{\color[rgb]{1,0,0}}%
      \expandafter\def\csname LT1\endcsname{\color[rgb]{0,1,0}}%
      \expandafter\def\csname LT2\endcsname{\color[rgb]{0,0,1}}%
      \expandafter\def\csname LT3\endcsname{\color[rgb]{1,0,1}}%
      \expandafter\def\csname LT4\endcsname{\color[rgb]{0,1,1}}%
      \expandafter\def\csname LT5\endcsname{\color[rgb]{1,1,0}}%
      \expandafter\def\csname LT6\endcsname{\color[rgb]{0,0,0}}%
      \expandafter\def\csname LT7\endcsname{\color[rgb]{1,0.3,0}}%
      \expandafter\def\csname LT8\endcsname{\color[rgb]{0.5,0.5,0.5}}%
    \else
      \def\colorrgb#1{\color{black}}%
      \def\colorgray#1{\color[gray]{#1}}%
      \expandafter\def\csname LTw\endcsname{\color{white}}%
      \expandafter\def\csname LTb\endcsname{\color{black}}%
      \expandafter\def\csname LTa\endcsname{\color{black}}%
      \expandafter\def\csname LT0\endcsname{\color{black}}%
      \expandafter\def\csname LT1\endcsname{\color{black}}%
      \expandafter\def\csname LT2\endcsname{\color{black}}%
      \expandafter\def\csname LT3\endcsname{\color{black}}%
      \expandafter\def\csname LT4\endcsname{\color{black}}%
      \expandafter\def\csname LT5\endcsname{\color{black}}%
      \expandafter\def\csname LT6\endcsname{\color{black}}%
      \expandafter\def\csname LT7\endcsname{\color{black}}%
      \expandafter\def\csname LT8\endcsname{\color{black}}%
    \fi
  \fi
  \setlength{\unitlength}{0.0500bp}%
  \begin{picture}(7200.00,7560.00)%
    \gplgaddtomacro\gplbacktext{%
      \csname LTb\endcsname%
      \put(1107,5253){\makebox(0,0)[r]{\strut{} 1}}%
      \put(1107,5778){\makebox(0,0)[r]{\strut{} 1.5}}%
      \put(1107,6303){\makebox(0,0)[r]{\strut{} 2}}%
      \put(1107,6827){\makebox(0,0)[r]{\strut{} 2.5}}%
      \put(1107,7352){\makebox(0,0)[r]{\strut{} 3}}%
      \put(1239,7572){\makebox(0,0){\strut{}-1}}%
      \put(1984,7572){\makebox(0,0){\strut{}-0.5}}%
      \put(2729,7572){\makebox(0,0){\strut{} 0}}%
      \put(3473,7572){\makebox(0,0){\strut{} 0.5}}%
      \put(4218,7572){\makebox(0,0){\strut{} 1}}%
      \put(601,6302){\rotatebox{-270}{\makebox(0,0){\strut{}$\psi$}}}%
      \put(2728,7901){\makebox(0,0){\strut{}$x$}}%
    }%
    \gplgaddtomacro\gplfronttext{%
    }%
    \gplgaddtomacro\gplbacktext{%
      \csname LTb\endcsname%
      \put(1100,2761){\makebox(0,0)[r]{\strut{} 0}}%
      \put(1100,3176){\makebox(0,0)[r]{\strut{} 0.5}}%
      \put(1100,3592){\makebox(0,0)[r]{\strut{} 1}}%
      \put(1100,4007){\makebox(0,0)[r]{\strut{} 1.5}}%
      \put(1100,4423){\makebox(0,0)[r]{\strut{} 2}}%
      \put(1100,4838){\makebox(0,0)[r]{\strut{} 2.5}}%
      \put(594,4006){\rotatebox{-270}{\makebox(0,0){\strut{}$X^x \cdot 10^6$}}}%
    }%
    \gplgaddtomacro\gplfronttext{%
    }%
    \gplgaddtomacro\gplbacktext{%
      \csname LTb\endcsname%
      \put(7206,5253){\makebox(0,0)[l]{\strut{}-2}}%
      \put(7206,5553){\makebox(0,0)[l]{\strut{} 0}}%
      \put(7206,5853){\makebox(0,0)[l]{\strut{} 2}}%
      \put(7206,6153){\makebox(0,0)[l]{\strut{} 4}}%
      \put(7206,6452){\makebox(0,0)[l]{\strut{} 6}}%
      \put(7206,6752){\makebox(0,0)[l]{\strut{} 8}}%
      \put(7206,7052){\makebox(0,0)[l]{\strut{} 10}}%
      \put(7206,7352){\makebox(0,0)[l]{\strut{} 12}}%
      \put(4218,7572){\makebox(0,0){\strut{}-1}}%
      \put(4932,7572){\makebox(0,0){\strut{}-0.5}}%
      \put(5646,7572){\makebox(0,0){\strut{} 0}}%
      \put(6360,7572){\makebox(0,0){\strut{} 0.5}}%
      \put(7074,7572){\makebox(0,0){\strut{} 1}}%
      \put(7843,6302){\rotatebox{-270}{\makebox(0,0){\strut{}$e_{\psi} \cdot 10^6$}}}%
      \put(5646,7901){\makebox(0,0){\strut{}$x$}}%
    }%
    \gplgaddtomacro\gplfronttext{%
    }%
    \gplgaddtomacro\gplbacktext{%
      \csname LTb\endcsname%
      \put(7206,2893){\makebox(0,0)[l]{\strut{}-2.5}}%
      \put(7206,3365){\makebox(0,0)[l]{\strut{}-2}}%
      \put(7206,3837){\makebox(0,0)[l]{\strut{}-1.5}}%
      \put(7206,4310){\makebox(0,0)[l]{\strut{}-1}}%
      \put(7206,4782){\makebox(0,0)[l]{\strut{}-0.5}}%
      \put(7843,4006){\rotatebox{-270}{\makebox(0,0){\strut{}$e_{X^x} \cdot 10^6$}}}%
    }%
    \gplgaddtomacro\gplfronttext{%
    }%
    \gplgaddtomacro\gplbacktext{%
      \csname LTb\endcsname%
      \put(1102,856){\makebox(0,0)[r]{\strut{}$10^{-6}$}}%
      \put(1102,1128){\makebox(0,0)[r]{\strut{}$10^{-5}$}}%
      \put(1102,1400){\makebox(0,0)[r]{\strut{}$10^{-4}$}}%
      \put(1102,1672){\makebox(0,0)[r]{\strut{}$10^{-3}$}}%
      \put(1102,1943){\makebox(0,0)[r]{\strut{}$10^{-2}$}}%
      \put(1102,2215){\makebox(0,0)[r]{\strut{}$10^{-1}$}}%
      \put(1102,2487){\makebox(0,0)[r]{\strut{}$10^{0}$}}%
      \put(1234,484){\makebox(0,0){\strut{} 0}}%
      \put(2071,484){\makebox(0,0){\strut{} 100}}%
      \put(2907,484){\makebox(0,0){\strut{} 200}}%
      \put(3744,484){\makebox(0,0){\strut{} 300}}%
      \put(4580,484){\makebox(0,0){\strut{} 400}}%
      \put(5417,484){\makebox(0,0){\strut{} 500}}%
      \put(6253,484){\makebox(0,0){\strut{} 600}}%
      \put(7090,484){\makebox(0,0){\strut{} 700}}%
      \put(464,1731){\rotatebox{-270}{\makebox(0,0){\strut{}$|e|$}}}%
      \put(4162,154){\makebox(0,0){\strut{}Iteration}}%
    }%
    \gplgaddtomacro\gplfronttext{%
      \csname LTb\endcsname%
      \put(6103,2586){\makebox(0,0)[r]{\strut{}$|e_{\psi}|$}}%
      \csname LTb\endcsname%
      \put(6103,2366){\makebox(0,0)[r]{\strut{}$|e_{X^x}|$}}%
    }%
    \gplbacktext
    \put(0,0){\includegraphics{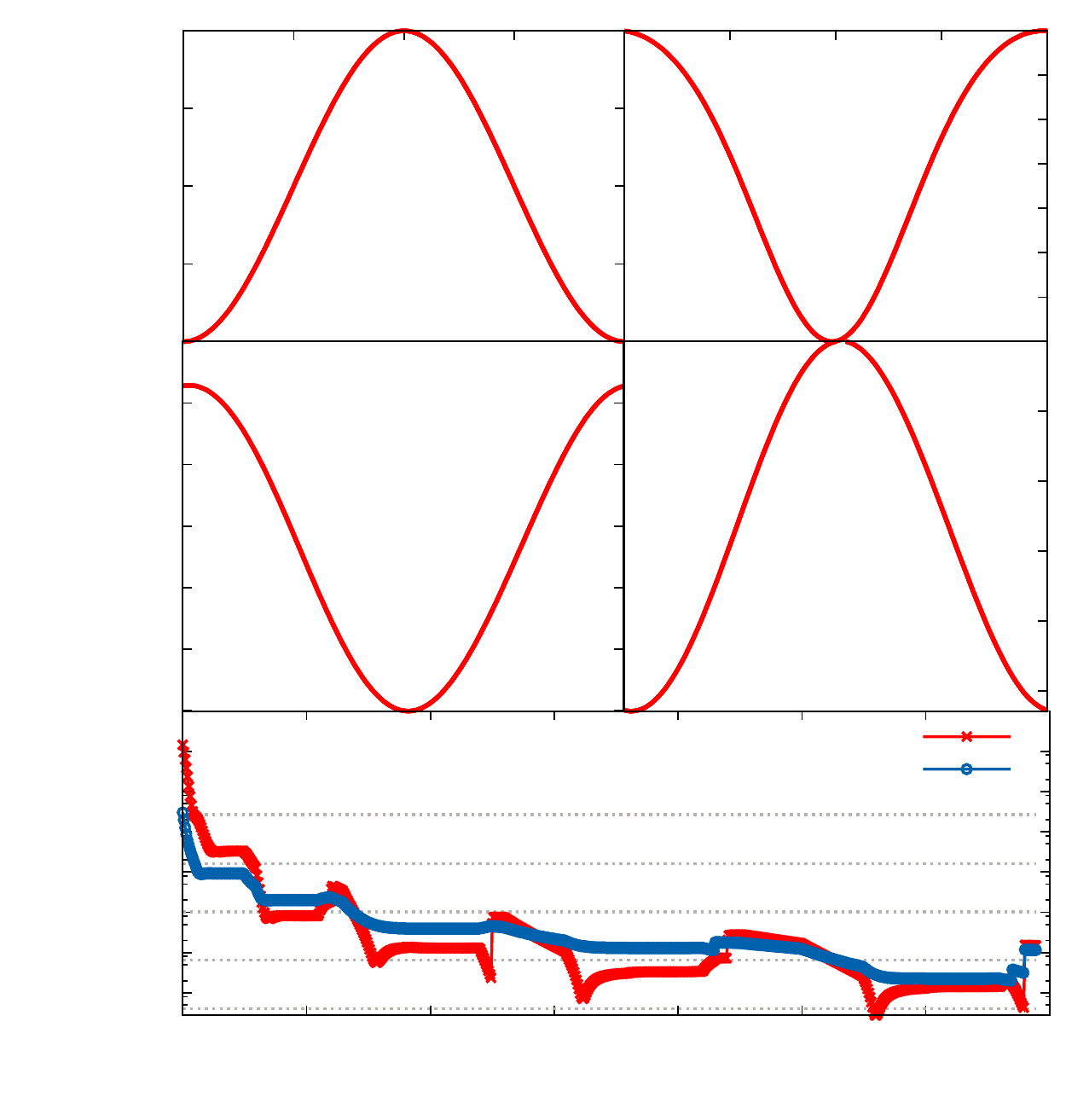}}%
    \gplfronttext
  \end{picture}%
\endgroup

\caption{Top: final solution and associated error for the Hamiltonian constraint equation 
with periodic boundary conditions and a spherical source. Middle: final solution and associated error for the $x$-component
of the momentum constraint equation. Bottom: $L_1$ norm of the solution error for both quantities.
The dashed grey lines denote the truncation error on the five grids (the topmost
corresponding to the coarsest grid).
\label{fig:SP}}
\ece
\efi

\subsubsection{Two punctures}
\label{sec:TP}
As a last test, I will illustrate the behavior of the solver in a familiar
case: that of two non-spinning black holes, in the puncture representation and in 
a quasicircular orbit around each other. Whilst an exact solution to this problem
does not exist, I will compare the result with an existing component of the Einstein
Toolkit, the \texttt{TwoPunctures} initial-data generator~\cite{Ansorg:2004ds}.

In this case, the metric is assumed to be conformally flat:
\beq
\gamma_{ij} = \psi^4 \delta_{ij}
\eeq
and the extrinsic curvature is purely traceless ($K=0$) and given by:
\beq
A_{ij} = \sum_{N=1}^2 \frac{3}{2r_{N}^2}(P^{N}_i n^{N}_j+P^{N}_j n^{N}_i+(\delta_{ij}-n^{N}_in^{N}_j)n_{N}^kP^{N}_k) 
\eeq
where $r_N$ is the distance to puncture $N$, $n_N^i$ the outgoing unit radial vector 
from this point, and $P_N^i$ is its linear momentum. The only equation to solve is 
the Hamiltonian constraint for $\psi$:
\beq
\Delta \psi + \frac{1}{8} {A}_{ij} {A}^{ij} \psi^{-7} = 0
\eeq
The conformal factor $\psi$ is also typically decomposed in a regular part $u$ and a singular,
$1/r$ term according to:
\beq
\psi = u + \sum_N \frac{m_N}{r_N}
\eeq
where $m_N$ represent the \emph{bare masses} of the black holes.
The equation to solve becomes:
\beq
\label{eq:Htp}
\Delta u + \frac{1}{8} {A}_{ij} {A}^{ij} \psi^{-7} = 0
\eeq
Since the spatial domain has a finite extent, I use Dirichlet boundary conditions obtained
by setting $u$ at the boundaries equal to the solution of \texttt{TwoPunctures}.
 
For this test, I have chosen to solve for the QC0 configuration~\cite{Baker:2002qf}, in which the black holes 
have bare masses equal to $0.453$, are located on the $x$-axis at coordinates $x=\pm 1.168642873$, and have
linear momenta parallel to the $y$-axis and equal to $\pm 0.3331917498$. In test A1,
the grid structure contains a coarse grid C4 extending out to $x=y=z=\pm50$ with spacing $0.625$, 
and three finer refinement levels C5, C6 and C7 of radii $25$, $12$ and $6$ respectively, each with spacing equal to half
that of the previous level. In addition, extra coarse grids C3 to C0, covering the whole domain with spacings 
equal to $1.25$, $2.5$, $5$ and $10$ respectively, are added in order to speed-up the convergence according
to the multigrid prescription. In this case, a canonical FMG cycle is performed on the 
domain-covering levels, whilst the local refinement levels are relaxed one by one, each providing
the initial data and the boundary conditions for the next. In each V cycle, the bottow level 
is relaxed 50 times, the top one 10, and the others 50 times in the downward part and 10
in the upward. Test A2 has the same grid structure as A1, but with all the grid spacings 
multiplied by 1.25.

The results for this test are illustrated in Figure~\ref{fig:TP}. Here, the final conformal
factor is shown for $A1$ and $A2$.
These are also compared to the solution from \texttt{TwoPunctures}, obtained using the
``Evaluation'' option (TP1) on the same grid structure of A1; TP1 will be considered the ``exact'' solution to test
the convergence of the multigrid solver. The results agree to within a few percent;
the difference can be accounted for by observing that the magnitude of the truncation
error on C4 can be as high as $10^{-2}$ 
(to obtain this estimate, I have calculated the numerical laplacian of the ``exact''
solution from \texttt{TwoPunctures} on C3 and C4, and evaluated the difference,
shown in Figure~\ref{fig:TPerr}).
Figure~\ref{fig:TP} supports this claim by showing that the errors $e_{A*} \equiv 
u_{\rm TP1}-u_{\rm A*}$ scale quartically
with the grid spacing, as would be expected of the truncation error in a fourth-order 
finite-differencing scheme.

For reference, in Figure~\ref{fig:TP} I also show the Hamiltonian-constraint violation
obtained by tests A1 and A2, by the reference run TP1 and by a run TP2 in which
\texttt{TwoPunctures} is used in the ``Taylor expansion'' mode.
I include this run for fairness of comparison: A1 and A2, run on 48 and 24 processing
cores respectively, take about one minute to complete, whilst TP1 requires about fifty times
as long on 48 cores~\footnote{In~\cite{Paschalidis:2013fk}, the performance of 
\texttt{TwoPunctures} in the ``Evaluation'' mode has recently been improved by roughly 
an order of magnitude. This would bring the walltime requirement from fifty times that of
A1 down to about five times.}. TP2, using a faster interpolation method on the same number of cores, requires 
about half a minute.

\bfi
\bce
\begingroup
  \makeatletter
  \providecommand\color[2][]{%
    \GenericError{(gnuplot) \space\space\space\@spaces}{%
      Package color not loaded in conjunction with
      terminal option `colourtext'%
    }{See the gnuplot documentation for explanation.%
    }{Either use 'blacktext' in gnuplot or load the package
      color.sty in LaTeX.}%
    \renewcommand\color[2][]{}%
  }%
  \providecommand\includegraphics[2][]{%
    \GenericError{(gnuplot) \space\space\space\@spaces}{%
      Package graphicx or graphics not loaded%
    }{See the gnuplot documentation for explanation.%
    }{The gnuplot epslatex terminal needs graphicx.sty or graphics.sty.}%
    \renewcommand\includegraphics[2][]{}%
  }%
  \providecommand\rotatebox[2]{#2}%
  \@ifundefined{ifGPcolor}{%
    \newif\ifGPcolor
    \GPcolortrue
  }{}%
  \@ifundefined{ifGPblacktext}{%
    \newif\ifGPblacktext
    \GPblacktexttrue
  }{}%
  \let\gplgaddtomacro\g@addto@macro
  \gdef\gplbacktext{}%
  \gdef\gplfronttext{}%
  \makeatother
  \ifGPblacktext
    \def\colorrgb#1{}%
    \def\colorgray#1{}%
  \else
    \ifGPcolor
      \def\colorrgb#1{\color[rgb]{#1}}%
      \def\colorgray#1{\color[gray]{#1}}%
      \expandafter\def\csname LTw\endcsname{\color{white}}%
      \expandafter\def\csname LTb\endcsname{\color{black}}%
      \expandafter\def\csname LTa\endcsname{\color{black}}%
      \expandafter\def\csname LT0\endcsname{\color[rgb]{1,0,0}}%
      \expandafter\def\csname LT1\endcsname{\color[rgb]{0,1,0}}%
      \expandafter\def\csname LT2\endcsname{\color[rgb]{0,0,1}}%
      \expandafter\def\csname LT3\endcsname{\color[rgb]{1,0,1}}%
      \expandafter\def\csname LT4\endcsname{\color[rgb]{0,1,1}}%
      \expandafter\def\csname LT5\endcsname{\color[rgb]{1,1,0}}%
      \expandafter\def\csname LT6\endcsname{\color[rgb]{0,0,0}}%
      \expandafter\def\csname LT7\endcsname{\color[rgb]{1,0.3,0}}%
      \expandafter\def\csname LT8\endcsname{\color[rgb]{0.5,0.5,0.5}}%
    \else
      \def\colorrgb#1{\color{black}}%
      \def\colorgray#1{\color[gray]{#1}}%
      \expandafter\def\csname LTw\endcsname{\color{white}}%
      \expandafter\def\csname LTb\endcsname{\color{black}}%
      \expandafter\def\csname LTa\endcsname{\color{black}}%
      \expandafter\def\csname LT0\endcsname{\color{black}}%
      \expandafter\def\csname LT1\endcsname{\color{black}}%
      \expandafter\def\csname LT2\endcsname{\color{black}}%
      \expandafter\def\csname LT3\endcsname{\color{black}}%
      \expandafter\def\csname LT4\endcsname{\color{black}}%
      \expandafter\def\csname LT5\endcsname{\color{black}}%
      \expandafter\def\csname LT6\endcsname{\color{black}}%
      \expandafter\def\csname LT7\endcsname{\color{black}}%
      \expandafter\def\csname LT8\endcsname{\color{black}}%
    \fi
  \fi
  \setlength{\unitlength}{0.0500bp}%
  \begin{picture}(7056.00,3528.00)%
    \gplgaddtomacro\gplbacktext{%
      \csname LTb\endcsname%
      \put(1210,704){\makebox(0,0)[r]{\strut{} 0.001}}%
      \put(1210,1984){\makebox(0,0)[r]{\strut{} 0.01}}%
      \put(1210,3263){\makebox(0,0)[r]{\strut{} 0.1}}%
      \put(1874,484){\makebox(0,0){\strut{}-40}}%
      \put(2937,484){\makebox(0,0){\strut{}-20}}%
      \put(4001,484){\makebox(0,0){\strut{} 0}}%
      \put(5064,484){\makebox(0,0){\strut{} 20}}%
      \put(6127,484){\makebox(0,0){\strut{} 40}}%
      \put(176,1983){\rotatebox{-270}{\makebox(0,0){\strut{}$u-1$}}}%
      \put(4000,154){\makebox(0,0){\strut{}$x$}}%
    }%
    \gplgaddtomacro\gplfronttext{%
      \csname LTb\endcsname%
      \put(5672,3090){\makebox(0,0)[r]{\strut{}A1}}%
      \csname LTb\endcsname%
      \put(5672,2870){\makebox(0,0)[r]{\strut{}A2}}%
      \csname LTb\endcsname%
      \put(5672,2650){\makebox(0,0)[r]{\strut{}TP1}}%
    }%
    \gplbacktext
    \put(0,0){\includegraphics{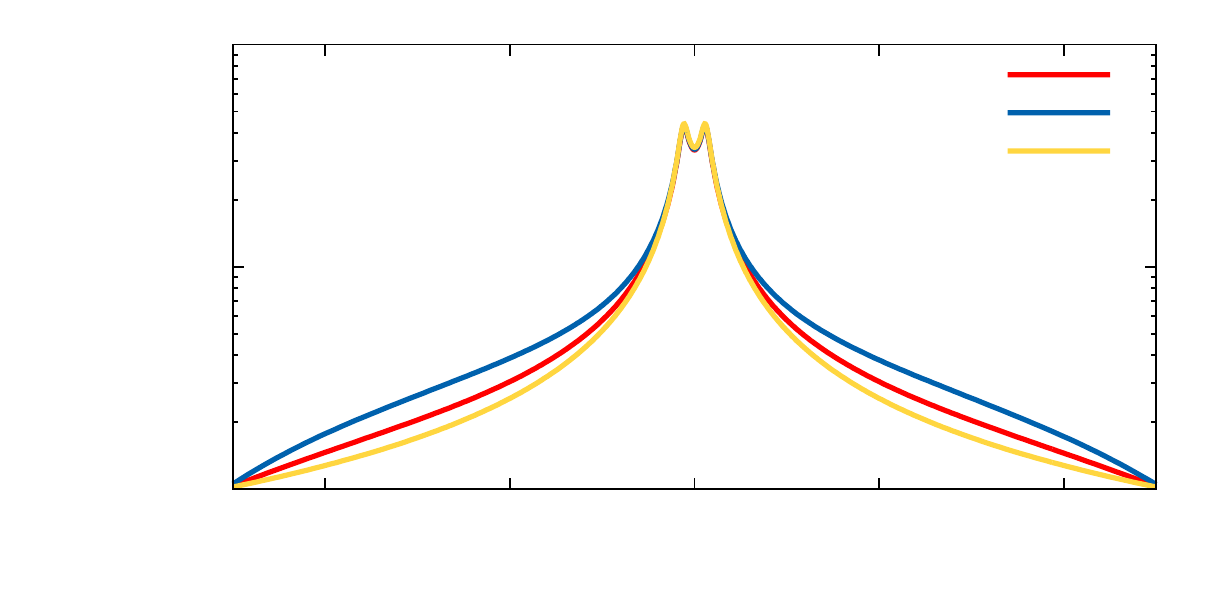}}%
    \gplfronttext
  \end{picture}%
\endgroup

\begingroup
  \makeatletter
  \providecommand\color[2][]{%
    \GenericError{(gnuplot) \space\space\space\@spaces}{%
      Package color not loaded in conjunction with
      terminal option `colourtext'%
    }{See the gnuplot documentation for explanation.%
    }{Either use 'blacktext' in gnuplot or load the package
      color.sty in LaTeX.}%
    \renewcommand\color[2][]{}%
  }%
  \providecommand\includegraphics[2][]{%
    \GenericError{(gnuplot) \space\space\space\@spaces}{%
      Package graphicx or graphics not loaded%
    }{See the gnuplot documentation for explanation.%
    }{The gnuplot epslatex terminal needs graphicx.sty or graphics.sty.}%
    \renewcommand\includegraphics[2][]{}%
  }%
  \providecommand\rotatebox[2]{#2}%
  \@ifundefined{ifGPcolor}{%
    \newif\ifGPcolor
    \GPcolortrue
  }{}%
  \@ifundefined{ifGPblacktext}{%
    \newif\ifGPblacktext
    \GPblacktexttrue
  }{}%
  \let\gplgaddtomacro\g@addto@macro
  \gdef\gplbacktext{}%
  \gdef\gplfronttext{}%
  \makeatother
  \ifGPblacktext
    \def\colorrgb#1{}%
    \def\colorgray#1{}%
  \else
    \ifGPcolor
      \def\colorrgb#1{\color[rgb]{#1}}%
      \def\colorgray#1{\color[gray]{#1}}%
      \expandafter\def\csname LTw\endcsname{\color{white}}%
      \expandafter\def\csname LTb\endcsname{\color{black}}%
      \expandafter\def\csname LTa\endcsname{\color{black}}%
      \expandafter\def\csname LT0\endcsname{\color[rgb]{1,0,0}}%
      \expandafter\def\csname LT1\endcsname{\color[rgb]{0,1,0}}%
      \expandafter\def\csname LT2\endcsname{\color[rgb]{0,0,1}}%
      \expandafter\def\csname LT3\endcsname{\color[rgb]{1,0,1}}%
      \expandafter\def\csname LT4\endcsname{\color[rgb]{0,1,1}}%
      \expandafter\def\csname LT5\endcsname{\color[rgb]{1,1,0}}%
      \expandafter\def\csname LT6\endcsname{\color[rgb]{0,0,0}}%
      \expandafter\def\csname LT7\endcsname{\color[rgb]{1,0.3,0}}%
      \expandafter\def\csname LT8\endcsname{\color[rgb]{0.5,0.5,0.5}}%
    \else
      \def\colorrgb#1{\color{black}}%
      \def\colorgray#1{\color[gray]{#1}}%
      \expandafter\def\csname LTw\endcsname{\color{white}}%
      \expandafter\def\csname LTb\endcsname{\color{black}}%
      \expandafter\def\csname LTa\endcsname{\color{black}}%
      \expandafter\def\csname LT0\endcsname{\color{black}}%
      \expandafter\def\csname LT1\endcsname{\color{black}}%
      \expandafter\def\csname LT2\endcsname{\color{black}}%
      \expandafter\def\csname LT3\endcsname{\color{black}}%
      \expandafter\def\csname LT4\endcsname{\color{black}}%
      \expandafter\def\csname LT5\endcsname{\color{black}}%
      \expandafter\def\csname LT6\endcsname{\color{black}}%
      \expandafter\def\csname LT7\endcsname{\color{black}}%
      \expandafter\def\csname LT8\endcsname{\color{black}}%
    \fi
  \fi
  \setlength{\unitlength}{0.0500bp}%
  \begin{picture}(7200.00,3528.00)%
    \gplgaddtomacro\gplbacktext{%
      \csname LTb\endcsname%
      \put(946,704){\makebox(0,0)[r]{\strut{}$10^{-6}$}}%
      \put(946,1557){\makebox(0,0)[r]{\strut{}$10^{-5}$}}%
      \put(946,2410){\makebox(0,0)[r]{\strut{}$10^{-4}$}}%
      \put(946,3263){\makebox(0,0)[r]{\strut{}$10^{-3}$}}%
      \put(1651,484){\makebox(0,0){\strut{}-40}}%
      \put(2796,484){\makebox(0,0){\strut{}-20}}%
      \put(3941,484){\makebox(0,0){\strut{} 0}}%
      \put(5086,484){\makebox(0,0){\strut{} 20}}%
      \put(6231,484){\makebox(0,0){\strut{} 40}}%
      \put(176,1983){\rotatebox{-270}{\makebox(0,0){\strut{}$|e|$}}}%
      \put(3940,154){\makebox(0,0){\strut{}$x$}}%
    }%
    \gplgaddtomacro\gplfronttext{%
      \csname LTb\endcsname%
      \put(5816,1537){\makebox(0,0)[r]{\strut{}$|e_{\rm A1}|$}}%
      \csname LTb\endcsname%
      \put(5816,1317){\makebox(0,0)[r]{\strut{}$ c_2|e_{\rm A2}|$}}%
      \csname LTb\endcsname%
      \put(5816,1097){\makebox(0,0)[r]{\strut{}$ c_4|e_{\rm A2}|$}}%
      \csname LTb\endcsname%
      \put(5816,877){\makebox(0,0)[r]{\strut{}$ c_6|e_{\rm A2}|$}}%
    }%
    \gplbacktext
    \put(0,0){\includegraphics{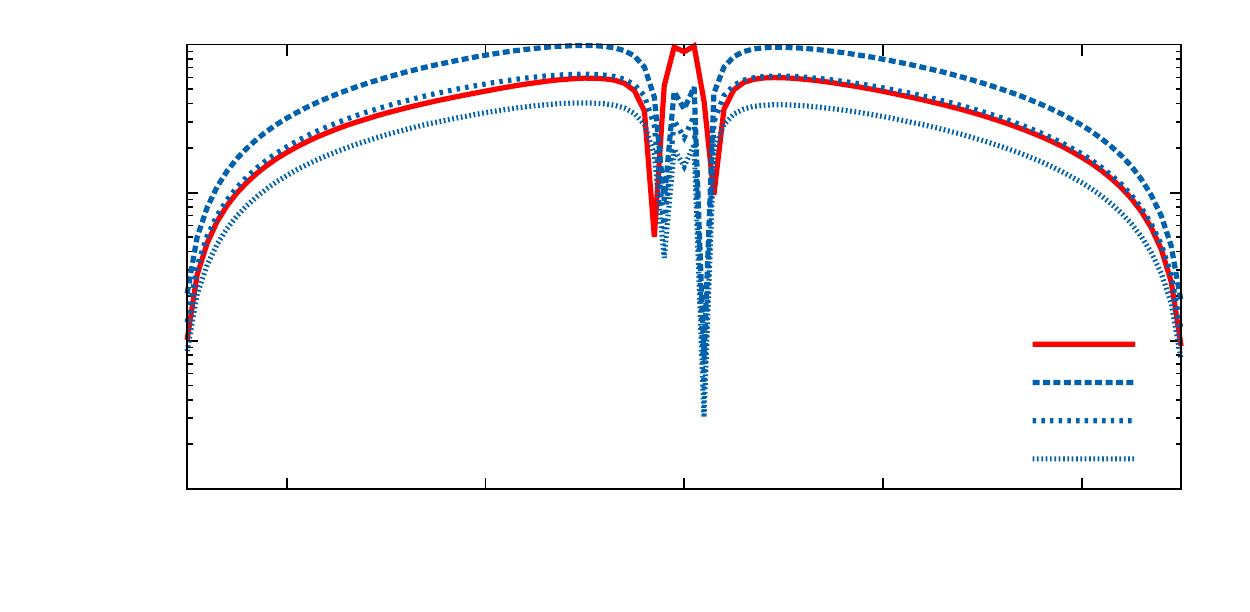}}%
    \gplfronttext
  \end{picture}%
\endgroup

\begingroup
  \makeatletter
  \providecommand\color[2][]{%
    \GenericError{(gnuplot) \space\space\space\@spaces}{%
      Package color not loaded in conjunction with
      terminal option `colourtext'%
    }{See the gnuplot documentation for explanation.%
    }{Either use 'blacktext' in gnuplot or load the package
      color.sty in LaTeX.}%
    \renewcommand\color[2][]{}%
  }%
  \providecommand\includegraphics[2][]{%
    \GenericError{(gnuplot) \space\space\space\@spaces}{%
      Package graphicx or graphics not loaded%
    }{See the gnuplot documentation for explanation.%
    }{The gnuplot epslatex terminal needs graphicx.sty or graphics.sty.}%
    \renewcommand\includegraphics[2][]{}%
  }%
  \providecommand\rotatebox[2]{#2}%
  \@ifundefined{ifGPcolor}{%
    \newif\ifGPcolor
    \GPcolortrue
  }{}%
  \@ifundefined{ifGPblacktext}{%
    \newif\ifGPblacktext
    \GPblacktexttrue
  }{}%
  \let\gplgaddtomacro\g@addto@macro
  \gdef\gplbacktext{}%
  \gdef\gplfronttext{}%
  \makeatother
  \ifGPblacktext
    \def\colorrgb#1{}%
    \def\colorgray#1{}%
  \else
    \ifGPcolor
      \def\colorrgb#1{\color[rgb]{#1}}%
      \def\colorgray#1{\color[gray]{#1}}%
      \expandafter\def\csname LTw\endcsname{\color{white}}%
      \expandafter\def\csname LTb\endcsname{\color{black}}%
      \expandafter\def\csname LTa\endcsname{\color{black}}%
      \expandafter\def\csname LT0\endcsname{\color[rgb]{1,0,0}}%
      \expandafter\def\csname LT1\endcsname{\color[rgb]{0,1,0}}%
      \expandafter\def\csname LT2\endcsname{\color[rgb]{0,0,1}}%
      \expandafter\def\csname LT3\endcsname{\color[rgb]{1,0,1}}%
      \expandafter\def\csname LT4\endcsname{\color[rgb]{0,1,1}}%
      \expandafter\def\csname LT5\endcsname{\color[rgb]{1,1,0}}%
      \expandafter\def\csname LT6\endcsname{\color[rgb]{0,0,0}}%
      \expandafter\def\csname LT7\endcsname{\color[rgb]{1,0.3,0}}%
      \expandafter\def\csname LT8\endcsname{\color[rgb]{0.5,0.5,0.5}}%
    \else
      \def\colorrgb#1{\color{black}}%
      \def\colorgray#1{\color[gray]{#1}}%
      \expandafter\def\csname LTw\endcsname{\color{white}}%
      \expandafter\def\csname LTb\endcsname{\color{black}}%
      \expandafter\def\csname LTa\endcsname{\color{black}}%
      \expandafter\def\csname LT0\endcsname{\color{black}}%
      \expandafter\def\csname LT1\endcsname{\color{black}}%
      \expandafter\def\csname LT2\endcsname{\color{black}}%
      \expandafter\def\csname LT3\endcsname{\color{black}}%
      \expandafter\def\csname LT4\endcsname{\color{black}}%
      \expandafter\def\csname LT5\endcsname{\color{black}}%
      \expandafter\def\csname LT6\endcsname{\color{black}}%
      \expandafter\def\csname LT7\endcsname{\color{black}}%
      \expandafter\def\csname LT8\endcsname{\color{black}}%
    \fi
  \fi
  \setlength{\unitlength}{0.0500bp}%
  \begin{picture}(7200.00,3528.00)%
    \gplgaddtomacro\gplbacktext{%
      \csname LTb\endcsname%
      \put(1078,704){\makebox(0,0)[r]{\strut{}$10^{-14}$}}%
      \put(1078,1024){\makebox(0,0)[r]{\strut{}$10^{-12}$}}%
      \put(1078,1344){\makebox(0,0)[r]{\strut{}$10^{-10}$}}%
      \put(1078,1664){\makebox(0,0)[r]{\strut{}$10^{-8}$}}%
      \put(1078,1984){\makebox(0,0)[r]{\strut{}$10^{-6}$}}%
      \put(1078,2303){\makebox(0,0)[r]{\strut{}$10^{-4}$}}%
      \put(1078,2623){\makebox(0,0)[r]{\strut{}$10^{-2}$}}%
      \put(1078,2943){\makebox(0,0)[r]{\strut{}$10^{0}$}}%
      \put(1078,3263){\makebox(0,0)[r]{\strut{}$10^{2}$}}%
      \put(1769,484){\makebox(0,0){\strut{}-40}}%
      \put(2888,484){\makebox(0,0){\strut{}-20}}%
      \put(4007,484){\makebox(0,0){\strut{} 0}}%
      \put(5125,484){\makebox(0,0){\strut{} 20}}%
      \put(6244,484){\makebox(0,0){\strut{} 40}}%
      \put(176,1983){\rotatebox{-270}{\makebox(0,0){\strut{}$|H|$}}}%
      \put(4006,154){\makebox(0,0){\strut{}$x$}}%
    }%
    \gplgaddtomacro\gplfronttext{%
      \csname LTb\endcsname%
      \put(5816,3090){\makebox(0,0)[r]{\strut{}A1}}%
      \csname LTb\endcsname%
      \put(5816,2870){\makebox(0,0)[r]{\strut{}A2}}%
      \csname LTb\endcsname%
      \put(5816,2650){\makebox(0,0)[r]{\strut{}TP1}}%
      \csname LTb\endcsname%
      \put(5816,2430){\makebox(0,0)[r]{\strut{}TP2}}%
    }%
    \gplbacktext
    \put(0,0){\includegraphics{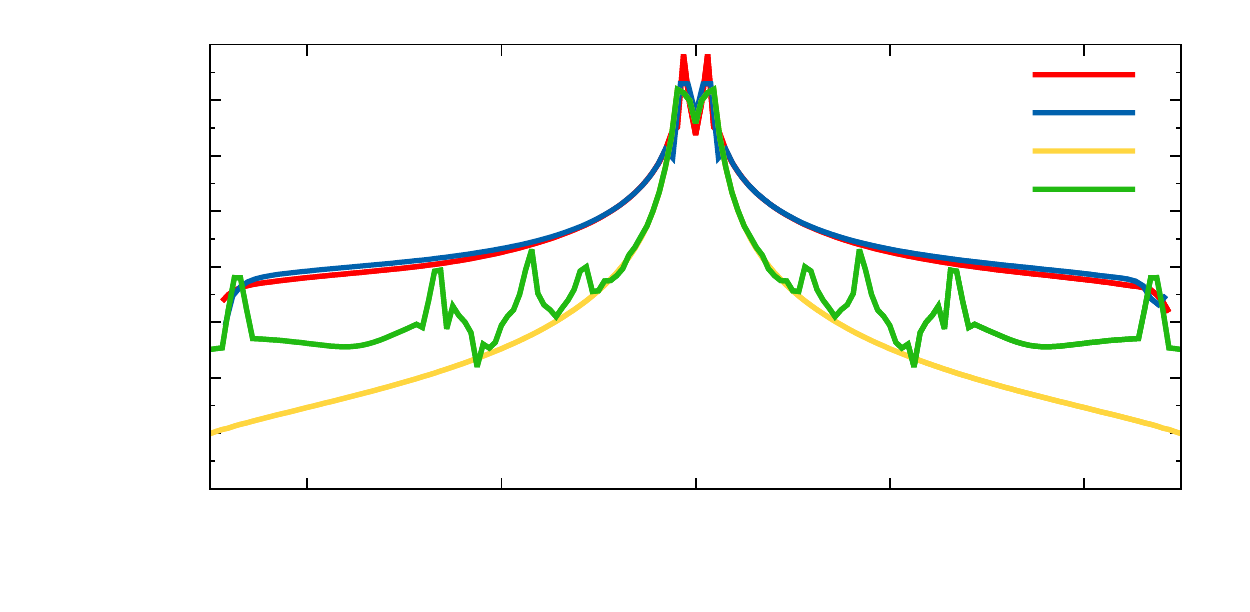}}%
    \gplfronttext
  \end{picture}%
\endgroup

\caption{Conformal factor for the QC0 configuration. Top: $u-1$ for two different resolutions,
from runs A1 and A1, plotted against the TwoPuncture solution TP1. 
Middle: error $e_{\rm A*} \equiv u_{\rm TP1}-u_{\rm A*}$. The error of the coarse resolution A2 has been scaled
according to the second-, fourth- and sixth-order expectation. The data shows fourth-order
scaling, consistently with the spatial differencing scheme used by the solver. 
Bottom: Hamiltonian constraint violation for runs A1, A2, TP1 and TP2.
\label{fig:TP}}
\ece
\efi

\bfi
\bce
\begingroup
  \makeatletter
  \providecommand\color[2][]{%
    \GenericError{(gnuplot) \space\space\space\@spaces}{%
      Package color not loaded in conjunction with
      terminal option `colourtext'%
    }{See the gnuplot documentation for explanation.%
    }{Either use 'blacktext' in gnuplot or load the package
      color.sty in LaTeX.}%
    \renewcommand\color[2][]{}%
  }%
  \providecommand\includegraphics[2][]{%
    \GenericError{(gnuplot) \space\space\space\@spaces}{%
      Package graphicx or graphics not loaded%
    }{See the gnuplot documentation for explanation.%
    }{The gnuplot epslatex terminal needs graphicx.sty or graphics.sty.}%
    \renewcommand\includegraphics[2][]{}%
  }%
  \providecommand\rotatebox[2]{#2}%
  \@ifundefined{ifGPcolor}{%
    \newif\ifGPcolor
    \GPcolortrue
  }{}%
  \@ifundefined{ifGPblacktext}{%
    \newif\ifGPblacktext
    \GPblacktexttrue
  }{}%
  \let\gplgaddtomacro\g@addto@macro
  \gdef\gplbacktext{}%
  \gdef\gplfronttext{}%
  \makeatother
  \ifGPblacktext
    \def\colorrgb#1{}%
    \def\colorgray#1{}%
  \else
    \ifGPcolor
      \def\colorrgb#1{\color[rgb]{#1}}%
      \def\colorgray#1{\color[gray]{#1}}%
      \expandafter\def\csname LTw\endcsname{\color{white}}%
      \expandafter\def\csname LTb\endcsname{\color{black}}%
      \expandafter\def\csname LTa\endcsname{\color{black}}%
      \expandafter\def\csname LT0\endcsname{\color[rgb]{1,0,0}}%
      \expandafter\def\csname LT1\endcsname{\color[rgb]{0,1,0}}%
      \expandafter\def\csname LT2\endcsname{\color[rgb]{0,0,1}}%
      \expandafter\def\csname LT3\endcsname{\color[rgb]{1,0,1}}%
      \expandafter\def\csname LT4\endcsname{\color[rgb]{0,1,1}}%
      \expandafter\def\csname LT5\endcsname{\color[rgb]{1,1,0}}%
      \expandafter\def\csname LT6\endcsname{\color[rgb]{0,0,0}}%
      \expandafter\def\csname LT7\endcsname{\color[rgb]{1,0.3,0}}%
      \expandafter\def\csname LT8\endcsname{\color[rgb]{0.5,0.5,0.5}}%
    \else
      \def\colorrgb#1{\color{black}}%
      \def\colorgray#1{\color[gray]{#1}}%
      \expandafter\def\csname LTw\endcsname{\color{white}}%
      \expandafter\def\csname LTb\endcsname{\color{black}}%
      \expandafter\def\csname LTa\endcsname{\color{black}}%
      \expandafter\def\csname LT0\endcsname{\color{black}}%
      \expandafter\def\csname LT1\endcsname{\color{black}}%
      \expandafter\def\csname LT2\endcsname{\color{black}}%
      \expandafter\def\csname LT3\endcsname{\color{black}}%
      \expandafter\def\csname LT4\endcsname{\color{black}}%
      \expandafter\def\csname LT5\endcsname{\color{black}}%
      \expandafter\def\csname LT6\endcsname{\color{black}}%
      \expandafter\def\csname LT7\endcsname{\color{black}}%
      \expandafter\def\csname LT8\endcsname{\color{black}}%
    \fi
  \fi
  \setlength{\unitlength}{0.0500bp}%
  \begin{picture}(7200.00,3024.00)%
    \gplgaddtomacro\gplbacktext{%
      \csname LTb\endcsname%
      \put(814,704){\makebox(0,0)[r]{\strut{}$10^{-4}$}}%
      \put(814,1389){\makebox(0,0)[r]{\strut{}$10^{-3}$}}%
      \put(814,2074){\makebox(0,0)[r]{\strut{}$10^{-2}$}}%
      \put(814,2759){\makebox(0,0)[r]{\strut{}$10^{-1}$}}%
      \put(946,484){\makebox(0,0){\strut{} 0}}%
      \put(1759,484){\makebox(0,0){\strut{} 100}}%
      \put(2573,484){\makebox(0,0){\strut{} 200}}%
      \put(3386,484){\makebox(0,0){\strut{} 300}}%
      \put(4200,484){\makebox(0,0){\strut{} 400}}%
      \put(5013,484){\makebox(0,0){\strut{} 500}}%
      \put(5827,484){\makebox(0,0){\strut{} 600}}%
      \put(6640,484){\makebox(0,0){\strut{} 700}}%
      \put(176,1731){\rotatebox{-270}{\makebox(0,0){\strut{}$|e|$}}}%
      \put(3874,154){\makebox(0,0){\strut{}Iteration}}%
    }%
    \gplgaddtomacro\gplfronttext{%
    }%
    \gplbacktext
    \put(0,0){\includegraphics{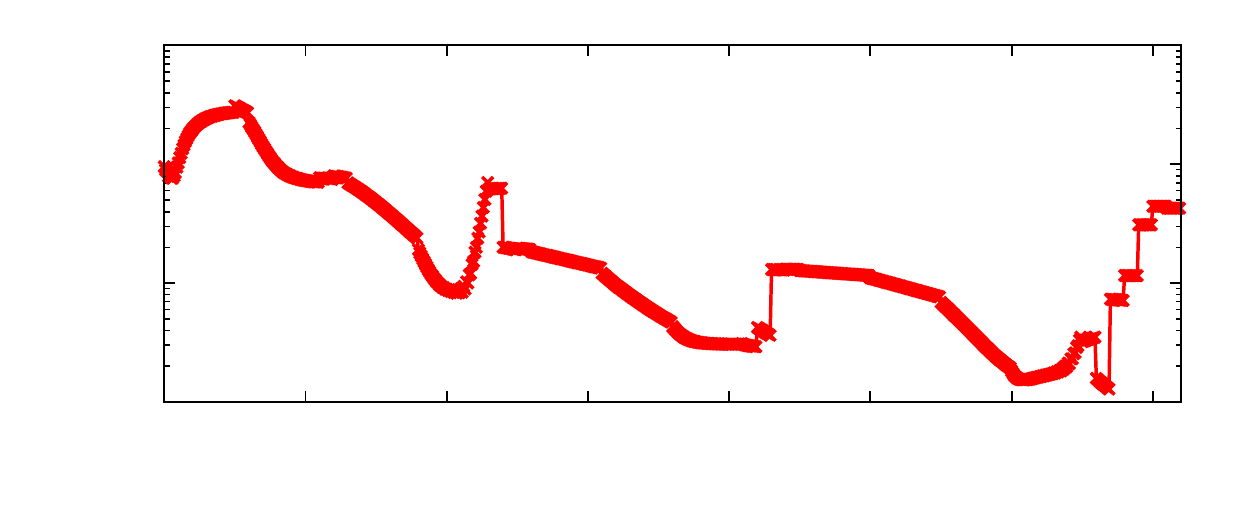}}%
    \gplfronttext
  \end{picture}%
\endgroup

\begingroup
  \makeatletter
  \providecommand\color[2][]{%
    \GenericError{(gnuplot) \space\space\space\@spaces}{%
      Package color not loaded in conjunction with
      terminal option `colourtext'%
    }{See the gnuplot documentation for explanation.%
    }{Either use 'blacktext' in gnuplot or load the package
      color.sty in LaTeX.}%
    \renewcommand\color[2][]{}%
  }%
  \providecommand\includegraphics[2][]{%
    \GenericError{(gnuplot) \space\space\space\@spaces}{%
      Package graphicx or graphics not loaded%
    }{See the gnuplot documentation for explanation.%
    }{The gnuplot epslatex terminal needs graphicx.sty or graphics.sty.}%
    \renewcommand\includegraphics[2][]{}%
  }%
  \providecommand\rotatebox[2]{#2}%
  \@ifundefined{ifGPcolor}{%
    \newif\ifGPcolor
    \GPcolortrue
  }{}%
  \@ifundefined{ifGPblacktext}{%
    \newif\ifGPblacktext
    \GPblacktexttrue
  }{}%
  \let\gplgaddtomacro\g@addto@macro
  \gdef\gplbacktext{}%
  \gdef\gplfronttext{}%
  \makeatother
  \ifGPblacktext
    \def\colorrgb#1{}%
    \def\colorgray#1{}%
  \else
    \ifGPcolor
      \def\colorrgb#1{\color[rgb]{#1}}%
      \def\colorgray#1{\color[gray]{#1}}%
      \expandafter\def\csname LTw\endcsname{\color{white}}%
      \expandafter\def\csname LTb\endcsname{\color{black}}%
      \expandafter\def\csname LTa\endcsname{\color{black}}%
      \expandafter\def\csname LT0\endcsname{\color[rgb]{1,0,0}}%
      \expandafter\def\csname LT1\endcsname{\color[rgb]{0,1,0}}%
      \expandafter\def\csname LT2\endcsname{\color[rgb]{0,0,1}}%
      \expandafter\def\csname LT3\endcsname{\color[rgb]{1,0,1}}%
      \expandafter\def\csname LT4\endcsname{\color[rgb]{0,1,1}}%
      \expandafter\def\csname LT5\endcsname{\color[rgb]{1,1,0}}%
      \expandafter\def\csname LT6\endcsname{\color[rgb]{0,0,0}}%
      \expandafter\def\csname LT7\endcsname{\color[rgb]{1,0.3,0}}%
      \expandafter\def\csname LT8\endcsname{\color[rgb]{0.5,0.5,0.5}}%
    \else
      \def\colorrgb#1{\color{black}}%
      \def\colorgray#1{\color[gray]{#1}}%
      \expandafter\def\csname LTw\endcsname{\color{white}}%
      \expandafter\def\csname LTb\endcsname{\color{black}}%
      \expandafter\def\csname LTa\endcsname{\color{black}}%
      \expandafter\def\csname LT0\endcsname{\color{black}}%
      \expandafter\def\csname LT1\endcsname{\color{black}}%
      \expandafter\def\csname LT2\endcsname{\color{black}}%
      \expandafter\def\csname LT3\endcsname{\color{black}}%
      \expandafter\def\csname LT4\endcsname{\color{black}}%
      \expandafter\def\csname LT5\endcsname{\color{black}}%
      \expandafter\def\csname LT6\endcsname{\color{black}}%
      \expandafter\def\csname LT7\endcsname{\color{black}}%
      \expandafter\def\csname LT8\endcsname{\color{black}}%
    \fi
  \fi
  \setlength{\unitlength}{0.0500bp}%
  \begin{picture}(7200.00,3528.00)%
    \gplgaddtomacro\gplbacktext{%
      \csname LTb\endcsname%
      \put(858,765){\makebox(0,0)[r]{\strut{}$10^{-12}$}}%
      \put(858,1224){\makebox(0,0)[r]{\strut{}$10^{-10}$}}%
      \put(858,1682){\makebox(0,0)[r]{\strut{}$10^{-8}$}}%
      \put(858,2141){\makebox(0,0)[r]{\strut{}$10^{-6}$}}%
      \put(858,2599){\makebox(0,0)[r]{\strut{}$10^{-4}$}}%
      \put(858,3058){\makebox(0,0)[r]{\strut{}$10^{-2}$}}%
      \put(990,484){\makebox(0,0){\strut{} 0}}%
      \put(2153,484){\makebox(0,0){\strut{} 10}}%
      \put(3315,484){\makebox(0,0){\strut{} 20}}%
      \put(4478,484){\makebox(0,0){\strut{} 30}}%
      \put(5640,484){\makebox(0,0){\strut{} 40}}%
      \put(6803,484){\makebox(0,0){\strut{} 50}}%
      \put(3896,154){\makebox(0,0){\strut{}$x$}}%
    }%
    \gplgaddtomacro\gplfronttext{%
      \csname LTb\endcsname%
      \put(5816,3090){\makebox(0,0)[r]{\strut{}$|\Delta u_{\rm TP1}^{\rm C3}-\Delta u_{\rm TP1}^{\rm C4}|$}}%
      \csname LTb\endcsname%
      \put(5816,2870){\makebox(0,0)[r]{\strut{}$|e_{\rm A1}|$}}%
    }%
    \gplbacktext
    \put(0,0){\includegraphics{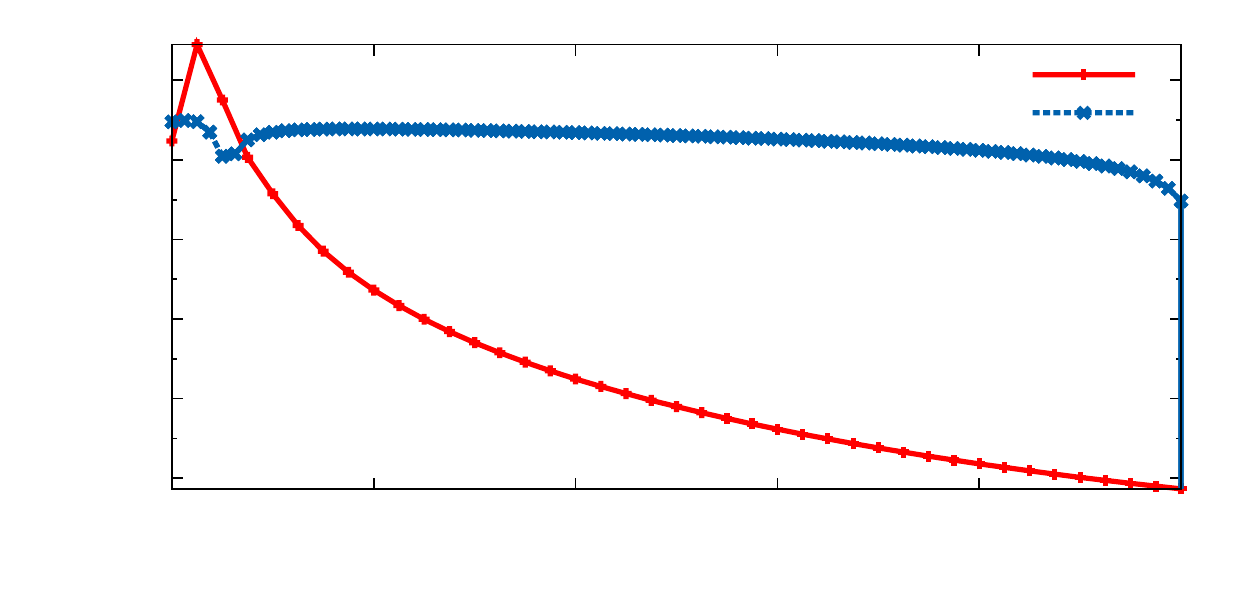}}%
    \gplfronttext
  \end{picture}%
\endgroup

\caption{Top: solution error for run A1. Notice that it is not possible to estimate the 
truncation error here as in the previous examples, due to the lack of an analytical
exact solution. One can, however, take the absolute value of the difference 
in the numerical laplacian $\Delta u_{\rm TP1}$ computed on level C3 and C4
and compare it with the solution error $e_{\rm A1} = u_{\rm TP1} - u_{\rm A1}$ on C4
(bottom): this shows that the large truncation error due to finite-differencing around
the puncture sets a limit for the convergence of $u_{\rm A1}$ to $u_{\rm TP1}$
over the whole domain. 
\label{fig:TPerr}}
\ece
\efi
\section{Puncture lattice}
\label{sec:lattice}
In this section, I will use the solver described and tested above to solve the constraint system
for a space filled with a cubic black-hole lattice. The solution will in practice be carried
out on a single cell of this lattice, with periodic boundary conditions at the cell faces.
Each cell will contain a single black hole at its center.

As discussed in~\cite{Bentivegna:2012ei}, if one is interested in conformally flat initial data,
the extrinsic curvature cannot be zero. A possible construction for a lattice of non-spinning,
non-maximally-sliced black holes has been proposed in~\cite{Yoo:2012jz}: according to this
recipe, one starts with the following \emph{ansatz} for the conformal factor:
\beq
\psi = u + \frac{m}{2 r} \left ( 1 - W(r) \right)
\eeq
and for the trace of the extrinsic curvature:
\beq
K=K_c W(r)
\eeq
where $K_c$ is a constant and $W(r)$ is a function which is equal to zero inside a sphere
around $r=0$, equal to one at large distances from $r=0$, and has a
transition region in between, e.g.:
\bea
\label{eq:W}
W(r) &=& \left\{
  \begin{array}{ll}
  0 & \textrm{for } 0 \le r \le \ell \\
  ((r-\ell-\sigma)^6 \sigma^{-6}-1)^6&\textrm{for } \ell \le r \le \ell + \sigma \\
  1 & \textrm{for } \ell + \sigma \le r
  \end{array} 
\right.
\eea
This has the advantage of starting out close to familiar solutions (the Schwarzschild
solution for $r<\ell$ and a Friedmann-Lema\^itre-Robertson-Walker model for $r>\ell+\sigma$) 
on most of the domain.

The system to solve is then:
\begin{eqnarray}
 \Delta u - \Delta \left( \frac{m}{2r} W(r)\right)- \frac{K^2}{12}\,\psi^5 + \frac{1}{8} {A}_{ij} {A}^{ij} \psi^{-7} = 0 \\
 \Delta X^{i} + \partial^i \partial_j X^j - \frac{2}{3} \psi^6 \partial^i K = 0
\end{eqnarray}
with:
\beq
A_{ij}=\partial_i X_j + \partial_j X_i - \frac{2}{3} \delta_{ij} \partial_k X^k
\eeq
for $u$ and $X^i$. As in section~\ref{sec:spherp}, an integral condition, which now
reads:
\beq
\label{eq:integr}
K_c^2 \int W^2 \psi^5 - 2 \pi ( m + \frac{1}{8}\int A_{ij} A^{ij} \psi^{-7} ) = 0 
\eeq
has to be respected by all the solutions. Again, I find that enforcing this
condition explicitly is not necessary, but doing so greatly speeds up the
convergence. In~\cite{Yoo:2012jz}, the condition is imposed
by solving for $K_c$, and updating this parameter in the Hamiltonian constraint
equation at the next iteration. However, resetting the scale of $\psi$ as
in section~\ref{sec:spherp} leads to an even faster convergence. 
Here, we need to find a root of:
\beq
\label{eq:f}
f(A) = K_c^2 \int W^2 (\psi+A)^5 - 2 \pi ( m + \frac{1}{8}\int A_{ij} A^{ij} (\psi+A)^{-7} )
\eeq 
which again is solved with the Newton-Raphson method.
For the same reasons described in~\ref{sec:spherp}, there is no condition associated to the
momentum constraint.

I first solve for a configuration L1 where the outer boundaries of 
the periodic cell are located at $x=\pm 5$, $y=\pm 5$, and $z=\pm 5$,
the black hole at the origin has a mass $m=1$, the trace of the 
extrinsic curvature $K_c=-0.21$ initially, and the parameters defining
the transition region are $\ell=0.5$ and $\sigma=4$. The cell is covered 
by five refinement levels (all extending out to the outer boundaries),
with coarse spacing $\Delta_0=1$.
This time, the Gauss-Seidel solver operates 100 times on the coarsest level,
10 on the finest, and 100 and 10 times on the intermediate ones on the downward
and upward portion of each V-cycle, respectively.

In Figure~\ref{fig:LPerr}, it is shown how the errors $u$ and $X^x$ converge
up to the discretization error, in a manner similar to what was presented
in the code-test section above. Notice, however, that the definitions of 
solution error and truncation error adopted here are rather different, 
given to the lack of an exact solution to take as reference. Therefore,
$e$ in Figure~\ref{fig:LPerr} is not the difference between the numerical
solution and the (unknown) exact solution, but the estimate of this error
obtained in the multigrid scheme, i.e. by relaxing equation (\ref{eq:MGerr}).
Likewise, the truncation error is obtained by subtracting the analytic
derivative of $W$ in (\ref{eq:W}) from the numerical one; this is reasonable
given that the largest contribution to the finite-differencing error is
likely to come from computing derivatives in the transition region. However,
it has to be kept in mind that this is not the truncation error associated 
to any of the solution functions. 

\bfi
\bce
\begingroup
  \makeatletter
  \providecommand\color[2][]{%
    \GenericError{(gnuplot) \space\space\space\@spaces}{%
      Package color not loaded in conjunction with
      terminal option `colourtext'%
    }{See the gnuplot documentation for explanation.%
    }{Either use 'blacktext' in gnuplot or load the package
      color.sty in LaTeX.}%
    \renewcommand\color[2][]{}%
  }%
  \providecommand\includegraphics[2][]{%
    \GenericError{(gnuplot) \space\space\space\@spaces}{%
      Package graphicx or graphics not loaded%
    }{See the gnuplot documentation for explanation.%
    }{The gnuplot epslatex terminal needs graphicx.sty or graphics.sty.}%
    \renewcommand\includegraphics[2][]{}%
  }%
  \providecommand\rotatebox[2]{#2}%
  \@ifundefined{ifGPcolor}{%
    \newif\ifGPcolor
    \GPcolortrue
  }{}%
  \@ifundefined{ifGPblacktext}{%
    \newif\ifGPblacktext
    \GPblacktexttrue
  }{}%
  \let\gplgaddtomacro\g@addto@macro
  \gdef\gplbacktext{}%
  \gdef\gplfronttext{}%
  \makeatother
  \ifGPblacktext
    \def\colorrgb#1{}%
    \def\colorgray#1{}%
  \else
    \ifGPcolor
      \def\colorrgb#1{\color[rgb]{#1}}%
      \def\colorgray#1{\color[gray]{#1}}%
      \expandafter\def\csname LTw\endcsname{\color{white}}%
      \expandafter\def\csname LTb\endcsname{\color{black}}%
      \expandafter\def\csname LTa\endcsname{\color{black}}%
      \expandafter\def\csname LT0\endcsname{\color[rgb]{1,0,0}}%
      \expandafter\def\csname LT1\endcsname{\color[rgb]{0,1,0}}%
      \expandafter\def\csname LT2\endcsname{\color[rgb]{0,0,1}}%
      \expandafter\def\csname LT3\endcsname{\color[rgb]{1,0,1}}%
      \expandafter\def\csname LT4\endcsname{\color[rgb]{0,1,1}}%
      \expandafter\def\csname LT5\endcsname{\color[rgb]{1,1,0}}%
      \expandafter\def\csname LT6\endcsname{\color[rgb]{0,0,0}}%
      \expandafter\def\csname LT7\endcsname{\color[rgb]{1,0.3,0}}%
      \expandafter\def\csname LT8\endcsname{\color[rgb]{0.5,0.5,0.5}}%
    \else
      \def\colorrgb#1{\color{black}}%
      \def\colorgray#1{\color[gray]{#1}}%
      \expandafter\def\csname LTw\endcsname{\color{white}}%
      \expandafter\def\csname LTb\endcsname{\color{black}}%
      \expandafter\def\csname LTa\endcsname{\color{black}}%
      \expandafter\def\csname LT0\endcsname{\color{black}}%
      \expandafter\def\csname LT1\endcsname{\color{black}}%
      \expandafter\def\csname LT2\endcsname{\color{black}}%
      \expandafter\def\csname LT3\endcsname{\color{black}}%
      \expandafter\def\csname LT4\endcsname{\color{black}}%
      \expandafter\def\csname LT5\endcsname{\color{black}}%
      \expandafter\def\csname LT6\endcsname{\color{black}}%
      \expandafter\def\csname LT7\endcsname{\color{black}}%
      \expandafter\def\csname LT8\endcsname{\color{black}}%
    \fi
  \fi
  \setlength{\unitlength}{0.0500bp}%
  \begin{picture}(7200.00,3528.00)%
    \gplgaddtomacro\gplbacktext{%
      \csname LTb\endcsname%
      \put(814,704){\makebox(0,0)[r]{\strut{}$10^{-7}$}}%
      \put(814,1086){\makebox(0,0)[r]{\strut{}$10^{-6}$}}%
      \put(814,1468){\makebox(0,0)[r]{\strut{}$10^{-5}$}}%
      \put(814,1850){\makebox(0,0)[r]{\strut{}$10^{-4}$}}%
      \put(814,2232){\makebox(0,0)[r]{\strut{}$10^{-3}$}}%
      \put(814,2614){\makebox(0,0)[r]{\strut{}$10^{-2}$}}%
      \put(814,2996){\makebox(0,0)[r]{\strut{}$10^{-1}$}}%
      \put(973,484){\makebox(0,0){\strut{} 0}}%
      \put(2073,484){\makebox(0,0){\strut{} 200}}%
      \put(3173,484){\makebox(0,0){\strut{} 400}}%
      \put(4273,484){\makebox(0,0){\strut{} 600}}%
      \put(5373,484){\makebox(0,0){\strut{} 800}}%
      \put(6473,484){\makebox(0,0){\strut{} 1000}}%
      \put(176,1983){\rotatebox{-270}{\makebox(0,0){\strut{}$|e|$}}}%
      \put(3874,154){\makebox(0,0){\strut{}Iteration}}%
    }%
    \gplgaddtomacro\gplfronttext{%
      \csname LTb\endcsname%
      \put(5816,3090){\makebox(0,0)[r]{\strut{}$|e_{\psi}|$}}%
      \csname LTb\endcsname%
      \put(5816,2870){\makebox(0,0)[r]{\strut{}$|e_{X^x}|$}}%
    }%
    \gplbacktext
    \put(0,0){\includegraphics{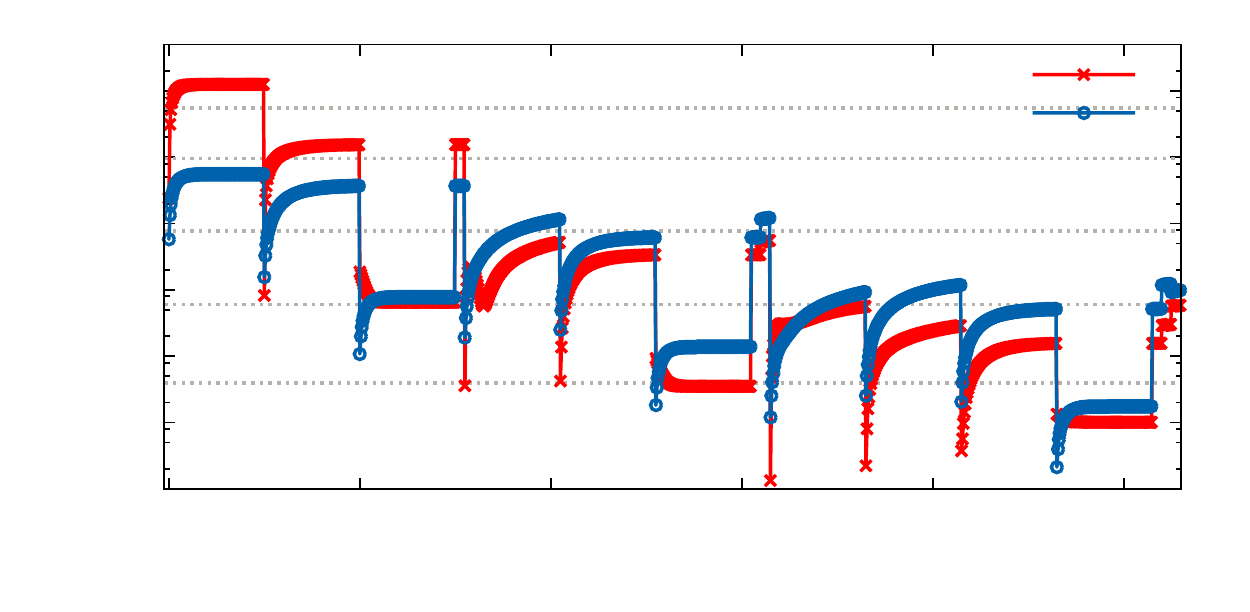}}%
    \gplfronttext
  \end{picture}%
\endgroup

\caption{Norm of the solution errors corresponding to $u$ (red) and $X^x$ (blue)
for run L1.
The dashed grey lines denote the truncation error on the five grids (the topmost
corresponding to the coarsest grid).
\label{fig:LPerr}}
\ece
\efi

Figure~\ref{fig:LP} shows the final solution for $u$ and $X^x$
(the other two components of $X^i$ can be obtained by remembering 
the symmetry of this system under permutations of the three Cartesian directions).
On the same plot, three other variants L4, L5 and L6, obtained with the
same parameters but changing $m$ to $0.5$, $2$ and $5$ respectively, are displayed.
Qualitatively, runs with different $m$ appear similar, with a change in scale
of both $u$ and $X^x$ as the major difference. This is quantified in Figure~\ref{fig:uvsM},
where the value of $u$ at the center of the cell and at a vertex, and
the maximum of $X^x$ on the $x$ axis are shown as a function of $m$.

\bfi
\bce
\begingroup
  \makeatletter
  \providecommand\color[2][]{%
    \GenericError{(gnuplot) \space\space\space\@spaces}{%
      Package color not loaded in conjunction with
      terminal option `colourtext'%
    }{See the gnuplot documentation for explanation.%
    }{Either use 'blacktext' in gnuplot or load the package
      color.sty in LaTeX.}%
    \renewcommand\color[2][]{}%
  }%
  \providecommand\includegraphics[2][]{%
    \GenericError{(gnuplot) \space\space\space\@spaces}{%
      Package graphicx or graphics not loaded%
    }{See the gnuplot documentation for explanation.%
    }{The gnuplot epslatex terminal needs graphicx.sty or graphics.sty.}%
    \renewcommand\includegraphics[2][]{}%
  }%
  \providecommand\rotatebox[2]{#2}%
  \@ifundefined{ifGPcolor}{%
    \newif\ifGPcolor
    \GPcolortrue
  }{}%
  \@ifundefined{ifGPblacktext}{%
    \newif\ifGPblacktext
    \GPblacktexttrue
  }{}%
  \let\gplgaddtomacro\g@addto@macro
  \gdef\gplbacktext{}%
  \gdef\gplfronttext{}%
  \makeatother
  \ifGPblacktext
    \def\colorrgb#1{}%
    \def\colorgray#1{}%
  \else
    \ifGPcolor
      \def\colorrgb#1{\color[rgb]{#1}}%
      \def\colorgray#1{\color[gray]{#1}}%
      \expandafter\def\csname LTw\endcsname{\color{white}}%
      \expandafter\def\csname LTb\endcsname{\color{black}}%
      \expandafter\def\csname LTa\endcsname{\color{black}}%
      \expandafter\def\csname LT0\endcsname{\color[rgb]{1,0,0}}%
      \expandafter\def\csname LT1\endcsname{\color[rgb]{0,1,0}}%
      \expandafter\def\csname LT2\endcsname{\color[rgb]{0,0,1}}%
      \expandafter\def\csname LT3\endcsname{\color[rgb]{1,0,1}}%
      \expandafter\def\csname LT4\endcsname{\color[rgb]{0,1,1}}%
      \expandafter\def\csname LT5\endcsname{\color[rgb]{1,1,0}}%
      \expandafter\def\csname LT6\endcsname{\color[rgb]{0,0,0}}%
      \expandafter\def\csname LT7\endcsname{\color[rgb]{1,0.3,0}}%
      \expandafter\def\csname LT8\endcsname{\color[rgb]{0.5,0.5,0.5}}%
    \else
      \def\colorrgb#1{\color{black}}%
      \def\colorgray#1{\color[gray]{#1}}%
      \expandafter\def\csname LTw\endcsname{\color{white}}%
      \expandafter\def\csname LTb\endcsname{\color{black}}%
      \expandafter\def\csname LTa\endcsname{\color{black}}%
      \expandafter\def\csname LT0\endcsname{\color{black}}%
      \expandafter\def\csname LT1\endcsname{\color{black}}%
      \expandafter\def\csname LT2\endcsname{\color{black}}%
      \expandafter\def\csname LT3\endcsname{\color{black}}%
      \expandafter\def\csname LT4\endcsname{\color{black}}%
      \expandafter\def\csname LT5\endcsname{\color{black}}%
      \expandafter\def\csname LT6\endcsname{\color{black}}%
      \expandafter\def\csname LT7\endcsname{\color{black}}%
      \expandafter\def\csname LT8\endcsname{\color{black}}%
    \fi
  \fi
  \setlength{\unitlength}{0.0500bp}%
  \begin{picture}(7200.00,3528.00)%
    \gplgaddtomacro\gplbacktext{%
      \csname LTb\endcsname%
      \put(946,704){\makebox(0,0)[r]{\strut{} 0.8}}%
      \put(946,1005){\makebox(0,0)[r]{\strut{} 1}}%
      \put(946,1306){\makebox(0,0)[r]{\strut{} 1.2}}%
      \put(946,1607){\makebox(0,0)[r]{\strut{} 1.4}}%
      \put(946,1908){\makebox(0,0)[r]{\strut{} 1.6}}%
      \put(946,2209){\makebox(0,0)[r]{\strut{} 1.8}}%
      \put(946,2510){\makebox(0,0)[r]{\strut{} 2}}%
      \put(946,2811){\makebox(0,0)[r]{\strut{} 2.2}}%
      \put(946,3112){\makebox(0,0)[r]{\strut{} 2.4}}%
      \put(1651,484){\makebox(0,0){\strut{}-4}}%
      \put(2796,484){\makebox(0,0){\strut{}-2}}%
      \put(3941,484){\makebox(0,0){\strut{} 0}}%
      \put(5086,484){\makebox(0,0){\strut{} 2}}%
      \put(6231,484){\makebox(0,0){\strut{} 4}}%
      \put(176,1983){\rotatebox{-270}{\makebox(0,0){\strut{}$u$}}}%
      \put(3940,154){\makebox(0,0){\strut{}$x$}}%
    }%
    \gplgaddtomacro\gplfronttext{%
      \csname LTb\endcsname%
      \put(2459,3090){\makebox(0,0)[r]{\strut{}L1}}%
      \csname LTb\endcsname%
      \put(3578,3090){\makebox(0,0)[r]{\strut{}L4}}%
      \csname LTb\endcsname%
      \put(4697,3090){\makebox(0,0)[r]{\strut{}L5}}%
      \csname LTb\endcsname%
      \put(5816,3090){\makebox(0,0)[r]{\strut{}L6}}%
    }%
    \gplbacktext
    \put(0,0){\includegraphics{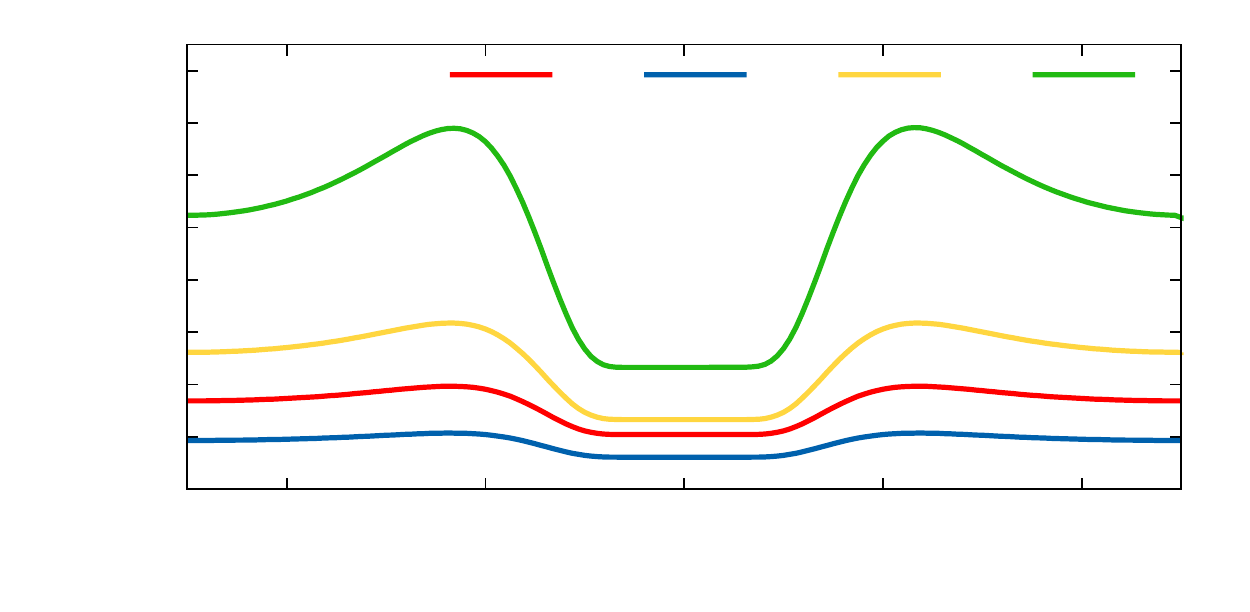}}%
    \gplfronttext
  \end{picture}%
\endgroup

\begingroup
  \makeatletter
  \providecommand\color[2][]{%
    \GenericError{(gnuplot) \space\space\space\@spaces}{%
      Package color not loaded in conjunction with
      terminal option `colourtext'%
    }{See the gnuplot documentation for explanation.%
    }{Either use 'blacktext' in gnuplot or load the package
      color.sty in LaTeX.}%
    \renewcommand\color[2][]{}%
  }%
  \providecommand\includegraphics[2][]{%
    \GenericError{(gnuplot) \space\space\space\@spaces}{%
      Package graphicx or graphics not loaded%
    }{See the gnuplot documentation for explanation.%
    }{The gnuplot epslatex terminal needs graphicx.sty or graphics.sty.}%
    \renewcommand\includegraphics[2][]{}%
  }%
  \providecommand\rotatebox[2]{#2}%
  \@ifundefined{ifGPcolor}{%
    \newif\ifGPcolor
    \GPcolortrue
  }{}%
  \@ifundefined{ifGPblacktext}{%
    \newif\ifGPblacktext
    \GPblacktexttrue
  }{}%
  \let\gplgaddtomacro\g@addto@macro
  \gdef\gplbacktext{}%
  \gdef\gplfronttext{}%
  \makeatother
  \ifGPblacktext
    \def\colorrgb#1{}%
    \def\colorgray#1{}%
  \else
    \ifGPcolor
      \def\colorrgb#1{\color[rgb]{#1}}%
      \def\colorgray#1{\color[gray]{#1}}%
      \expandafter\def\csname LTw\endcsname{\color{white}}%
      \expandafter\def\csname LTb\endcsname{\color{black}}%
      \expandafter\def\csname LTa\endcsname{\color{black}}%
      \expandafter\def\csname LT0\endcsname{\color[rgb]{1,0,0}}%
      \expandafter\def\csname LT1\endcsname{\color[rgb]{0,1,0}}%
      \expandafter\def\csname LT2\endcsname{\color[rgb]{0,0,1}}%
      \expandafter\def\csname LT3\endcsname{\color[rgb]{1,0,1}}%
      \expandafter\def\csname LT4\endcsname{\color[rgb]{0,1,1}}%
      \expandafter\def\csname LT5\endcsname{\color[rgb]{1,1,0}}%
      \expandafter\def\csname LT6\endcsname{\color[rgb]{0,0,0}}%
      \expandafter\def\csname LT7\endcsname{\color[rgb]{1,0.3,0}}%
      \expandafter\def\csname LT8\endcsname{\color[rgb]{0.5,0.5,0.5}}%
    \else
      \def\colorrgb#1{\color{black}}%
      \def\colorgray#1{\color[gray]{#1}}%
      \expandafter\def\csname LTw\endcsname{\color{white}}%
      \expandafter\def\csname LTb\endcsname{\color{black}}%
      \expandafter\def\csname LTa\endcsname{\color{black}}%
      \expandafter\def\csname LT0\endcsname{\color{black}}%
      \expandafter\def\csname LT1\endcsname{\color{black}}%
      \expandafter\def\csname LT2\endcsname{\color{black}}%
      \expandafter\def\csname LT3\endcsname{\color{black}}%
      \expandafter\def\csname LT4\endcsname{\color{black}}%
      \expandafter\def\csname LT5\endcsname{\color{black}}%
      \expandafter\def\csname LT6\endcsname{\color{black}}%
      \expandafter\def\csname LT7\endcsname{\color{black}}%
      \expandafter\def\csname LT8\endcsname{\color{black}}%
    \fi
  \fi
  \setlength{\unitlength}{0.0500bp}%
  \begin{picture}(7200.00,3528.00)%
    \gplgaddtomacro\gplbacktext{%
      \csname LTb\endcsname%
      \put(814,704){\makebox(0,0)[r]{\strut{}-25}}%
      \put(814,960){\makebox(0,0)[r]{\strut{}-20}}%
      \put(814,1216){\makebox(0,0)[r]{\strut{}-15}}%
      \put(814,1472){\makebox(0,0)[r]{\strut{}-10}}%
      \put(814,1728){\makebox(0,0)[r]{\strut{}-5}}%
      \put(814,1984){\makebox(0,0)[r]{\strut{} 0}}%
      \put(814,2239){\makebox(0,0)[r]{\strut{} 5}}%
      \put(814,2495){\makebox(0,0)[r]{\strut{} 10}}%
      \put(814,2751){\makebox(0,0)[r]{\strut{} 15}}%
      \put(814,3007){\makebox(0,0)[r]{\strut{} 20}}%
      \put(814,3263){\makebox(0,0)[r]{\strut{} 25}}%
      \put(1532,484){\makebox(0,0){\strut{}-4}}%
      \put(2703,484){\makebox(0,0){\strut{}-2}}%
      \put(3875,484){\makebox(0,0){\strut{} 0}}%
      \put(5046,484){\makebox(0,0){\strut{} 2}}%
      \put(6217,484){\makebox(0,0){\strut{} 4}}%
      \put(176,1983){\rotatebox{-270}{\makebox(0,0){\strut{}$X^x$}}}%
      \put(3874,154){\makebox(0,0){\strut{}$x$}}%
    }%
    \gplgaddtomacro\gplfronttext{%
      \csname LTb\endcsname%
      \put(1342,3090){\makebox(0,0)[r]{\strut{}L1}}%
      \csname LTb\endcsname%
      \put(1342,2870){\makebox(0,0)[r]{\strut{}L4}}%
      \csname LTb\endcsname%
      \put(1342,2650){\makebox(0,0)[r]{\strut{}L5}}%
      \csname LTb\endcsname%
      \put(1342,2430){\makebox(0,0)[r]{\strut{}L6}}%
    }%
    \gplgaddtomacro\gplbacktext{%
      \csname LTb\endcsname%
      \put(4818,1047){\makebox(0,0)[r]{\strut{} 0}}%
      \put(4818,1256){\makebox(0,0)[r]{\strut{} 0.25}}%
      \put(4818,1464){\makebox(0,0)[r]{\strut{} 0.5}}%
      \put(4818,1672){\makebox(0,0)[r]{\strut{} 0.75}}%
      \put(4950,824){\makebox(0,0){\strut{} 0}}%
      \put(5249,824){\makebox(0,0){\strut{} 1}}%
      \put(5547,824){\makebox(0,0){\strut{} 2}}%
      \put(5846,824){\makebox(0,0){\strut{} 3}}%
      \put(6144,824){\makebox(0,0){\strut{} 4}}%
      \put(6443,824){\makebox(0,0){\strut{} 5}}%
    }%
    \gplgaddtomacro\gplfronttext{%
    }%
    \gplbacktext
    \put(0,0){\includegraphics{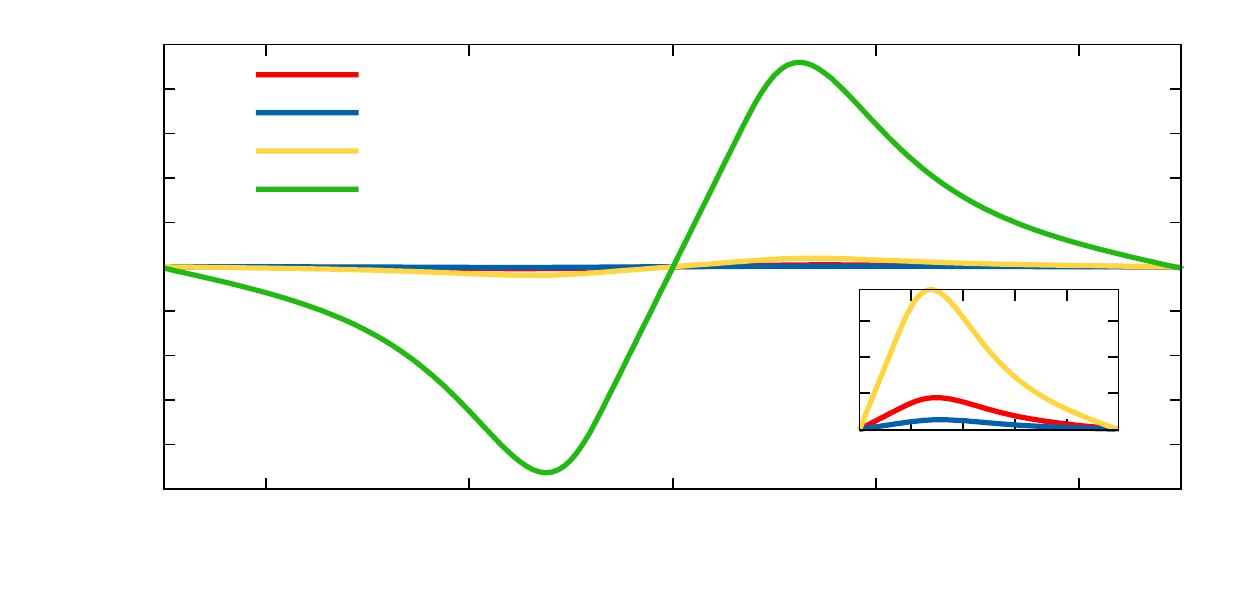}}%
    \gplfronttext
  \end{picture}%
\endgroup

\caption{Regular part of the conformal factor $u$ (top) and $x$-component of
the vector $X^i$ for a black-hole lattice with $m=1$ at the origin, cell faces
at $x=\pm 5$, $y=\pm 5$, and $z=\pm 5$, and parameters $\ell=0.5$ and $\sigma=4$.
\label{fig:LP}}
\ece
\efi

\bfi
\bce
\begingroup
  \makeatletter
  \providecommand\color[2][]{%
    \GenericError{(gnuplot) \space\space\space\@spaces}{%
      Package color not loaded in conjunction with
      terminal option `colourtext'%
    }{See the gnuplot documentation for explanation.%
    }{Either use 'blacktext' in gnuplot or load the package
      color.sty in LaTeX.}%
    \renewcommand\color[2][]{}%
  }%
  \providecommand\includegraphics[2][]{%
    \GenericError{(gnuplot) \space\space\space\@spaces}{%
      Package graphicx or graphics not loaded%
    }{See the gnuplot documentation for explanation.%
    }{The gnuplot epslatex terminal needs graphicx.sty or graphics.sty.}%
    \renewcommand\includegraphics[2][]{}%
  }%
  \providecommand\rotatebox[2]{#2}%
  \@ifundefined{ifGPcolor}{%
    \newif\ifGPcolor
    \GPcolortrue
  }{}%
  \@ifundefined{ifGPblacktext}{%
    \newif\ifGPblacktext
    \GPblacktexttrue
  }{}%
  \let\gplgaddtomacro\g@addto@macro
  \gdef\gplbacktext{}%
  \gdef\gplfronttext{}%
  \makeatother
  \ifGPblacktext
    \def\colorrgb#1{}%
    \def\colorgray#1{}%
  \else
    \ifGPcolor
      \def\colorrgb#1{\color[rgb]{#1}}%
      \def\colorgray#1{\color[gray]{#1}}%
      \expandafter\def\csname LTw\endcsname{\color{white}}%
      \expandafter\def\csname LTb\endcsname{\color{black}}%
      \expandafter\def\csname LTa\endcsname{\color{black}}%
      \expandafter\def\csname LT0\endcsname{\color[rgb]{1,0,0}}%
      \expandafter\def\csname LT1\endcsname{\color[rgb]{0,1,0}}%
      \expandafter\def\csname LT2\endcsname{\color[rgb]{0,0,1}}%
      \expandafter\def\csname LT3\endcsname{\color[rgb]{1,0,1}}%
      \expandafter\def\csname LT4\endcsname{\color[rgb]{0,1,1}}%
      \expandafter\def\csname LT5\endcsname{\color[rgb]{1,1,0}}%
      \expandafter\def\csname LT6\endcsname{\color[rgb]{0,0,0}}%
      \expandafter\def\csname LT7\endcsname{\color[rgb]{1,0.3,0}}%
      \expandafter\def\csname LT8\endcsname{\color[rgb]{0.5,0.5,0.5}}%
    \else
      \def\colorrgb#1{\color{black}}%
      \def\colorgray#1{\color[gray]{#1}}%
      \expandafter\def\csname LTw\endcsname{\color{white}}%
      \expandafter\def\csname LTb\endcsname{\color{black}}%
      \expandafter\def\csname LTa\endcsname{\color{black}}%
      \expandafter\def\csname LT0\endcsname{\color{black}}%
      \expandafter\def\csname LT1\endcsname{\color{black}}%
      \expandafter\def\csname LT2\endcsname{\color{black}}%
      \expandafter\def\csname LT3\endcsname{\color{black}}%
      \expandafter\def\csname LT4\endcsname{\color{black}}%
      \expandafter\def\csname LT5\endcsname{\color{black}}%
      \expandafter\def\csname LT6\endcsname{\color{black}}%
      \expandafter\def\csname LT7\endcsname{\color{black}}%
      \expandafter\def\csname LT8\endcsname{\color{black}}%
    \fi
  \fi
  \setlength{\unitlength}{0.0500bp}%
  \begin{picture}(7920.00,3024.00)%
    \gplgaddtomacro\gplbacktext{%
      \csname LTb\endcsname%
      \put(594,704){\makebox(0,0)[r]{\strut{}0.9}}%
      \put(594,961){\makebox(0,0)[r]{\strut{}1.0}}%
      \put(594,1218){\makebox(0,0)[r]{\strut{}1.1}}%
      \put(594,1475){\makebox(0,0)[r]{\strut{}1.2}}%
      \put(594,1732){\makebox(0,0)[r]{\strut{}1.3}}%
      \put(594,1988){\makebox(0,0)[r]{\strut{}1.4}}%
      \put(594,2245){\makebox(0,0)[r]{\strut{}1.5}}%
      \put(594,2502){\makebox(0,0)[r]{\strut{}1.6}}%
      \put(594,2759){\makebox(0,0)[r]{\strut{}1.7}}%
      \put(726,484){\makebox(0,0){\strut{} 0.5}}%
      \put(1041,484){\makebox(0,0){\strut{} 1}}%
      \put(1356,484){\makebox(0,0){\strut{} 1.5}}%
      \put(1672,484){\makebox(0,0){\strut{} 2}}%
      \put(1987,484){\makebox(0,0){\strut{} 2.5}}%
      \put(2302,484){\makebox(0,0){\strut{} 3}}%
      \put(2617,484){\makebox(0,0){\strut{} 3.5}}%
      \put(2933,484){\makebox(0,0){\strut{} 4}}%
      \put(3248,484){\makebox(0,0){\strut{} 4.5}}%
      \put(3563,484){\makebox(0,0){\strut{} 5}}%
      \put(2144,154){\makebox(0,0){\strut{}$m$}}%
    }%
    \gplgaddtomacro\gplfronttext{%
      \csname LTb\endcsname%
      \put(1386,2586){\makebox(0,0)[r]{\strut{}$u_{\rm C}$}}%
      \csname LTb\endcsname%
      \put(1386,2366){\makebox(0,0)[r]{\strut{}$u_{\rm V}$}}%
    }%
    \gplgaddtomacro\gplbacktext{%
      \csname LTb\endcsname%
      \put(4326,704){\makebox(0,0)[r]{\strut{}0.01}}%
      \put(4326,1218){\makebox(0,0)[r]{\strut{}0.1}}%
      \put(4326,1732){\makebox(0,0)[r]{\strut{}1}}%
      \put(4326,2245){\makebox(0,0)[r]{\strut{}10}}%
      \put(4326,2759){\makebox(0,0)[r]{\strut{}100}}%
      \put(4458,484){\makebox(0,0){\strut{} 0.5}}%
      \put(4758,484){\makebox(0,0){\strut{} 1}}%
      \put(5059,484){\makebox(0,0){\strut{} 1.5}}%
      \put(5359,484){\makebox(0,0){\strut{} 2}}%
      \put(5660,484){\makebox(0,0){\strut{} 2.5}}%
      \put(5960,484){\makebox(0,0){\strut{} 3}}%
      \put(6261,484){\makebox(0,0){\strut{} 3.5}}%
      \put(6561,484){\makebox(0,0){\strut{} 4}}%
      \put(6862,484){\makebox(0,0){\strut{} 4.5}}%
      \put(7162,484){\makebox(0,0){\strut{} 5}}%
      \put(3952,1731){\rotatebox{-270}{\makebox(0,0){\strut{}max($X^x$)}}}%
      \put(5810,154){\makebox(0,0){\strut{}$m$}}%
    }%
    \gplgaddtomacro\gplfronttext{%
    }%
    \gplbacktext
    \put(0,0){\includegraphics{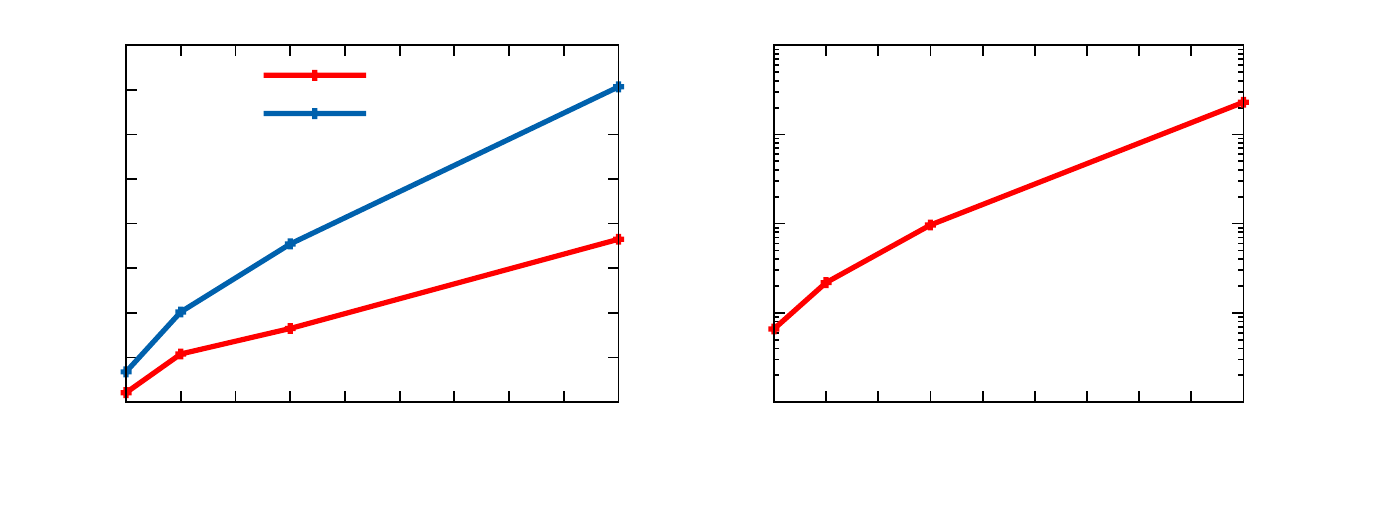}}%
    \gplfronttext
  \end{picture}%
\endgroup

\caption{Left: values of the regular part of the conformal factor $u$ at the 
center of the cell and at a vertex, as a function of $m$. Right:
maximum of $X^x$ on the $x$ axis as a function of $m$.
\label{fig:uvsM}}
\ece
\efi

Given the lack of an exact solution, I next perform a three-point resolution
study with runs L1, L2 (with $\Delta_0=0.8\bar3$) and L3 (with $\Delta_0=0.625$).
The results are illustrated in Figure~\ref{fig:LPC}, where the ratio
of the difference between solutions at different resolutions:
\bea
c_u &=& \frac{|u_{\rm L1}-u_{\rm L2}|}{|u_{\rm L2}-u_{\rm L3}|} \\
c_X &=& \frac{|X^x_{\rm L1}-X^x_{\rm L2}|}{|X^x_{\rm L2}-X^x_{\rm L3}|}
\eea 
is plotted against the expected value for a second-, fourth- and sixth-order
numerical error. Whilst the values of $c_u$ and $c_X$ support a scaling in this
range, the analysis is clearly much less straightforward than in the 
asymptotically-flat example of Figure~\ref{fig:TP}, especially with respect
to the convergence of $u$ near the boundaries, where the difference 
between solutions at different resolutions seems to decrease much more
rapidly than elsewhere. Coupling the relaxation process with the
the update of $\psi$ (a step which entails the calculation of the 
two integrals in (\ref{eq:f}), with the associated error) could be 
responsible for this effect.

\bfi
\bce
\begingroup
  \makeatletter
  \providecommand\color[2][]{%
    \GenericError{(gnuplot) \space\space\space\@spaces}{%
      Package color not loaded in conjunction with
      terminal option `colourtext'%
    }{See the gnuplot documentation for explanation.%
    }{Either use 'blacktext' in gnuplot or load the package
      color.sty in LaTeX.}%
    \renewcommand\color[2][]{}%
  }%
  \providecommand\includegraphics[2][]{%
    \GenericError{(gnuplot) \space\space\space\@spaces}{%
      Package graphicx or graphics not loaded%
    }{See the gnuplot documentation for explanation.%
    }{The gnuplot epslatex terminal needs graphicx.sty or graphics.sty.}%
    \renewcommand\includegraphics[2][]{}%
  }%
  \providecommand\rotatebox[2]{#2}%
  \@ifundefined{ifGPcolor}{%
    \newif\ifGPcolor
    \GPcolortrue
  }{}%
  \@ifundefined{ifGPblacktext}{%
    \newif\ifGPblacktext
    \GPblacktexttrue
  }{}%
  \let\gplgaddtomacro\g@addto@macro
  \gdef\gplbacktext{}%
  \gdef\gplfronttext{}%
  \makeatother
  \ifGPblacktext
    \def\colorrgb#1{}%
    \def\colorgray#1{}%
  \else
    \ifGPcolor
      \def\colorrgb#1{\color[rgb]{#1}}%
      \def\colorgray#1{\color[gray]{#1}}%
      \expandafter\def\csname LTw\endcsname{\color{white}}%
      \expandafter\def\csname LTb\endcsname{\color{black}}%
      \expandafter\def\csname LTa\endcsname{\color{black}}%
      \expandafter\def\csname LT0\endcsname{\color[rgb]{1,0,0}}%
      \expandafter\def\csname LT1\endcsname{\color[rgb]{0,1,0}}%
      \expandafter\def\csname LT2\endcsname{\color[rgb]{0,0,1}}%
      \expandafter\def\csname LT3\endcsname{\color[rgb]{1,0,1}}%
      \expandafter\def\csname LT4\endcsname{\color[rgb]{0,1,1}}%
      \expandafter\def\csname LT5\endcsname{\color[rgb]{1,1,0}}%
      \expandafter\def\csname LT6\endcsname{\color[rgb]{0,0,0}}%
      \expandafter\def\csname LT7\endcsname{\color[rgb]{1,0.3,0}}%
      \expandafter\def\csname LT8\endcsname{\color[rgb]{0.5,0.5,0.5}}%
    \else
      \def\colorrgb#1{\color{black}}%
      \def\colorgray#1{\color[gray]{#1}}%
      \expandafter\def\csname LTw\endcsname{\color{white}}%
      \expandafter\def\csname LTb\endcsname{\color{black}}%
      \expandafter\def\csname LTa\endcsname{\color{black}}%
      \expandafter\def\csname LT0\endcsname{\color{black}}%
      \expandafter\def\csname LT1\endcsname{\color{black}}%
      \expandafter\def\csname LT2\endcsname{\color{black}}%
      \expandafter\def\csname LT3\endcsname{\color{black}}%
      \expandafter\def\csname LT4\endcsname{\color{black}}%
      \expandafter\def\csname LT5\endcsname{\color{black}}%
      \expandafter\def\csname LT6\endcsname{\color{black}}%
      \expandafter\def\csname LT7\endcsname{\color{black}}%
      \expandafter\def\csname LT8\endcsname{\color{black}}%
    \fi
  \fi
  \setlength{\unitlength}{0.0500bp}%
  \begin{picture}(7920.00,3024.00)%
    \gplgaddtomacro\gplbacktext{%
      \csname LTb\endcsname%
      \put(462,704){\makebox(0,0)[r]{\strut{}0}}%
      \put(462,961){\makebox(0,0)[r]{\strut{}2}}%
      \put(462,1218){\makebox(0,0)[r]{\strut{}4}}%
      \put(462,1475){\makebox(0,0)[r]{\strut{}6}}%
      \put(462,1732){\makebox(0,0)[r]{\strut{}8}}%
      \put(462,1988){\makebox(0,0)[r]{\strut{}10}}%
      \put(462,2245){\makebox(0,0)[r]{\strut{}12}}%
      \put(462,2502){\makebox(0,0)[r]{\strut{}14}}%
      \put(462,2759){\makebox(0,0)[r]{\strut{}16}}%
      \put(905,484){\makebox(0,0){\strut{}-4}}%
      \put(1528,484){\makebox(0,0){\strut{}-2}}%
      \put(2151,484){\makebox(0,0){\strut{} 0}}%
      \put(2773,484){\makebox(0,0){\strut{} 2}}%
      \put(3396,484){\makebox(0,0){\strut{} 4}}%
      \put(2150,154){\makebox(0,0){\strut{}$x$}}%
    }%
    \gplgaddtomacro\gplfronttext{%
      \csname LTb\endcsname%
      \put(1205,2586){\makebox(0,0)[r]{\strut{}$c_u$}}%
      \csname LTb\endcsname%
      \put(1205,2366){\makebox(0,0)[r]{\strut{}$2^{\rm nd}$}}%
      \csname LTb\endcsname%
      \put(2720,2586){\makebox(0,0)[r]{\strut{}$4^{\rm th}$}}%
      \csname LTb\endcsname%
      \put(2720,2366){\makebox(0,0)[r]{\strut{}$6^{\rm th}$}}%
    }%
    \gplgaddtomacro\gplbacktext{%
      \csname LTb\endcsname%
      \put(4194,704){\makebox(0,0)[r]{\strut{}0.5}}%
      \put(4194,1115){\makebox(0,0)[r]{\strut{}1}}%
      \put(4194,1526){\makebox(0,0)[r]{\strut{}1.5}}%
      \put(4194,1937){\makebox(0,0)[r]{\strut{}2}}%
      \put(4194,2348){\makebox(0,0)[r]{\strut{}2.5}}%
      \put(4194,2759){\makebox(0,0)[r]{\strut{}3}}%
      \put(4624,484){\makebox(0,0){\strut{}-4}}%
      \put(5220,484){\makebox(0,0){\strut{}-2}}%
      \put(5816,484){\makebox(0,0){\strut{} 0}}%
      \put(6412,484){\makebox(0,0){\strut{} 2}}%
      \put(7008,484){\makebox(0,0){\strut{} 4}}%
      \put(5816,154){\makebox(0,0){\strut{}$x$}}%
    }%
    \gplgaddtomacro\gplfronttext{%
      \csname LTb\endcsname%
      \put(4804,2586){\makebox(0,0)[r]{\strut{}$c_X$}}%
      \csname LTb\endcsname%
      \put(4804,2366){\makebox(0,0)[r]{\strut{}$2^{\rm nd}$}}%
      \csname LTb\endcsname%
      \put(6319,2586){\makebox(0,0)[r]{\strut{}$4^{\rm th}$}}%
      \csname LTb\endcsname%
      \put(6319,2366){\makebox(0,0)[r]{\strut{}$6^{\rm th}$}}%
    }%
    \gplbacktext
    \put(0,0){\includegraphics{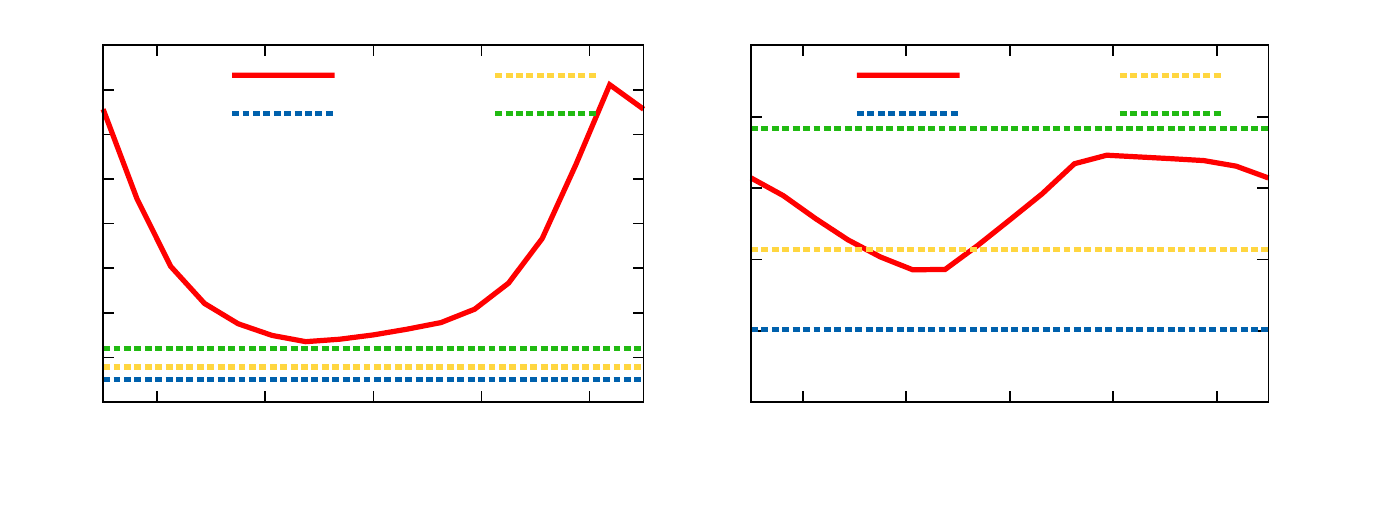}}%
    \gplfronttext
  \end{picture}%
\endgroup

\caption{
Convergence factors $c_u$ and $c_X$ along the $x$-axis.
\label{fig:LPC}}
\ece
\efi

In Figure~\ref{fig:HLP}, the value of the Hamiltonian constraint and 
of the $x$-component of the momentum constraint are plotted for $L1$,
$L2$ and $L3$, with the two finest resolutions rescaled for fourth-order
convergence. Here, again, one observes that the results are acceptable
in the central region, where finite-differencing dominates the error,
but degrade towards the boundaries, where the finite-differencing error
decreases below the error introduced by the volume integrals.
\bfi
\bce
\begingroup
  \makeatletter
  \providecommand\color[2][]{%
    \GenericError{(gnuplot) \space\space\space\@spaces}{%
      Package color not loaded in conjunction with
      terminal option `colourtext'%
    }{See the gnuplot documentation for explanation.%
    }{Either use 'blacktext' in gnuplot or load the package
      color.sty in LaTeX.}%
    \renewcommand\color[2][]{}%
  }%
  \providecommand\includegraphics[2][]{%
    \GenericError{(gnuplot) \space\space\space\@spaces}{%
      Package graphicx or graphics not loaded%
    }{See the gnuplot documentation for explanation.%
    }{The gnuplot epslatex terminal needs graphicx.sty or graphics.sty.}%
    \renewcommand\includegraphics[2][]{}%
  }%
  \providecommand\rotatebox[2]{#2}%
  \@ifundefined{ifGPcolor}{%
    \newif\ifGPcolor
    \GPcolortrue
  }{}%
  \@ifundefined{ifGPblacktext}{%
    \newif\ifGPblacktext
    \GPblacktexttrue
  }{}%
  \let\gplgaddtomacro\g@addto@macro
  \gdef\gplbacktext{}%
  \gdef\gplfronttext{}%
  \makeatother
  \ifGPblacktext
    \def\colorrgb#1{}%
    \def\colorgray#1{}%
  \else
    \ifGPcolor
      \def\colorrgb#1{\color[rgb]{#1}}%
      \def\colorgray#1{\color[gray]{#1}}%
      \expandafter\def\csname LTw\endcsname{\color{white}}%
      \expandafter\def\csname LTb\endcsname{\color{black}}%
      \expandafter\def\csname LTa\endcsname{\color{black}}%
      \expandafter\def\csname LT0\endcsname{\color[rgb]{1,0,0}}%
      \expandafter\def\csname LT1\endcsname{\color[rgb]{0,1,0}}%
      \expandafter\def\csname LT2\endcsname{\color[rgb]{0,0,1}}%
      \expandafter\def\csname LT3\endcsname{\color[rgb]{1,0,1}}%
      \expandafter\def\csname LT4\endcsname{\color[rgb]{0,1,1}}%
      \expandafter\def\csname LT5\endcsname{\color[rgb]{1,1,0}}%
      \expandafter\def\csname LT6\endcsname{\color[rgb]{0,0,0}}%
      \expandafter\def\csname LT7\endcsname{\color[rgb]{1,0.3,0}}%
      \expandafter\def\csname LT8\endcsname{\color[rgb]{0.5,0.5,0.5}}%
    \else
      \def\colorrgb#1{\color{black}}%
      \def\colorgray#1{\color[gray]{#1}}%
      \expandafter\def\csname LTw\endcsname{\color{white}}%
      \expandafter\def\csname LTb\endcsname{\color{black}}%
      \expandafter\def\csname LTa\endcsname{\color{black}}%
      \expandafter\def\csname LT0\endcsname{\color{black}}%
      \expandafter\def\csname LT1\endcsname{\color{black}}%
      \expandafter\def\csname LT2\endcsname{\color{black}}%
      \expandafter\def\csname LT3\endcsname{\color{black}}%
      \expandafter\def\csname LT4\endcsname{\color{black}}%
      \expandafter\def\csname LT5\endcsname{\color{black}}%
      \expandafter\def\csname LT6\endcsname{\color{black}}%
      \expandafter\def\csname LT7\endcsname{\color{black}}%
      \expandafter\def\csname LT8\endcsname{\color{black}}%
    \fi
  \fi
  \setlength{\unitlength}{0.0500bp}%
  \begin{picture}(7200.00,3528.00)%
    \gplgaddtomacro\gplbacktext{%
      \csname LTb\endcsname%
      \put(1078,704){\makebox(0,0)[r]{\strut{}$10^{-10}$}}%
      \put(1078,968){\makebox(0,0)[r]{\strut{}$10^{-9}$}}%
      \put(1078,1232){\makebox(0,0)[r]{\strut{}$10^{-8}$}}%
      \put(1078,1496){\makebox(0,0)[r]{\strut{}$10^{-7}$}}%
      \put(1078,1759){\makebox(0,0)[r]{\strut{}$10^{-6}$}}%
      \put(1078,2023){\makebox(0,0)[r]{\strut{}$10^{-5}$}}%
      \put(1078,2287){\makebox(0,0)[r]{\strut{}$10^{-4}$}}%
      \put(1078,2551){\makebox(0,0)[r]{\strut{}$10^{-3}$}}%
      \put(1078,2815){\makebox(0,0)[r]{\strut{}$10^{-2}$}}%
      \put(1078,3079){\makebox(0,0)[r]{\strut{}$10^{-1}$}}%
      \put(1769,484){\makebox(0,0){\strut{}-4}}%
      \put(2888,484){\makebox(0,0){\strut{}-2}}%
      \put(4007,484){\makebox(0,0){\strut{} 0}}%
      \put(5125,484){\makebox(0,0){\strut{} 2}}%
      \put(6244,484){\makebox(0,0){\strut{} 4}}%
      \put(176,1983){\rotatebox{-270}{\makebox(0,0){\strut{}$|H|$}}}%
      \put(4006,154){\makebox(0,0){\strut{}$x$}}%
    }%
    \gplgaddtomacro\gplfronttext{%
      \csname LTb\endcsname%
      \put(5816,3090){\makebox(0,0)[r]{\strut{}L1}}%
      \csname LTb\endcsname%
      \put(5816,2870){\makebox(0,0)[r]{\strut{}L2 (scaled)}}%
      \csname LTb\endcsname%
      \put(5816,2650){\makebox(0,0)[r]{\strut{}L3 (scaled)}}%
    }%
    \gplbacktext
    \put(0,0){\includegraphics{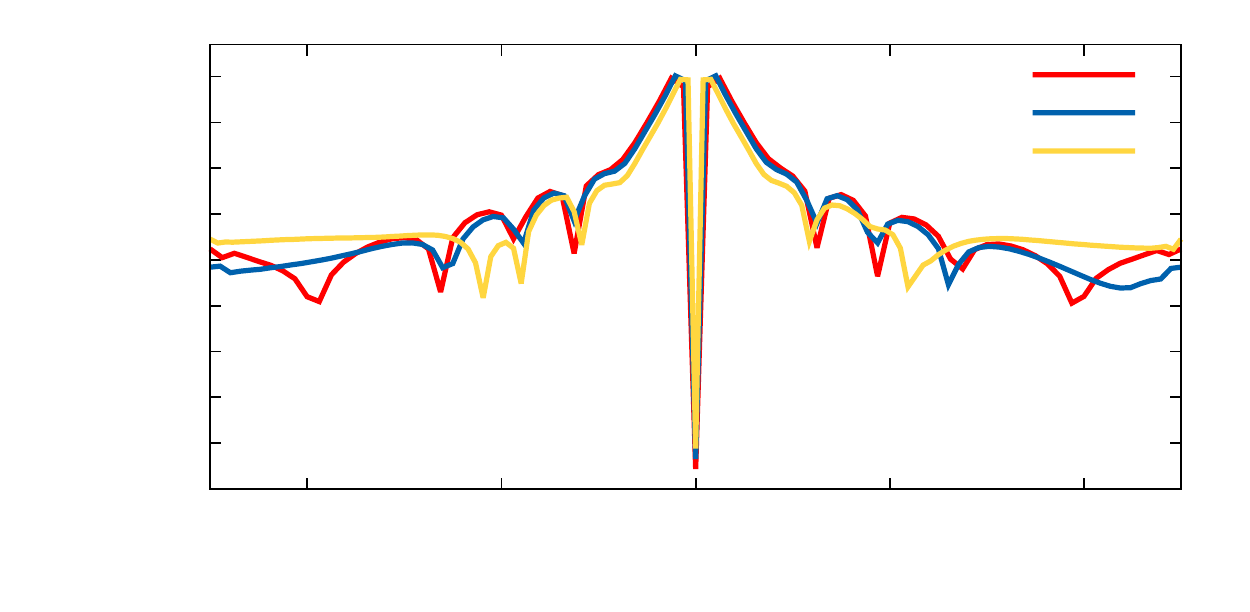}}%
    \gplfronttext
  \end{picture}%
\endgroup

\begingroup
  \makeatletter
  \providecommand\color[2][]{%
    \GenericError{(gnuplot) \space\space\space\@spaces}{%
      Package color not loaded in conjunction with
      terminal option `colourtext'%
    }{See the gnuplot documentation for explanation.%
    }{Either use 'blacktext' in gnuplot or load the package
      color.sty in LaTeX.}%
    \renewcommand\color[2][]{}%
  }%
  \providecommand\includegraphics[2][]{%
    \GenericError{(gnuplot) \space\space\space\@spaces}{%
      Package graphicx or graphics not loaded%
    }{See the gnuplot documentation for explanation.%
    }{The gnuplot epslatex terminal needs graphicx.sty or graphics.sty.}%
    \renewcommand\includegraphics[2][]{}%
  }%
  \providecommand\rotatebox[2]{#2}%
  \@ifundefined{ifGPcolor}{%
    \newif\ifGPcolor
    \GPcolortrue
  }{}%
  \@ifundefined{ifGPblacktext}{%
    \newif\ifGPblacktext
    \GPblacktexttrue
  }{}%
  \let\gplgaddtomacro\g@addto@macro
  \gdef\gplbacktext{}%
  \gdef\gplfronttext{}%
  \makeatother
  \ifGPblacktext
    \def\colorrgb#1{}%
    \def\colorgray#1{}%
  \else
    \ifGPcolor
      \def\colorrgb#1{\color[rgb]{#1}}%
      \def\colorgray#1{\color[gray]{#1}}%
      \expandafter\def\csname LTw\endcsname{\color{white}}%
      \expandafter\def\csname LTb\endcsname{\color{black}}%
      \expandafter\def\csname LTa\endcsname{\color{black}}%
      \expandafter\def\csname LT0\endcsname{\color[rgb]{1,0,0}}%
      \expandafter\def\csname LT1\endcsname{\color[rgb]{0,1,0}}%
      \expandafter\def\csname LT2\endcsname{\color[rgb]{0,0,1}}%
      \expandafter\def\csname LT3\endcsname{\color[rgb]{1,0,1}}%
      \expandafter\def\csname LT4\endcsname{\color[rgb]{0,1,1}}%
      \expandafter\def\csname LT5\endcsname{\color[rgb]{1,1,0}}%
      \expandafter\def\csname LT6\endcsname{\color[rgb]{0,0,0}}%
      \expandafter\def\csname LT7\endcsname{\color[rgb]{1,0.3,0}}%
      \expandafter\def\csname LT8\endcsname{\color[rgb]{0.5,0.5,0.5}}%
    \else
      \def\colorrgb#1{\color{black}}%
      \def\colorgray#1{\color[gray]{#1}}%
      \expandafter\def\csname LTw\endcsname{\color{white}}%
      \expandafter\def\csname LTb\endcsname{\color{black}}%
      \expandafter\def\csname LTa\endcsname{\color{black}}%
      \expandafter\def\csname LT0\endcsname{\color{black}}%
      \expandafter\def\csname LT1\endcsname{\color{black}}%
      \expandafter\def\csname LT2\endcsname{\color{black}}%
      \expandafter\def\csname LT3\endcsname{\color{black}}%
      \expandafter\def\csname LT4\endcsname{\color{black}}%
      \expandafter\def\csname LT5\endcsname{\color{black}}%
      \expandafter\def\csname LT6\endcsname{\color{black}}%
      \expandafter\def\csname LT7\endcsname{\color{black}}%
      \expandafter\def\csname LT8\endcsname{\color{black}}%
    \fi
  \fi
  \setlength{\unitlength}{0.0500bp}%
  \begin{picture}(7200.00,3528.00)%
    \gplgaddtomacro\gplbacktext{%
      \csname LTb\endcsname%
      \put(1078,704){\makebox(0,0)[r]{\strut{}$10^{-10}$}}%
      \put(1078,1070){\makebox(0,0)[r]{\strut{}$10^{-9}$}}%
      \put(1078,1435){\makebox(0,0)[r]{\strut{}$10^{-8}$}}%
      \put(1078,1801){\makebox(0,0)[r]{\strut{}$10^{-7}$}}%
      \put(1078,2166){\makebox(0,0)[r]{\strut{}$10^{-6}$}}%
      \put(1078,2532){\makebox(0,0)[r]{\strut{}$10^{-5}$}}%
      \put(1078,2897){\makebox(0,0)[r]{\strut{}$10^{-4}$}}%
      \put(1078,3263){\makebox(0,0)[r]{\strut{}$10^{-3}$}}%
      \put(1769,484){\makebox(0,0){\strut{}-4}}%
      \put(2888,484){\makebox(0,0){\strut{}-2}}%
      \put(4007,484){\makebox(0,0){\strut{} 0}}%
      \put(5125,484){\makebox(0,0){\strut{} 2}}%
      \put(6244,484){\makebox(0,0){\strut{} 4}}%
      \put(176,1983){\rotatebox{-270}{\makebox(0,0){\strut{}$|P^x|$}}}%
      \put(4006,154){\makebox(0,0){\strut{}$x$}}%
    }%
    \gplgaddtomacro\gplfronttext{%
      \csname LTb\endcsname%
      \put(5816,1317){\makebox(0,0)[r]{\strut{}L1}}%
      \csname LTb\endcsname%
      \put(5816,1097){\makebox(0,0)[r]{\strut{}L2 (scaled)}}%
      \csname LTb\endcsname%
      \put(5816,877){\makebox(0,0)[r]{\strut{}L3 (scaled)}}%
    }%
    \gplbacktext
    \put(0,0){\includegraphics{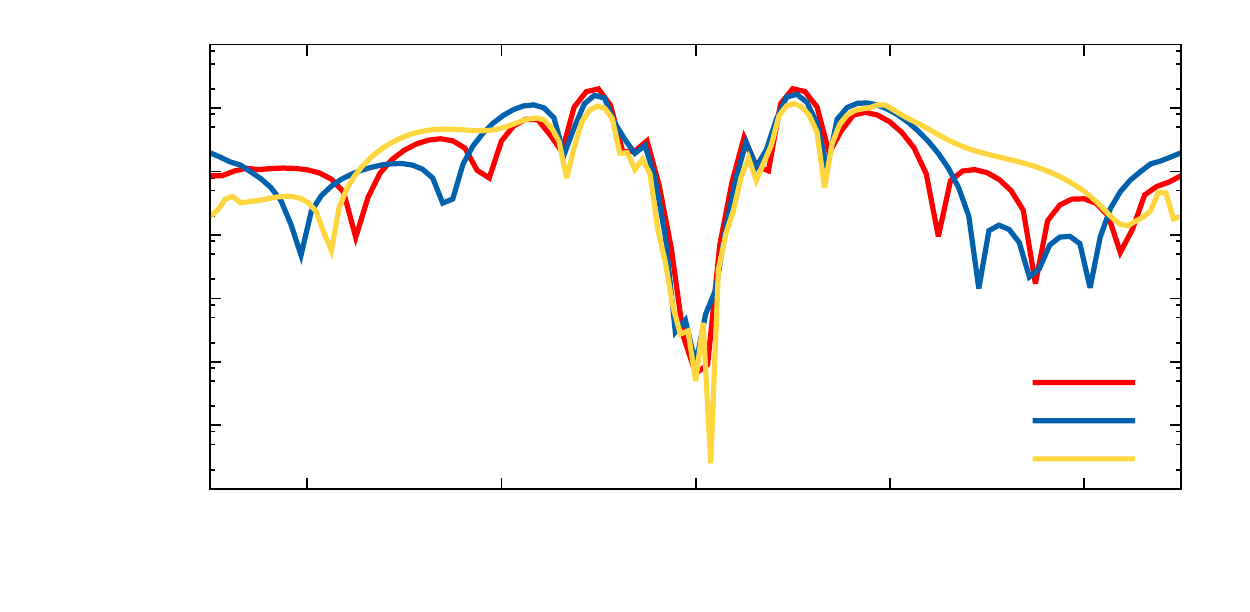}}%
    \gplfronttext
  \end{picture}%
\endgroup

\caption{Hamiltonian (top) and momentum (bottom) constraint violations
for runs L1, L2 and L3. Runs L2 and L3 have been rescaled by the factor
expected for fourth-order convergence.
\label{fig:HLP}}
\ece
\efi

Finally, I also search for the black hole apparent horizon and calculate 
the corresponding mass in each case. All these results are summarized 
in Table~\ref{tab:series}.

\bt
\bce
\caption{Summary of the lattice-puncture simulations discussed in
section~\ref{sec:lattice}. The missing horizon mass for run L4
is due to the fact that it is not possible to locate such a small
horizon at this resolution.
\label{tab:series}}
\btb{ccccccccc}
Run & $m$ & $\ell$ & $\sigma$ & $m_{\rm hor}$ & $L_{\rm p}$ & $\Delta_0$     & $|H|$                  & Reset \\
\hline
L1  &   1 &  0.5   &  4       &  1.007962     &  12.2607    &   1            & $1.29 \cdot 10^{-4}$    & $\psi$ \\
L2  &   1 &  0.5   &  4       &  1.008002     &  12.2607    &   $0.8\bar3$   & $7.03 \cdot 10^{-5}$    & $\psi$ \\
L3  &   1 &  0.5   &  4       &  1.008005     &  12.2604    &   0.625        & $2.72 \cdot 10^{-5}$    & $\psi$ \\
L4  & 0.5 &  0.5   &  4       &  --           &  9.41388    &   1            & $1.61 \cdot 10^{-3}$    & $\psi$ \\
L5  &   2 &  0.5   &  4       &  2.130278     &  15.9943    &   1            & $2.36 \cdot 10^{-5}$    & $\psi$ \\
L6  &   5 &  0.5   &  4       &  6.332967     &  26.9543    &   1            & $8.99 \cdot 10^{-4}$    & $\psi$ \\
L7  &   1 &  0.1   &  4.8     &  1.003605     &  12.1985    &   1            & $1.30 \cdot 10^{-4}$    & $K$    \\
\hline
\etb
\ece
\et

Notice that L7, the last run in Table~\ref{tab:series}, has the same physical parameters of L1, but 
a slightly larger transition region, and is obtained by updating $K_c$ through
the integral condition (and resetting $\psi$ to one).
This provides a comparison with an independent code, as this run 
is identical to the second configuration presented in~\cite{Yoo:2012jz} (see Figures 4 and 6 there, but 
notice the typo immediately after equation (38), where $\ell$ is attributed the value of 
$\sigma$, and viceversa).

\bfi
\bce
\begingroup
  \makeatletter
  \providecommand\color[2][]{%
    \GenericError{(gnuplot) \space\space\space\@spaces}{%
      Package color not loaded in conjunction with
      terminal option `colourtext'%
    }{See the gnuplot documentation for explanation.%
    }{Either use 'blacktext' in gnuplot or load the package
      color.sty in LaTeX.}%
    \renewcommand\color[2][]{}%
  }%
  \providecommand\includegraphics[2][]{%
    \GenericError{(gnuplot) \space\space\space\@spaces}{%
      Package graphicx or graphics not loaded%
    }{See the gnuplot documentation for explanation.%
    }{The gnuplot epslatex terminal needs graphicx.sty or graphics.sty.}%
    \renewcommand\includegraphics[2][]{}%
  }%
  \providecommand\rotatebox[2]{#2}%
  \@ifundefined{ifGPcolor}{%
    \newif\ifGPcolor
    \GPcolortrue
  }{}%
  \@ifundefined{ifGPblacktext}{%
    \newif\ifGPblacktext
    \GPblacktexttrue
  }{}%
  \let\gplgaddtomacro\g@addto@macro
  \gdef\gplbacktext{}%
  \gdef\gplfronttext{}%
  \makeatother
  \ifGPblacktext
    \def\colorrgb#1{}%
    \def\colorgray#1{}%
  \else
    \ifGPcolor
      \def\colorrgb#1{\color[rgb]{#1}}%
      \def\colorgray#1{\color[gray]{#1}}%
      \expandafter\def\csname LTw\endcsname{\color{white}}%
      \expandafter\def\csname LTb\endcsname{\color{black}}%
      \expandafter\def\csname LTa\endcsname{\color{black}}%
      \expandafter\def\csname LT0\endcsname{\color[rgb]{1,0,0}}%
      \expandafter\def\csname LT1\endcsname{\color[rgb]{0,1,0}}%
      \expandafter\def\csname LT2\endcsname{\color[rgb]{0,0,1}}%
      \expandafter\def\csname LT3\endcsname{\color[rgb]{1,0,1}}%
      \expandafter\def\csname LT4\endcsname{\color[rgb]{0,1,1}}%
      \expandafter\def\csname LT5\endcsname{\color[rgb]{1,1,0}}%
      \expandafter\def\csname LT6\endcsname{\color[rgb]{0,0,0}}%
      \expandafter\def\csname LT7\endcsname{\color[rgb]{1,0.3,0}}%
      \expandafter\def\csname LT8\endcsname{\color[rgb]{0.5,0.5,0.5}}%
    \else
      \def\colorrgb#1{\color{black}}%
      \def\colorgray#1{\color[gray]{#1}}%
      \expandafter\def\csname LTw\endcsname{\color{white}}%
      \expandafter\def\csname LTb\endcsname{\color{black}}%
      \expandafter\def\csname LTa\endcsname{\color{black}}%
      \expandafter\def\csname LT0\endcsname{\color{black}}%
      \expandafter\def\csname LT1\endcsname{\color{black}}%
      \expandafter\def\csname LT2\endcsname{\color{black}}%
      \expandafter\def\csname LT3\endcsname{\color{black}}%
      \expandafter\def\csname LT4\endcsname{\color{black}}%
      \expandafter\def\csname LT5\endcsname{\color{black}}%
      \expandafter\def\csname LT6\endcsname{\color{black}}%
      \expandafter\def\csname LT7\endcsname{\color{black}}%
      \expandafter\def\csname LT8\endcsname{\color{black}}%
    \fi
  \fi
  \setlength{\unitlength}{0.0500bp}%
  \begin{picture}(7920.00,3024.00)%
    \gplgaddtomacro\gplbacktext{%
      \csname LTb\endcsname%
      \put(748,704){\makebox(0,0)[r]{\strut{}0.98}}%
      \put(748,891){\makebox(0,0)[r]{\strut{}1}}%
      \put(748,1078){\makebox(0,0)[r]{\strut{}1.02}}%
      \put(748,1264){\makebox(0,0)[r]{\strut{}1.04}}%
      \put(748,1451){\makebox(0,0)[r]{\strut{}1.06}}%
      \put(748,1638){\makebox(0,0)[r]{\strut{}1.08}}%
      \put(748,1825){\makebox(0,0)[r]{\strut{}1.1}}%
      \put(748,2012){\makebox(0,0)[r]{\strut{}1.12}}%
      \put(748,2199){\makebox(0,0)[r]{\strut{}1.14}}%
      \put(748,2385){\makebox(0,0)[r]{\strut{}1.16}}%
      \put(748,2572){\makebox(0,0)[r]{\strut{}1.18}}%
      \put(748,2759){\makebox(0,0)[r]{\strut{}1.2}}%
      \put(1148,484){\makebox(0,0){\strut{}-4}}%
      \put(1685,484){\makebox(0,0){\strut{}-2}}%
      \put(2222,484){\makebox(0,0){\strut{}0}}%
      \put(2758,484){\makebox(0,0){\strut{}2}}%
      \put(3295,484){\makebox(0,0){\strut{}4}}%
      \put(176,1731){\rotatebox{-270}{\makebox(0,0){\strut{}$u$}}}%
      \put(2221,154){\makebox(0,0){\strut{}$x$}}%
    }%
    \gplgaddtomacro\gplfronttext{%
    }%
    \gplgaddtomacro\gplbacktext{%
      \csname LTb\endcsname%
      \put(4818,704){\makebox(0,0)[r]{\strut{}-0.25}}%
      \put(4818,910){\makebox(0,0)[r]{\strut{}-0.2}}%
      \put(4818,1115){\makebox(0,0)[r]{\strut{}-0.15}}%
      \put(4818,1321){\makebox(0,0)[r]{\strut{}-0.1}}%
      \put(4818,1526){\makebox(0,0)[r]{\strut{}-0.05}}%
      \put(4818,1732){\makebox(0,0)[r]{\strut{}0}}%
      \put(4818,1937){\makebox(0,0)[r]{\strut{}0.05}}%
      \put(4818,2143){\makebox(0,0)[r]{\strut{}0.1}}%
      \put(4818,2348){\makebox(0,0)[r]{\strut{}0.15}}%
      \put(4818,2554){\makebox(0,0)[r]{\strut{}0.2}}%
      \put(4818,2759){\makebox(0,0)[r]{\strut{}0.25}}%
      \put(5207,484){\makebox(0,0){\strut{}-4}}%
      \put(5722,484){\makebox(0,0){\strut{}-2}}%
      \put(6236,484){\makebox(0,0){\strut{}0}}%
      \put(6750,484){\makebox(0,0){\strut{}2}}%
      \put(7265,484){\makebox(0,0){\strut{}4}}%
      \put(4180,1731){\rotatebox{-270}{\makebox(0,0){\strut{}$X^x$}}}%
      \put(6236,154){\makebox(0,0){\strut{}$x$}}%
    }%
    \gplgaddtomacro\gplfronttext{%
    }%
    \gplbacktext
    \put(0,0){\includegraphics{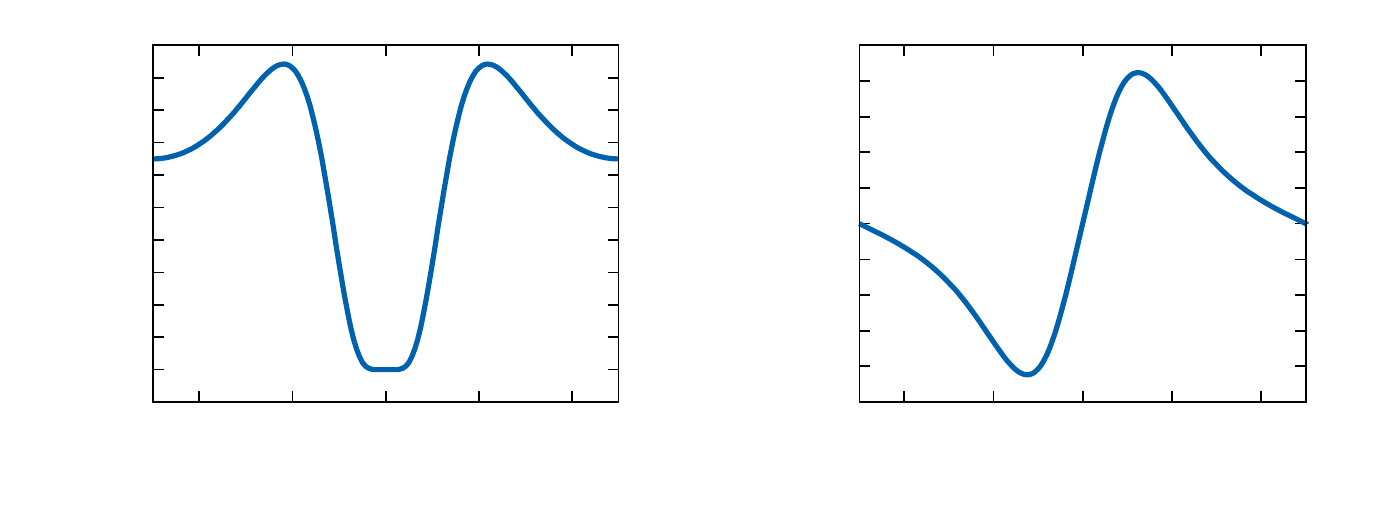}}%
    \gplfronttext
  \end{picture}%
\endgroup

\caption{Configuration L7. Left: regular part of the
conformal factor. Right: $x$-component of $X^i$. 
This is to be compared to Figure 4 and 6 of~\cite{Yoo:2012jz}.
\label{fig:LP7}}
\ece
\efi

\section{Conclusions}
In the first part of this work, I have presented a new multigrid code developed
as a module for the Cactus framework. The code is particularly suited to 
tackle elliptic PDEs in periodic spaces, as it implements the necessary
infrastructure to deal with the additional conditions that may be associated with this type of problems.
After a discussion of what these conditions may look like in a few simple scenarios,
I presented a series of tests, with the goal
of illustrating the behavior of relaxation methods in periodic spaces as well as
establishing the correct functioning of this implementation. 

In the second part, I applied the code to the solution of the Einstein constraints
in a periodic cell containing a single black hole at its center. The solution 
represents a spatial slice of an infinite, cubic black-hole lattice, and can be
used as initial data to study the evolution of such a system in the case of 
non-zero initial mean curvature. This result is of primary importance for the
study of black-hole lattices as models of inhomogeneous universes, as
their evolution has been obtained so far, in full General Relativity, only in the case of 
zero initial mean curvature~\cite{Bentivegna:2012ei}, where initial data 
could be constructed analytically.

This work lays the basis for the study of expanding black-hole lattices,
which is presented elsewhere~\cite{Bentivegna:2013jta}.
Further improvements on the solver, such as the implementation of higher-order
schemes or colored stencils, may also be the subject of future work.

\section*{Acknowledgements}
I acknowledge Lars Andersson, Miko\l{}aj Korzy\'nski, Ian Hinder, Bruno Mundim, Oliver Rinne and Erik
Schnetter for many clarifications and suggestions. This work was supported by a Marie Curie International 
Reintegration Grant (PIRG05-GA-2009-249290); computations were carried out on the MPI-GP Damiana and Datura 
clusters, as well as on SuperMUC at the Leibniz-Rechenzentrum in Munich.

\appendix
\section{Helmholtz's equation with a constant source}
\label{app:csource}
In this appendix, I explain the three statements quoted in~\ref{sec:wellpos}, and valid for 
Helmholtz's equation:
\beq
\Delta f + c f + d = 0
\eeq
with a constant coefficient $c$ and a constant source term $d$.
I will restrict myself to the one-dimensional case; the generalization
is obvious.
\bit
\item If $d$ is constant, the only solutions are Fourier modes
with frequency zero or $\sqrt{c}$. This can be proven by writing
a Taylor expansion for $f$:
\beq
f=\sum_{n=0}^{\infty} a_n x^n
\eeq
and substituting it into Helmholtz's equation, obtaining:
\beq
\sum_{n=2}^{\infty} a_n n (n-1) x^{n-2} + c \sum_{n=0}^{\infty} a_n x^n + d = 0
\eeq
which, for constant $d$, is only satisfied if:
\bea
c a_0 + d + 2 a_2 = 0\\
c a_1 + 6 a_3 = 0 \\
c a_n + (n+1)(n+2) a_{n+2} = 0
\eea
One is free to choose $a_0$ and $a_1$, and the other coefficients will
be given by:
\bea
a_{2p} = (-1)^p (d-ca_0) \frac{c^{p-1}}{(2p)!} \qquad p \geq 1 \\
a_{2p+1} = (-1)^p \frac{c^p a_1}{(2p+1)!} \qquad p \geq 1
\eea
Substituting back into $f$, and after a few manipulations, one
finally obtains:
\beq
f(x) = -\frac{d}{c} + \left(\frac{d}{c}+a_0 \right) \cos(\sqrt{c}x) + \frac{1}{\sqrt{c}} \sin(\sqrt{c}x)
\eeq
\item The amplitude of the zero-frequency mode of $f$ is fixed and equal
to $-d/c$; the other modes can have arbitrary amplitude.
This can be shown by working out the conditions under which $h=kf+J$ is
a solution, i.e.:
\beq
\Delta h + c h + d = 0
\eeq
One recovers the familiar relation:
\beq
k(J)=\frac{cJ}{d}+1
\eeq
so that a family of solutions is given by:
\bea
h&=&\left(\frac{cJ}{d}+1\right) f + J \\
&=& -\frac{d}{c} + \left(\frac{cJ}{d}+1\right) \left ( \left(\frac{d}{c}+a_0 \right) \cos(\sqrt{c}x) + \frac{1}{\sqrt{c}} \sin(\sqrt{c}x) \right ) \nonumber
\eea
with arbitrary $J$.
\eit

\section*{References}
\bibliographystyle{iopart-num}
\bibliography{references}

\end{document}